\RequirePackage{ifpdf}
\ifpdf 
\documentclass[pdftex]{sigma}
\else
\documentclass{sigma}
\fi

\def\Z{\mathbb Z}

\def\p{\partial}
\def\r{\rangle}
\def\l{\langle}
\def\p{\partial}

\numberwithin{equation}{section}

\begin{document}
\allowdisplaybreaks

\renewcommand{\PaperNumber}{002}

\FirstPageHeading

\ShortArticleName{$E$-Orbit Functions}

\ArticleName{$\boldsymbol{E}$-Orbit Functions}

\Author{Anatoliy U. KLIMYK~$^\dag$ and Jiri PATERA~$^\ddag$}
\AuthorNameForHeading{A.U. Klimyk and J. Patera}

\Address{$^\dag$~Bogolyubov Institute for Theoretical Physics,
       14-b Metrologichna Str., Kyiv 03680, Ukraine}
\EmailD{\href{mailto:aklimyk@bitp.kiev.ua}{aklimyk@bitp.kiev.ua}}

\Address{$^\ddag$~Centre de Recherches Math\'ematiques,
         Universit\'e de Montr\'eal,\\
$\phantom{^\ddag}$~C.P.6128-Centre ville,
         Montr\'eal, H3C\,3J7, Qu\'ebec, Canada}
\EmailD{\href{mailto:patera@crm.umontreal.ca}{patera@crm.umontreal.ca}}

\ArticleDates{Received December 20, 2007; Published on-line January 05, 2008}

\Abstract{We review and further develop the theory of $E$-orbit
functions. They are functions on the Euclidean space $E_n$
obtained from the multivariate exponential function by symmetrization
by means of an even part $W_{e}$ of a Weyl group $W$, corresponding to a~Coxeter--Dynkin diagram. Properties of such functions are described.
They are closely related to symmetric and antisymmetric orbit
functions which are received from exponential functions by symmetrization
and antisymmetrization procedure by means of a Weyl group~$W$.
The $E$-orbit functions, determined by integral parameters, are invariant
with respect to even part $W^{\rm af\/f}_{e}$ of the
af\/f\/ine Weyl group corresponding to $W$.
The $E$-orbit functions determine a
symmetrized Fourier transform, where these functions serve as a kernel
of the transform. They also determine a transform on
a f\/inite set of points of the fundamental domain $F^{e}$
of the group $W^{\rm af\/f}_{e}$ (the discrete $E$-orbit
function transform).}


\Classification{33-02; 33E99; 42B99; 42C15; 58C40}

\section{Introduction}
In \cite{P04} and \cite{P-SIG-05} it was initiated a study of orbit
functions which are closely related to f\/inite groups $W$ of
geometric symmetries generated by ref\/lection transformations $r_i$
(that is, such that $r_i^2=1$), $i=1,2,\dots ,n$, of the
$n$-dimensional Euclidean space $E_n$ with respect to
$(n{-}1)$-dimensional subspaces containing the origin. In fact,
orbit functions are multivariate exponential functions symmetrized
or antisymmetrized by means of a Weyl group $W$ of a semisimple Lie
algebra or symmetrized by means of its subgroup $W_e$ consisting of
even elements of $W$. Orbit functions on the 2-dimensional Euclidean
space $E_2$, invariant or anti-invariant with respect to $W$,  were
considered in detail in \cite{AP, Pat-Z-1, Pat-Z-2, PZ-06}. A
detailed description of symmetric and antisymmetric orbit functions
on any Euclidean space $E_n$ is given in \cite{KP06} and in
\cite{KP07}. Orbit functions on $E_2$, invariant with respect to
$W_e$ are studied in \cite{Kash-P}.

The important peculiarity of orbit functions is a possibility of
their discretization \cite{MP06}, which is made by using the results
of paper \cite{MP87}. This possibility makes orbit functions useful
for applications. In particular, on this way multivariate discrete
Fourier transform and multivariate discrete sine and cosine
transforms are received (see \cite{KP07, KP-JP,KP-JMP}).

 In order to
obtain a symmetric orbit function we take a point $\lambda \in E_n$
and act upon $\lambda$ by all elements of the group $W$. If
$O(\lambda)$ is the $W$-orbit of the point $\lambda$, that is, the
set of all dif\/ferent points of the form $w\lambda$, $w\in W$, then
the symmetric orbit function, determined by $\lambda$, coincides
with
\[
\phi_\lambda(x)=\sum_{\mu\in O(\lambda)} e^{2\pi{\rm i}\langle
\mu,x\rangle},
\]
where $\langle \mu,x\rangle$ is the scalar product on $E_n$. These
functions are invariant with respect to the action by elements of
the group $W$: $\phi_\lambda(wx)=\phi_\lambda(x)$, $w\in W$.
If $\lambda$ is an integral point of $E_n$, then $\phi_\lambda(x)$
is invariant with respect to the af\/f\/ine Weyl group $W^{\rm af\/f}$
corresponding to $W$.

Symmetry is the main property of symmetric orbit functions which
make them useful in applications. Being a modif\/ication of monomial
symmetric functions, they are directly related to the theory of
symmetric (Laurent) polynomials \cite{Mac1, Mac2, Mac3, KVl} (see
Section~11 in \cite{KP06}).

Symmetric orbit functions $\phi_\lambda(x)$ for integral
$\lambda$ are closely related to the representation
theory of compact groups $G$. In particular, they were ef\/fectively
used for dif\/ferent calculations in representation theory \cite{MP84,
MMP85, MMP86, PS, GP}. They are constituents of traces (characters)
of irreducible unitary representations of $G$.

Antisymmetric orbit functions are
given by
\[
\varphi_\lambda(x)=\sum_{w\in W} (\det w) e^{2\pi{\rm i}\langle
w\lambda,x\rangle},\qquad x\in E_n,
\]
where $\lambda$ is an element, which does not lie on a wall of a
Weyl chamber, and $\det w$ is a determinant of the transformation
$w$ (it is equal to 1 or $-1$, depending on either $w$ is a product
of even or odd number of ref\/lections). The orbit functions
$\varphi_\lambda$ have many properties that the symmetric orbit
functions $\phi_\lambda$ do. However, antisymmetry leads to some new
properties which are useful for applications \cite{PZ-06}. For
integral $\lambda$, antisymmetric orbit functions are closely
related to characters of irreducible representations of the
corresponding compact Lie group $G$. Namely, the character~$\chi_\lambda$ of the irreducible representation $T_\lambda$,
$\lambda\in P_+$, coincides with
$\varphi_{\lambda+\rho}/\varphi_{\rho}$, where $\rho$ is the
half-sum of positive roots related to the Weyl group~$W$.

A symmetric (antisymmetric) orbit function is the
exponential function $e^{2\pi{\rm i}\langle \lambda,x\rangle}$ on $E_n$
symmetrized (antisymmetrized) by means of the group $W$. For each
transformation group $W$, the symmetric (antisymmetric) orbit functions,
characterized by integral $\lambda$, form a
complete basis in the space of symmetric (antisymmetric)
with respect to $W$ polynomials in $e^{2\pi{\rm i}x_j}$, $j=1,2,\ldots,n$,
or an orthogonal basis in the Hilbert space
obtained by closing this space of polynomials with respect to an
appropriate scalar product.

Orbit functions $\phi_\lambda(x)$ (or $\varphi_\lambda(x)$), when $\lambda$
runs over integral elements, determine the so-called
symmetric (antisymmetric) orbit function transform,
which is a symmerization (antisymmetrization) of the usual Fourier series
expansion on $E_n$.
If $\lambda$ runs over the dominant Weyl chamber in
the space $E_n$, then $\phi_\lambda(x)$
(or $\varphi_\lambda(x)$) determine a symmetric (antisymmetric) orbit
function transform, which is a symmetrization (antisymmetrization) of the
usual continuous Fourier expansion in $E_n$ (that is, of the
Fourier integral).

The Fourier transform on $\mathbb{R}$ leads to
the discrete Fourier transform on grids. In the same way
the symmetric and
antisymmetric orbit function transforms lead to
discrete analogues of these transforms (which are  generalizations of
the discrete cosine and sine transforms, respectively,~\cite{R}).
These discrete transforms are useful in
many things related to discretization (see~\cite{AP, Pat-Z-1,
Pat-Z-2}). Construction of the discrete orbit function transforms
are fulf\/illed by means of the results of paper \cite{MP87}.

Symmetric orbit functions are a generalization of the cosine function,
whereas antisymmetric orbit functions are a generalization of the sine
function. There appears a natural question: What is a generalization of
the exponential function of one variable? This generalization is given by
orbit functions symmetric with respect to the subgroup $W_e$ of even
elements in $W$.

Our goal in this paper is to give in a full generality the theory
of orbit functions symmetric with respect to the group
$W_e$. We shall call these functions
$E$-orbit functions, since they are an analogue of the well-known
exponential function (since symmetric and antisymmetric
orbit functions $\phi_\lambda$ and $\varphi_\lambda$ are
generalizations of the cosine and sine functions, they are often called as
$C$-functions and $S$-functions, respectively). This paper is a natural
continuation of our papers \cite{KP06} and \cite{KP07}.
Under our exposition we use the results of
papers \cite{P-SIG-05} and \cite{MP06}, where $E$-orbit functions are
def\/ined.

Roughly speaking, $E$-orbit functions are related to symmetric and
antisymmetric orbit functions in the same way as the exponential
function in one variable is related to the sine and cosine functions.

$E$-orbit functions are symmetric with respect to the subgroup $W_e$
of the Weyl group $W$, that is, $E_\lambda(wx)=E_\lambda(x)$ for any
$w\in W_e$. The subgroup $W_e$ is of index 2 in $W$, that is
$|W/W_e|=2$, where $|X|$ denote a number of elements in the
corresponding set $X$. This means that $E$-orbit functions are
determined not only for $\lambda$ from dominant Weyl chamber $D_+$
(as in the case of symmetric orbit functions), but also for elements
from the set $r_iD_+$, where $r_i$ is a f\/ixed ref\/lection from $W$.

If $\lambda$ is an {\it integral} element,
then the corresponding
$E$-orbit function $E_\lambda(x)$ is symmetric also with respect
to elements of the af\/f\/ine Weyl group $W_e^{\rm af\/f}$,
corresponding to the
group $W_e$ (in fact, the group $W_e^{\rm af\/f}$ consists of
even elements of the whole af\/f\/ine Weyl group $W^{\rm af\/f}$,
corresponding to the Weyl group $W$).

Symmetry with respect to $W_e^{\rm af\/f}$ is a main property of
$E$-orbit functions with integral $\lambda$. Because of this
symmetry, it is enough to determine $E$-orbit functions only on the
fundamental domain $F(W_e^{\rm af\/f})$ of the group $W_e^{\rm af\/f}$
(if $\lambda$ is integral). This fundamental domain consists of two
fundamental domains of the whole af\/f\/ine Weyl group $W^{\rm af\/f}$.

When the group $W$ is a direct product of its
subgroups, say $W=W_1\times W_2$, then $W_e=(W_1)_e\times (W_2)_e$.
In this case $E$-orbit functions of $W_e$ are products
of $E$-orbit
functions of $(W_1)_e$ and $(W_2)_e$. Hence it suf\/f\/ices to carry out our
considerations for groups $W_e$ which cannot be represented as a
product of its subgroups (that is, for such $W$ for which a
corresponding Coxeter--Dynkin diagram is connected).

$E$-orbit functions with integral $\lambda$ determine the so-called
$E$-orbit function transform on the fundamental domain
$F(W_e^{\rm af\/f})$ of the group $W_e^{\rm af\/f}$. It is
an expansion of functions on $F(W_e^{\rm af\/f})$ in these
$E$-orbit functions. $E$-orbit functions $E_\lambda(x)$ with $\lambda$
from $E_n$ determine $E$-orbit function transform on the fundamental
domain $F(W_e)$ of the group $W_e$. It is an analogue of the
usual integral Fourier transform.

$E$-orbit functions determine also the discrete $E$-orbit function
transforms. They are transforms on grids of the domain
$F(W_e^{\rm af\/f})$. These transforms are an analogue of the
usual discrete Fourier transforms.

We need for our exposition a general information on Weyl groups, af\/f\/ine Weyl
groups and root systems. We have given this
information in \cite{KP06} and \cite{KP07}. In order to make this paper
self-contained we repeat shortly a part of that information in Section \ref{weyl}.
In this section we also describe even Weyl groups and af\/f\/ine
even Weyl groups.

In Section \ref{s-orbi} we def\/ine and study $W_e$-orbits. It is shown how
$W_e$-orbits are related to $W$-orbits. Each $W$-orbit is
a $W_e$-orbit or consists of two $W_e$-orbit.
To each $W_e$-orbit there corresponds an $E$-orbit function.
$W_e$-orbits are parametrized by elements of even dominant
Weyl chamber $D_+^e$.

We describe in Section \ref{s-orbi} all $W_e$-orbits for $A_2$ and $C_2$. A big
class of $W_e$-orbits for $G_2$ is also given. All $W_e$-orbits of
$A_3$, $B_3$ and $C_3$ are derived. It is proposed to describe
points of $W_e$-orbits of $A_n$, $B_n$, $C_n$ and $D_n$ by means of
orthogonal coordinates. Then elements of the group $W_e$ and
$W_e$-orbits are described in a simple way. Section \ref{s-orbi} contains also
a description of fundamental domains for the groups $W_e^{\rm
af\/f}(A_n)$, $W_e^{\rm af\/f}(B_n)$, $W_e^{\rm af\/f}(C_n)$, and
$W_e^{\rm af\/f}(D_n)$.

Section \ref{Orb} is devoted to description of $E$-orbit
functions. $E$-orbit functions, corresponding to
Coxeter--Dynkin diagrams, containing only two nodes, are
given in an explicit form. In this section we also give explicit
formulas for $E$-orbit functions, corresponding to
the cases $A_n$, $B_n$, $C_n$ and $D_n$, in
the corresponding orthogonal coordinate systems.

In Section~\ref{sec5}, properties of $E$-orbit functions are derived. If
$\lambda$ is integral, then a main property of the $E$-orbit
function $E_\lambda(x)$ is an invariance with respect to the af\/f\/ine
even Weyl group $W_e^{\rm af\/f}$. Relation of $E$-orbit functions to
symmetric and antisymmetric orbit functions (that is, to $W$-orbit
functions) is described.

$E$-orbit functions $E_\lambda(x)$ with integral
$\lambda$ are orthogonal on the closure of the
fundamental domain of the group $W_e^{\rm af\/f}$.
This orthogonality is given in Section~\ref{sec5}.

$E$-orbit functions are solutions of the corresponding Laplace
equation. This description is exposed in Section~\ref{sec5}.
$E$-orbit functions also are solutions of some other
dif\/ferential equations.

In Section~\ref{operat} we consider expansions of products of
$E$-orbit functions into a sum of
$E$-orbit functions. These expansions
are closely related to properties of $W_e$-orbits,
namely, the expansions are reduced to decomposition of
products of $W_e$-orbits into separate $W_e$-orbits.
In general, it is a complicated problem (especially, when
multiple $W_e$-orbits appear in the decomposition).
Several propositions, describing decomposition of
products of $W_e$-orbits, are given.
Many examp\-les for expansions in the case of Coxeter--Dynkin diagrams
$A_2$, $C_2$, and $G_2$ are considered.

Section \ref{sec7} is devoted to expansion of
$W_e$-orbit functions into a sum of
$W_e'$-orbit functions, where $W_e'$ is an even Weyl subgroup of
the even Weyl group $W_e$. The cases of restriction of $A_n$
to $A_{n-1}$, of $B_n$ to $B_{n-1}$, of $C_n$ to $C_{n-1}$,
and of $D_n$ to $D_{n-1}$ are described in detail.

In Section \ref{sec8} we expose $E$-orbit function transforms.
There are two types of such transforms. The f\/irst one is an analogue of
the expansion into Fourier series (it is an expansion
on the fundamental domain of the group $W^{\rm af\/f}_e$)
and the second one is an analogue of the Fourier integral transform
(it is an expansion on the even dominant Weyl chamber).

In Section \ref{sec9} a description of
a $W_e$-generalization of the multi-dimensional f\/inite
Fourier transforms is given. This generalization is connected with
grids on the corresponding fundamental domains for the af\/f\/ine even
Weyl groups $W_e^{\rm af\/f}$. These grids are determined by
a positive integer~$M$. To each such an integer there
corresponds a grid on the fundamental domain. Examples of
such grids for $A_2$, $C_2$ and $G_2$ are given.

Section \ref{sec10} is devoted to exposition of $W_e$-symmetric
functions, which are symmetric analogues of
special functions of mathematical physics or orthogonal polynomials.
In particular, we f\/ind eigenfunctions of the
$W_e$-orbit function transforms. These eigenfunctions are connected with
classical Hermite polynomials.

\section{Root systems and Weyl groups}
\label{weyl}

\subsection[Coxeter-Dynkin diagrams and simple roots]{Coxeter--Dynkin diagrams and simple roots}\label{DD}
We need f\/inite transformation groups $W$, acting
on the $n$-dimensional Euclidean space $E_n$, which are
generated by ref\/lections $r_i$, $i=1,2,\dots ,n$
(that is, $r^2_i=1$); the theory of such groups see, for
example, in \cite{Kane} and \cite{Hem-1}. We are interested
in those groups $W$ which are
Weyl groups of semisimple Lie groups (semisimple Lie algebras). It
is well-known that such Weyl groups together with the corresponding
systems of ref\/lections $r_i$, $i=1,2,\dots ,n$, are determined by
Coxeter--Dynkin diagrams. There are 4 series of simple Lie algebras and 5
separate simple Lie algebras, which uniquely determine their Weyl
groups $W$. These algebras are denoted as
\[
A_n\ (n\geq1),\ B_n\ (n\geq3),\ C_n\ (n\geq2),\ D_n\ (n\geq4), \
E_6,\ E_7,\ E_8,\ F_4,\ G_2 .
\]
To these simple Lie algebras there correspond connected Coxeter--Dynkin
diagrams.

To semisimple Lie algebras (they are direct sums of simple
Lie subalgebras) there correspond Coxeter--Dynkin diagrams, which
consist of connected parts, corresponding to simple Lie subalgebras;
these parts are not connected with each other (a description of the
correspondence between simple Lie algebras and Coxeter--Dynkin
diagrams see, for example, in \cite{Hem-2}). Thus, we describe only
Coxeter--Dynkin diagrams, corresponding to simple Lie algebras. They
are of the form

{\centering

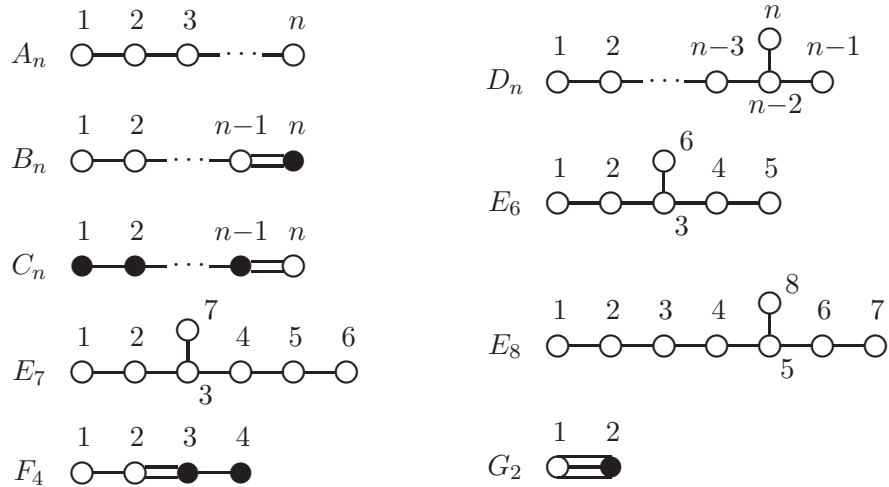
\parbox{.6\linewidth}{\setlength{\unitlength}{2pt}
\def\kr{\circle{4}}
\def\cr{\circle*{4}}
\thicklines
\begin{picture}(180,90)

\put(0,80){\makebox(0,0){${A_n}$}}
\put(10,80){\kr} \put(9,85){1}  \put(20,80){\kr} \put(19,85){2}
\put(30,80){\kr} \put(29,85){3}   \put(50,80){\kr} \put(49,85){$n$}
\put(12,80){\line(1,0){6}}
\put(22,80){\line(1,0){6}}
\put(32,80){\line(1,0){4}}  \put(37,80){$\dots$}
\put(44,80){\line(1,0){4}}

\put(0,60){\makebox(0,0){${B_n}$}}   
\put(10,60){\kr} \put(9,65){1}   \put(20,60){\kr} \put(19,65){2}
\put(40,60){\kr}  \put(35,65){$n{-}1$}  \put(50,60){\cr*} \put(49,65){$n$}
\put(12,60){\line(1,0){6}}
\put(22,60){\line(1,0){4}} \put(27,60){$\dots$}
\put(34,60){\line(1,0){4}}
\put(42,61){\line(1,0){6}}
\put(42,59){\line(1,0){6}}

\put(0,40){\makebox(0,0){${C_n}$}}   
\put(10,40){\cr*}  \put(9,45){1}   \put(20,40){\cr*} \put(19,45){2}
\put(40,40){\cr*}  \put(35,45){$n{-}1$}  \put(50,40){\kr}  \put(49,45){$n$}
\put(12,40){\line(1,0){6}}
\put(22,40){\line(1,0){4}}  \put(27,40){$\dots$}
\put(34,40){\line(1,0){4}}
\put(42,41){\line(1,0){6}}
\put(42,39){\line(1,0){6}}

\put(90,75){\makebox(0,0){${D_n}$}}    
\put(100,75){\kr}  \put(99,80){1}  \put(110,75){\kr} \put(109,80){2}
  \put(130,75){\kr} \put(125,80){$n{-}3$}
\put(140,75){\kr}  \put(136,69){$n{-}2$}  \put(150,75){\kr}  \put(147,80){$n{-}1$}
\put(140,83){\kr}  \put(139,87){$n$}
\put(102,75){\line(1,0){6}}
\put(112,75){\line(1,0){4}}   \put(117,75){$\dots$}
\put(124,75){\line(1,0){4}}
\put(132,75){\line(1,0){6}}
\put(142,75){\line(1,0){6}}
\put(140,77){\line(0,1){4}}

\put(90,52){\makebox(0,0){${E_6}$}}     
\put(100,52){\kr} \put(99,57){1}  \put(110,52){\kr}   \put(109,57){2}
\put(120,52){\kr} \put(122,46){3}   \put(130,52){\kr}  \put(129,57){4}
\put(140,52){\kr}  \put(139,57){5}  \put(120,60){\kr} \put(123,62){6}
\put(102,52){\line(1,0){6}}
\put(112,52){\line(1,0){6}}
\put(122,52){\line(1,0){6}}
\put(132,52){\line(1,0){6}}
\put(120,54){\line(0,1){4}}

\put(0,20){\makebox(0,0){${E_7}$}}   
\put(10,20){\kr} \put(9,25){1}  \put(20,20){\kr}  \put(19,25){2}
\put(30,20){\kr}  \put(32,14){3}  \put(40,20){\kr}  \put(39,25){4}
\put(50,20){\kr}  \put(49,25){5}  \put(60,20){\kr}  \put(59,25){6}
\put(30,28){\kr}   \put(33,30){7}
\put(12,20){\line(1,0){6}}
\put(22,20){\line(1,0){6}}
\put(32,20){\line(1,0){6}}
\put(42,20){\line(1,0){6}}
\put(52,20){\line(1,0){6}}
\put(30,22){\line(0,1){4}}

\put(90,25){\makebox(0,0){${E_8}$}}
\put(100,25){\kr}  \put(99,30){1}  \put(110,25){\kr}  \put(109,30){2}
\put(120,25){\kr}  \put(119,30){3}  \put(130,25){\kr}  \put(129,30){4}
\put(140,25){\kr}  \put(142,19){5}  \put(150,25){\kr}  \put(149,30){6}
\put(160,25){\kr}   \put(159,30){7}  \put(140,33){\kr}  \put(143,35){8}
\put(102,25){\line(1,0){6}}
\put(112,25){\line(1,0){6}}
\put(122,25){\line(1,0){6}}
\put(132,25){\line(1,0){6}}
\put(142,25){\line(1,0){6}}
\put(152,25){\line(1,0){6}}
\put(140,27){\line(0,1){4}}

\put(0,1){\makebox(0,0){${F_4}$}}   
\put(10,1){\kr}   \put(9,6){1}   \put(20,1){\kr}  \put(19,6){2}
\put(30,1){\cr*}  \put(29,6){3}   \put(40,1){\cr*} \put(39,6){4}
\put(12,1){\line(1,0){6}}
\put(22,2){\line(1,0){6}}
\put(22,0){\line(1,0){6}}
\put(32,1){\line(1,0){6}}
\put(90,2){\makebox(0,0){${G_2}$}}
\put(100,2){\kr} \put(99,7){1}   \put(110,2){\cr*}  \put(109,7){2}
\put(102,2){\line(1,0){6}}
\put(100,4){\line(1,0){10}}
\put(100,0){\line(1,0){10}}

\end{picture}}

}

\bigskip

A diagram determines a certain non-orthogonal basis
$\{\alpha_1,\alpha_2,\dots,\alpha_n\}$ in the Euclidean spa\-ce~$E_n$.
Each node of a diagram is associated with a basis vector $\alpha_k$, called a
{\it simple root}. A~direct link between two nodes indicates that
the corresponding basis vectors are not orthogonal. Conversely, an
absence of a direct link between nodes implies orthogonality of the
corresponding vectors. Single, double, and triple links indicate that
the relative angles between the two simple roots are $2\pi/3$,
$3\pi/4$, $5\pi/6$, respectively. There can be only two cases: all
simple roots are of the same length or there are only two dif\/ferent
lengths of simple roots. In the f\/irst case all simple roots are
denoted by white nodes. In the case of two lengths, shorter roots
are denoted by black nodes and longer ones by white nodes. Lengths
of roots are determined uniquely up to a common constant. For the
cases $B_n$, $C_n,$ and $F_4$, the squared longer root length is
double the squared shorter root length. For $G_2$, the squared
longer root length is triple the squared shorter root length.
Simple roots of the same length
are orthogonal to each other or an angle between them is $2\pi/3$.
A number of simple roots is called a {\it rank} of the corresponding
Lie algebra.

To each Coxeter--Dynkin diagram there corresponds a Cartan matrix
$M$, consisting of the entries
\begin{equation}\label{Mmatrix}
M_{jk}=\frac{2\l\alpha_j ,\alpha_k\r}
            {\l\alpha_k ,\alpha_k\r}\,,\qquad
            j,k\in\{1,2,\dots,n\},
\end{equation}
where $\l x, y\r$ denotes the scalar product of $x,y\in E_n$.
Cartan matrices of simple Lie algebras are given in many places
(see, for example, \cite{BMP}). For ranks $2$
and $3$ they are of the form:
\[
 A_2 :\ \left(
 \begin{array}{cc}
 2&-1\\ -1&2\
 \end{array} \right) ,\qquad
C_2 :\  \left(
 \begin{array}{cc}
 2&-1\\ -2&2\
 \end{array} \right) ,\qquad
 G_2 :\
 \left(
 \begin{array}{cc}
 2&-3\\ -1&2\
 \end{array} \right) ,
 \]  \[
 A_3 :\
\left(
 \begin{array}{ccc}
 2&-1&0\\ -1&2&-1\\ 0&-1&2
 \end{array} \right) ,\qquad
 B_3 :\
\left(
 \begin{array}{ccc}
 2&-1&0\\ -1&2&-2\\ 0&-1&2
 \end{array} \right) ,\qquad
 C_3 :\
\left(
 \begin{array}{ccc}
 2&-1&0\\ -1&2&-1\\ 0&-2&2
 \end{array} \right) .
\]

Lengths of the basis vectors $\alpha_i$ are f\/ixed by the
corresponding Coxeter--Dynkin diagram up to a constant. We adopt
the standard choice in the Lie theory, namely
\[
\l\alpha ,\alpha\r=2
\]
for all simple roots of $A_n$, $D_n$, $E_6$, $E_7$, $E_8$ and
for the longer simple roots of $B_n$, $C_n$, $F_4$,
$G_2$.

\subsection{Weyl group and even Weyl group}\label{reflect}
A Coxeter--Dynkin diagram determines uniquely the corresponding
transformation group $W$ of the Euclidean space $E_n$,
generated by ref\/lections $r_i$,
$i=1,2,\dots ,n$. These ref\/lections correspond to simple roots
$\alpha_i$, $i=1,2,\ldots,n$.
Namely, the transformation $r_i$ corresponds
to the simple root~$\alpha_i$ and is the ref\/lection with respect to
$(n-1)$-dimensional linear subspace (hyperplane) of~$E_n$
(containing the origin), orthogonal to $\alpha_i$.
Such ref\/lections are given by the formula
\begin{equation}\label{reflection}
r_ix=x-\frac{2\l x, \alpha_i\r}{\l\alpha_i, \alpha_i\r}\alpha_i,
\qquad i= 1,2,\ldots,n,\qquad x\in E_n .
\end{equation}
Each ref\/lection $r_i$ can be thought as attached to the
$i$-th node of the corresponding diagram.

A f\/inite group $W$, generated by the ref\/lections $r_i$,
$i=1,2,\dots ,n$, is called a {\it Weyl group}, corresponding to a
given Coxeter--Dynkin diagram. If a Weyl group $W$ corresponds to a
Coxeter--Dynkin diagram of a simple Lie algebra $L$, then this Weyl
group is often denoted by $W(L)$. Properties of Weyl groups are well
known (see \cite{Kane} and \cite{Hem-1}).
The orders (numbers of elements) of Weyl groups are given by
the formulas
\begin{alignat}{3}
&|W(A_n)|=(n+1)!,\quad &&|W(B_n)|=|W(C_n)|=2^nn!,\quad
                              &&|W(D_n)|=2^{n-1}n!,\notag\\
&|W(E_6)|=51\,840 ,\quad &&|W(E_7)|=2\ 903\,040,\quad
                                    &&|W(E_8)|=696\,729\ 600,\\
&|W(F_4)|=1\,152,\quad &&|W(G_2)|=12. &&\notag
\end{alignat}
In particular,
\[
|W(A_2)|=6,\qquad  |W(C_2)|=8,\qquad |W(A_3)|=24,\qquad |W(B_3)|=|W(C_3)|=48.
\]

Elements of the Weyl groups are linear transformations of the
Euclidean space $E_n$. To these transformations there
correspond in an orthonormal basis of $E_n$ the corresponding
$n\times n$ matrices. Since these transformations are orthogonal,
then determinants of these matrices are $+1$ or $-1$. We say that
a transformation $w\in W$ is even if $\det w=1$ and odd if
$\det w=-1$. Clearly, for ref\/lections $r_\alpha$ corresponding
to roots $\alpha$ we have $\det r_\alpha=-1$.
If $w\in W$ is a product of even (odd) number of ref\/lections,
then $\det w=1$ ($\det w=-1$).

The set of all elements $w\in W$ with $\det w=1$ constitute a
subgroup of $W$ which will be denoted by $W_e$. One says that it is
a subgroup of even elements of $W$. Moreover, $W_e$ is a normal
subgroup of $W$, that is, $wW_ew^{-1}=W_e$ for any $w\in W$. The
group $W_e$ is a basic group for def\/inition of $E$-orbit functions.

Elements of $W$, which do not belong to $W_e$, are called
odd. The number of even elements in~$W$ is equal to the number of
odd elements, that is, $|W/W_e|=2$. In particular, we have
\begin{gather*}
|W_e(A_2)|=3,\quad  |W_e(C_2)|=4,\quad |W_e(G_2)|=6,\quad
|W_e(A_3)|=12,\\ |W_e(B_3)|= |W_e(C_3)|=24.
\end{gather*}
The Weyl groups $W(B_3)$ and $W(C_3)$ are isomorphic. For
this reason, the even Weyl groups $W_e(B_3)$ and $W_e(C_3)$
are isomorphic.

Elements of $W_e$ are orthogonal transformations of $E_n$ with
a unit determinant. Therefore,  $W_e$ is a f\/inite subgroup of
the rotation group $SO(n)$ of $E_n$. That is, the group $W_e$
consists of rotations of the space $E_n$. In particular, for
rank 2 case the even Weyl groups consist of rotations of a plane:
\begin{gather*}
 W_e(A_2)=\{1, {\rm rot}(2\pi/3),{\rm rot}(4\pi/3)\} ,
\\
 W_e(C_2)=\{1, {\rm rot}(\pi/2),{\rm rot}(\pi), {\rm rot}(3\pi/2)\} ,
\\
 W_e(G_2)=\{1, {\rm rot}(k\pi/3),\ k=1,2,3,4,5 \} .
\end{gather*}

\subsection{Root and weight lattices}\label{roots}
A Coxeter--Dynkin diagram determines a system of simple roots in the
Euclidean space $E_n$. Acting by elements of the Weyl group $W$ upon
simple roots we obtain a f\/inite system of vectors, which is
invariant with respect to $W$. A set of all these vectors is called
a {\it system of roots} associated with a given Coxeter--Dynkin
diagram. It is denoted by $R$.

It is proved (see, for example, \cite{Hem-2}) that roots of $R$ are
linear combinations of simple roots with integral coef\/f\/icients.
Moreover, there exist no roots, which are linear combinations of
$\alpha_i$, $i=1,2,\dots ,n$, both with positive and negative
coef\/f\/icients. Therefore, the set of roots $R$ can be represented as
a union $R=R_+\cup R_-$, where $R_+$ (respectively $R_-$) is the set
of roots which are linear combinations of simple roots with positive
(negative) coef\/f\/icients. The set $R_+$ (the set $R_-$) is called a
{\it set of positive (negative) roots}.

As mentioned above, a set of roots $R$ is invariant under the action
of elements of the Weyl group $W(R)$. However, $wR_+\ne R_+$ if $w$
is not a trivial element of $W$.

Let $X_\alpha$ be the $(n-1)$-dimensional linear subspace
(hyperplane) of $E_n$ (containing the origin) which is orthogonal to
the root $\alpha$.
The hyperplane $X_\alpha$ consists of all points $x\in E_n$ such
that $\langle x,\alpha \rangle=0$.
Clearly, $X_\alpha=X_{-\alpha}$. The set of
ref\/lections with respect to $X_\alpha$, $\alpha\in R_+$, coincides
with the set of all ref\/lections of the corresponding Weyl group $W$.

The subspaces $X_\alpha$, $\alpha\in R_+$, split the Euclidean space
$E_n$ into connected parts which are called {\it Weyl chambers}.
(We assume that boundaries of Weyl chambers belong to the
corresponding chambers. Therefore, Weyl chambers can have common
points; they belong to boundaries of the corresponding chambers.)
A number of
Weyl chambers coincides with the number of elements of the Weyl
group $W$. Elements of the Weyl group permute Weyl chambers. A part
of a Weyl chamber, which belongs to some hyperplane $X_\alpha$ is
called a {\it wall} of this Weyl chamber. If for some element $x$ of
a Weyl chamber we have $\langle x,\alpha \rangle=0$ for some root
$\alpha$, then this point belongs to a wall. The Weyl chamber
consisting of points $x$ such that
\[
\langle x,\alpha_i \rangle\ge 0, \qquad i=1,2,\dots,n,
\]
is called the {\it dominant Weyl chamber}. It is denoted by $D_+$.
Elements of $D_+$ are called {\it dominant}. If
$\langle x,\alpha_i \rangle > 0$, $i=1,2,\dots,n$, then
$x$ is called {\it strictly dominant element}.

If we act by elements of the even subgroup $W_e$ of $W$ upon a f\/ixed
Weyl chamber, then we do not obtain all Weyl chambers. In order to
have transitive action of $W_e$ on parts of the Euclidean space
$E_n$, we have to split $E_n$ into a parts larger than Weyl
chambers. In order to obtain such parts, we take the dominant Weyl
chamber $D_+$ and act upon it by one of the ref\/lections $r_\alpha$,
where $\alpha$ is a root. Denote the union $D_+\cup r_\alpha D_+$
(where each point is taken only once) by $D^e_+$. Then acting upon
$D^e_+$ by elements of $W_e$ we cover the whole Euclidean space
$E_n$. The domains $w D^e_+$, $w\in W_e$, are called {\it even Weyl
chambers}. The procedure of splitting of $E_n$ into even Weyl
chambers is not unique. It depends on the ref\/lection $r_\alpha$
taken for obtaining the f\/irst even Weyl chambers. For dif\/ferent
roots $\alpha$ sets of even Weyl chambers are dif\/ferent. However,
for each f\/ixed root $\alpha$ the corresponding set of even Weyl
chambers is transitive for the group $W_e$. The set $D^e_+\equiv
D^e_+(\alpha)$ is called an {\it even dominant Weyl chamber}.

The set $Q$ of all linear combinations
\[
Q=\left\{ \sum_{i=1}^n a_i\alpha_i \ | \ a_i\in {\mathbb
Z}\right\}\equiv \bigoplus_i {\mathbb Z}\alpha_i
\]
is called a {\it root lattice} corresponding to a given
Coxeter--Dynkin diagram. Its subset{\samepage
\[
Q_+=\left\{ \sum_{i=1}^n a_i\alpha_i \ | \
a_i=0,1,2,\dots\right\}
\]
is called a {\it positive root lattice}.}

 To each root $\alpha\in R$
there corresponds a coroot $\alpha^\vee$ def\/ined by the formula
\[
\alpha^\vee =\frac{2\alpha}{\l \alpha,\alpha\r} .
\]
It is easy to see that $\alpha^{\vee\vee} =\alpha$. The set $Q^\vee$
of all linear combinations
\[
Q^\vee=\left\{ \sum_{i=1}^n a_i\alpha^\vee_i \ | \ a_i\in {\mathbb
Z}\right\}\equiv \bigoplus_i {\mathbb Z}\alpha^\vee_i
\]
is called a {\it coroot lattice} corresponding to a given
Coxeter--Dynkin diagram. The subset
\[
Q^\vee_+=\left\{ \sum_{i=1}^n a_i\alpha^\vee_i \ | \
a_i=0,1,2,\dots\right\}
\]
is called a {\it positive coroot lattice}.

As noted above, the set of simple roots $\alpha_i$, $i=1,2,\dots
,n$, form a basis of the space $E_n$. In addition to the
$\alpha$-basis, it is convenient to introduce the so-called
$\omega$-basis,
$\omega_1,\omega_2,\dots ,\omega_n$ (also called the {\it basis of
fundamental weights}). The two bases are dual to each other in the
following sense:
\begin{equation}\label{kronecker}
\frac{2\l\alpha_j ,\omega_k\r} {\l\alpha_j,\alpha_j\r}\equiv
\l\alpha^\vee_j ,\omega_k\r =\delta_{jk}\,,\qquad
j,k\in\{1,2,\dots,n\}\,.
\end{equation}
The $\omega$-basis (as well as the $\alpha$-basis) is not
orthogonal.

Note that the factor $2/\l\alpha_j,\alpha_j\r$ can take only three
values. Indeed, with the standard normalization of root lengths
(see Subsection~\ref{DD}), we
have
\begin{gather*}
\frac2{\l\alpha_k,\alpha_k\r}
 =1 \quad\text{for roots of}\quad
                A_n,\ D_n,\ E_6,\ E_7,\ E_8,
\\
\frac2{\l\alpha_k,\alpha_k\r}=1  \quad
 \text{for long roots of}\quad B_n,\ C_n,\ F_4,\ G_2,
\\
\frac2{\l\alpha_k,\alpha_k\r}=2
 \quad\text{ for short roots of}\quad B_n,\ C_n,\ F_4,
\\
\frac2{\l\alpha_k,\alpha_k\r}= 3 \quad\text{for short roots
of}\quad G_2 .
\end{gather*}
For this reason, we get
\begin{gather*}
\alpha^\vee_k=\alpha_k  \quad\text{for roots of}\quad
                A_n,\ D_n, \ E_6,\ E_7,\ E_8,
\\
\alpha^\vee_k=\alpha_k  \quad
 \text{for long roots of}\quad B_n,\ C_n,\ F_4,\ G_2,
\\
\alpha^\vee_k=2\alpha_k
 \quad\text{ for short roots of}\quad B_n,\ C_n,\ F_4,
\\
\alpha^\vee_k=3\alpha_k \quad\text{for short roots of}\quad G_2 .
\end{gather*}

The $\alpha$- and $\omega$-bases are related by the Cartan matrix
\eqref{Mmatrix} and by its inverse:
\begin{equation}\label{bases}
\alpha_j=\sum_{k=1}^nM_{jk}\,\omega_k\,,\qquad
\omega_j=\sum_{k=1}^n(M^{-1})_{jk}\,\alpha_k
\end{equation}
For ranks 2 and 3 the inverse Cartan matrices are of the form
\begin{gather*}
 A_2 :\ \frac 13 \left(
 \begin{array}{cc}
 2&1\\ 1&2\
 \end{array} \right) ,\qquad
C_2 :\  \left(
 \begin{array}{cc}
 1&1/2\\ 1&1\
 \end{array} \right) ,\qquad
 G_2 :\
 \left(
 \begin{array}{cc}
 2&3\\ 1&2\
 \end{array} \right) ,
\\
 A_3 :\ \frac 14
\left(
 \begin{array}{ccc}
 3&2&1\\ 2&4&2\\ 1&2&3
 \end{array} \right) ,\qquad
 B_3 :\  \frac 12
\left(
 \begin{array}{ccc}
 2&2&2\\ 2&4&4\\ 1&2&3
 \end{array} \right) ,\qquad
 C_3 :\ \frac 12
\left(
 \begin{array}{ccc}
 2&2&1\\ 2&4&2\\ 2&4&3
 \end{array} \right) .
 \end{gather*}

Later on we need to calculate the scalar product $\l x, y\r$ when
$x$ and $y$ are given by coordinates~$x_i$ and~$y_i$ in
$\omega$-basis. It is given by the formula
\begin{equation}\label{matr}
\l x, y\r
     =\frac12\sum_{j,k=1}^n
                    x_jy_k(M^{-1})_{jk}\l\alpha_k\mid\alpha_k\r
    = xM^{-1}Dy^{T}=xSy^{T},
\end{equation}
where $D$ is the diagonal matrix ${\rm diag}\, (\frac 12 \l
\alpha_1,\alpha_1 \r,\dots ,\frac 12 \l \alpha_n,\alpha_n \r)$.
Matrices $S$, called `quadratic form matrices', are shown in
\cite{BMP} for all connected Coxeter--Dynkin diagrams.

The sets $P$ and $P_+$, def\/ined as
\[
P=\Z\omega_1+\cdots+\Z\omega_n \ \supset\
P_+=\Z^{\geq0}\,\omega_1+\cdots+\Z^{\geq0}\,\omega_n,
\]
are called respectively the {\it weight lattice} and the {\it cone
of dominant weights}. The set $P$ can be characterized as a set of
all $\lambda\in E_n$ such that
\[
\frac{2\langle \alpha_j,\lambda \rangle}{\langle \alpha_j, \alpha_j
\rangle}= \langle \alpha^\vee_j,\lambda \rangle\in \Z
\]
for all simple roots $\alpha_j$. Clearly, $Q\subset P$. Below we
shall need also the set $P^+_+$ of dominant weights of $P_+$, which
do not belong to any Weyl chamber (the set of {\it integral strictly dominant
weights}). Then $\lambda\in P^+_+$ means that
$\langle \lambda,\alpha_i \rangle >0$ for all simple roots
$\alpha_i$. We have
\[
P^+_+=\Z^{{}>0}\omega_1+\Z^{{}>0}\omega_2+\cdots +\Z^{{}>0}\omega_n.
\]

The smallest dominant weights of $P_+$, dif\/ferent from zero, coincide with
the elements $\omega_1,\omega_2$, $\dots,\omega_n$ of the
$\omega$-basis. They are called {\it fundamental weights}.
They are highest weights of fundamental
irreducible representations of the corresponding simple Lie algebra $L$.

Through the paper we often use the
following notation for weights in $\omega$-basis:
\[
z=\sum_{j=1}^n a_j\omega_j=(a_1\ a_2\ \dots\ a_n),\qquad
a_1,\dots,a_n\in\Z\,.
\]
If $x=\sum\limits_{j=1}^n b_j\alpha^\vee_j$, then
\begin{equation}\label{weight}
\l z,x\r =\sum_{j=1}^n a_jb_j.
\end{equation}

\subsection[Highest root and affine root system]{Highest root and af\/f\/ine root system}\label{sec2.4} 

There exists a unique highest (long) root $\xi$ and a unique highest
short root $\xi_s$. The highest (long) root can be written as
\begin{equation}\label{highestroot}
\xi=\sum_{i=1}^nm_i\alpha_i=\sum_{i=1}^n m_i\frac{\langle
\alpha_i,\alpha_i\rangle}{2} \alpha_i^\vee\equiv \sum_{i=1}^n
q_i\alpha_i^\vee   .
\end{equation}
The coef\/f\/icients $m_i$ and $q_i$ can be viewed as attached to the $i$-th
node of the diagram. They are called {\it marks\/} and {\it comarks\/}
(see, for example, \cite{BMP}). In root systems
with two lengths of roots, that is, in $B_n$, $C_n$, $F_4$ and $G_2$,
the highest (long) root $\xi$ is of the form
\begin{alignat}{3} \label{highestroot-1}
B_n &:&\quad \xi &=&\ (0\,1\,0\,\dots\,0)
               &=\alpha_1+2\alpha_2+2\alpha_3+\cdots+2\alpha_n ,\\
C_n &:&\ \xi     &=&\ (2\,0\,\dots\,0)
               &=2\alpha_1+2\alpha_2+\cdots+2\alpha_{n-1}+\alpha_n ,\\
F_4 &:&\ \xi     &=&\ (1\,0\,0\,0)
               &=2\alpha_1+3\alpha_2+4\alpha_3+2\alpha_4 ,\\
G_2 &:&\ \xi     &=&\ (1\,0)   &= 2\alpha_1+3\alpha_2.
\end{alignat}
For $A_n$, $D_n$, and $E_n$, all roots are of the same length, hence
$\xi_s=\xi$. We have
\begin{alignat}{3} \label{highestroot-2}
A_n &:&\quad \xi &=&\ (1\,0\,\dots\,0\,1)
               &=\alpha_1+\alpha_2+\cdots+\alpha_n ,\\
D_n &:&\ \xi     &=&\ (0\,1\,0\,\dots\,0)
               &=\alpha_1+2\alpha_2+\cdots+2\alpha_{n-2}+
               \alpha_{n-1}+\alpha_n ,\\
E_6 &:&\ \xi     &=&\ (0\, 1\, 0\,\dots\,
0)&=\alpha_1+2\alpha_2+3\alpha_3+2\alpha_4+\alpha_5+2\alpha_6 ,
               \\
E_7 &:&\ \xi     &=&\ (1\, 0\,0\,\dots\,
0)&=2\alpha_1+3\alpha_2+4\alpha_3+3\alpha_4+2\alpha_5+\alpha_6+2\alpha_7,
      \\
E_8 &:&\ \xi     &=&\ (0\,0\,\dots\, 0\,
1)&=2\alpha_1+3\alpha_2+4\alpha_3+5\alpha_4+6\alpha_5+4\alpha_6+2\alpha_7+
3\alpha_8 .
\end{alignat}

For highest root $\xi$ we have
\begin{equation}\label{highest}
\xi^\vee =\xi .
\end{equation}
Moreover, if all simple roots are of the same length, then
 \[
\alpha_i^\vee =\alpha_i.
 \]
For this reason,
 \[
(q_1,q_2,\dots,q_n)=(m_1,m_2,\dots,m_n).
 \]
for $A_n$, $D_n$ and $E_n$. Formulas \eqref{highestroot-2}--\eqref{highest} determine these
numbers. For short roots $\alpha_i$ of $B_n$, $C_n$ and $F_4$ we have
$\alpha_i^\vee=2\alpha_i$.  For short root $\alpha_2$ of $G_2$ we
have $\alpha_2^\vee=3\alpha_2$. For this reason,
 \begin{gather*}
(q_1,q_2,\dots,q_n)=(1,2,\dots, 2,1)\qquad {\rm for}\quad  B_n,
\\
(q_1,q_2,\dots,q_n)=(1,1,\dots, 1,1)\qquad {\rm for}\quad  C_n,
\\
(q_1,q_2,q_3,q_4)=(2,3, 2,1)\qquad {\rm for}\quad  F_4,
\\
(q_1,q_2)=(2,1)\qquad {\rm for}\quad  G_2.
\end{gather*}

To each root system $R$ there corresponds an {\it extended root
system} (which is also called an {\it affine
root system}). It is constructed
with the help of the highest root $\xi$ of $R$.
Namely, if $\alpha_1,\alpha_2,\dots, \alpha_n$ is a
set of all simple roots, then {\it the roots
\[
\alpha_0:=-\xi,\alpha_1,\alpha_2,\dots, \alpha_n
\]
constitute a set of simple roots of the corresponding extended root
system}. Taking into account the orthogonality (non-orthogonality) of
the root $\alpha_0$ to other simple roots, a diagram of an
extended root system can be constructed (which is an extension of
the corresponding Coxeter--Dynkin diagram; see, for example,
\cite{K}). Note that for all simple Lie algebras (except for
$A_n$) only one simple root is orthogonal to the root $\alpha_0$. In
the case of $A_n$, the two simple roots $\alpha_1$ and $\alpha_n$
are not orthogonal to $\alpha_0$.

\subsection[Affine Weyl group and even affine Weyl group]{Af\/f\/ine Weyl group and even af\/f\/ine Weyl group}\label{Aff}

We are interested in $E$-orbit functions which are given
on the Euclidean space $E_n$. These functions are invariant
with respect to action by elements of an even Weyl group $W_e$, which is a~transformation group of $E_n$. However, $W_e$ does not describe all
symmetries of $E$-orbit functions corresponding to weights $\lambda\in
P^e_+\equiv P_+\cup r_\alpha P_+$. A whole group of symmetries of these $E$-orbit
functions is isomorphic to the even af\/f\/ine Weyl group $W_e^{\rm af\/f}$ which
is an extension of the even Weyl group $W$. To describe the group
$W_e^{\rm af\/f}$ we f\/irst def\/ine the af\/f\/ine Weyl group $W^{\rm af\/f}$.

Let $\alpha_1,\alpha_2,\dots ,\alpha_n$ be simple roots in the
Euclidean space $E_n$ and let $W$ be the corresponding Weyl group.
The group $W$ is generated by ref\/lections $r_{\alpha_i}$,
$i=1,2,\dots ,n$. In order to construct the af\/f\/ine Weyl group
$W^{\rm af\/f}$, corresponding to $W$, we have to add an additional
ref\/lection. This ref\/lection is constructed as follows.

We consider the ref\/lection $r_{\xi}$ with respect to the
$(n-1)$-dimensional subspace (hyperplane) $X_{n-1}$ containing the
origin and orthogonal to the highest (long) root $\xi$, given in
\eqref{highestroot}:
\begin{equation} \label{refl-s}
r_{\xi}x=x-\frac{2\langle x,\xi\rangle}{\langle \xi,\xi\rangle}
\xi .
\end{equation}
Clearly, $r_{\xi} \in W$.
We shift the hyperplane $X_{n-1}$ by the vector $\xi^\vee/2$,
where $\xi^\vee =2\xi/\langle\xi,\xi \rangle$. (Note that by
\eqref{highest} we have $\xi^\vee=\xi$. However, it is convenient
here to use $\xi^\vee$.) The ref\/lection with
respect to the hyperplane $X_{n-1}+\xi^\vee/2$ will be denoted by~$r_0$. In order to fulf\/ill the transformation~$r_0$ we have
to fulf\/ill the transformation $r_\xi$ and then to shift the result
by $\xi^\vee$, that~is,
\begin{gather*}
r_0x=r_\xi x+\xi^\vee .
\end{gather*}
We have $r_00=\xi^\vee$ and it follows from \eqref{refl-s} that
$r_0$ maps $x+\xi^\vee/2$ to
\[
r_\xi(x+ \xi^\vee/2)+\xi^\vee=x+\xi^\vee/2-\langle x,\xi^\vee
\rangle \xi
\]
Therefore,
\begin{alignat}{2}
r_0(x+\xi^\vee/2)&=x+\xi^\vee/2 - \frac{2\langle
x,\xi\rangle}{\langle \xi,\xi\rangle} \xi =x+\xi^\vee/2
-\frac{2\langle x,\xi^\vee\rangle}{\langle
\xi^\vee,\xi^\vee\rangle} \xi^\vee
\notag\\
&=x+\xi^\vee/2-\frac{2\langle x+\xi^\vee/2,\xi^\vee\rangle}{\langle
\xi^\vee,\xi^\vee\rangle} \xi^\vee +\frac{2\langle
\xi^\vee/2,\xi^\vee\rangle}{\langle \xi^\vee,\xi^\vee\rangle}
\xi^\vee .
\notag
\end{alignat}
Denoting $x+\xi^\vee/2$ by $y$ we obtain that $r_0$ is given also
by the formula
 \begin{equation} \label{refl-0}
r_0y=y+\left( 1- \frac{2\langle y,\xi^\vee\rangle}{\langle
\xi^\vee,\xi^\vee\rangle}\right) \xi^\vee=\xi^\vee +r_\xi y .
 \end{equation}
The element $r_0$ does not belongs to $W$ since elements of $W$
do not move the point $0\in E_n$.

The hyperplane $X_{n-1}+\xi^\vee/2$ coincides with the set of
points $y$ such that $r_0y=y$. It follows from \eqref{refl-0} that
this hyperplane is given by the equation
 \begin{equation} \label{hyperpl}
1=\frac{2\langle y,\xi^\vee\rangle}{\langle
\xi^\vee,\xi^\vee\rangle}  =\langle y,\xi\rangle =\sum _{k=1}^n
a_kq_k,
 \end{equation}
where
\[
y=\sum _{k=1}^n a_k\omega_k,\qquad \xi=\sum _{k=1}^n q_k
\alpha^\vee_k
\]
(see \eqref{weight}).

A group of transformations of the Euclidean space $E_n$ generated
by ref\/lections $r_0,r_{\alpha_1},\dots ,r_{\alpha_n}$ is
called the {\it affine Weyl group} of the root system $R$ and is
denoted by $W^{\rm af\/f}$ or by $W^{\rm af\/f}_R$ (if is necessary to
indicate the initial root system), see \cite{K}.
Adjoining the ref\/lection $r_0$ to the Weyl group $W$ completely
changes properties of the group $W^{\rm af\/f}$.

Due to \eqref{refl-s} and \eqref{refl-0} for any
$x\in E_n$ we have
\[
r_0r_{\xi}x=r_0(r_{\xi}x)=\xi^\vee+r_\xi r_\xi x =x+\xi^\vee .
\]
Clearly, $(r_0r_{\xi})^kx=x+k\xi^\vee$, $k= 0,\pm 1,\pm 2,\dots $,
that is, the set of elements $(r_0r_{\xi})^k$, $k= 0,\pm 1,\pm
2,\dots $, is an inf\/inite commutative subgroup of $W^{\rm af\/f}$.
This means that (unlike to the Weyl group $W$) $W^{\rm af\/f}$ {\it is
an infinite group}.

Since $r_00=\xi^\vee$, for any $w\in W$ we have
\[
wr_00=w\xi^\vee=\xi^\vee_w,
\]
where $\xi^\vee_w$ is a coroot of the same length as the coroot
$\xi^\vee$. For this reason, $wr_0$ is the ref\/lection with respect
to the $(n-1)$-hyperplane perpendicular to the root $\xi^\vee_w$
and containing the point~$\xi^\vee_w/2$. Moreover,
\[
(wr_0)r_{\xi^\vee_w} x=x+\xi^\vee_w
\]
We also have $((wr_0)r_{\xi^\vee_w})^kx=x+k\xi^\vee_w$, $k=0,\pm
1,\pm 2,\dots$. Since $w$ is any element of $W$, then the set
$w\xi^\vee$, $w\in W$, coincides with the set of coroots of
$R$, corresponding to all long roots of the root system $R$. Thus,
{\it the set $W^{\rm af\/f}\cdot 0$ coincides with the lattice
$Q^\vee_l$ generated by coroots $\alpha^\vee$ taken for all long
roots $\alpha$ from $R$.}

It is checked for each type of root systems that
each coroot $\xi_s^\vee$ for a short root $\xi_s$ of $R$ is a~linear
combination of coroots $w\xi^\vee\equiv \xi_w$, $w\in W$, with
integral coef\/f\/icients, that is, $Q^\vee =Q_l^\vee$. Therefore, {\it
The set $W^{\rm af\/f}\cdot 0$ coincides with the coroot lattice
$Q^\vee$ of $R$.}

Let $\hat Q^\vee$ be the subgroup of $W^{\rm af\/f}$ generated by
the elements
 \begin{equation} \label{ref-ow}
 (wr_0)r_w, \qquad  w\in W,
 \end{equation}
where $r_w\equiv r_{\xi^\vee_w}$ for $w\in W$. Since elements
\eqref{ref-ow} pairwise commute with each other (since they are
shifts), $\hat Q^\vee$ is a commutative group. The subgroup $\hat
Q^\vee$ can be identif\/ied with the coroot lattice $Q^\vee$.
Namely, if for $g\in \hat Q^\vee$ we have $g\cdot 0=\gamma\in
Q^\vee$, then $g$ is identif\/ied with $\gamma$. This correspondence
is one-to-one.

The subgroups $W$ and $\hat Q^\vee$ generate $W^{\rm af\/f}$ since a
subgroup of $W^{\rm af\/f}$, generated by $W$ and $\hat Q^\vee$,
contains the element $r_0$. {\it The group $W^{\rm af\/f}$ is a
semidirect product of its subgroups $W$ and $\hat Q^\vee$, where
$\hat Q^\vee$ is an invariant subgroup} (see Section~5.2 in
\cite{KP06} for details).

We shall need not the whole af\/f\/ine Weyl group $W^{\rm af\/f}$
but only its subgroup, constructed on the base of the even
Weyl group $W_e$. The subgroup $W_e^{\rm af\/f}$, coinciding with the semidirect
product of the even Weyl group $W_e$ and $\hat Q^\vee$, will
be called an {\it even affine Weyl group}. This subgroup does
not contain the ref\/lection $r_0\in W^{\rm af\/f}$. One says that
$W_e^{\rm af\/f}$ consists of even elements of $W^{\rm af\/f}$.

\subsection[$W^{\rm aff}$-fundamental domain and
$W_e^{\rm aff}$-fundamental domain]{$\boldsymbol{W^{\rm af\/f}}$-fundamental domain and
$\boldsymbol{W_e^{\rm af\/f}}$-fundamental domain}\label{Fund}

An open connected simply connected set $F(G)\subset E_n$ is called a
{\it fundamental domain} for the group $G$ ($G=W,W^{\rm af\/f},W_e,
W_e^{\rm af\/f}$) if it does not contains
equivalent points (that is, points $x$
and $x'$ such that $x=wx$, $w\in G$) and if its closure contains at least one
point from each $G$-orbit. It is
evident that {\it the dominant Weyl chamber $D_+$ without walls of this
chamber is a fundamental domain for the Weyl group $W$}. Recall
that this domain consists of all points $x=a_1\omega_1+\cdots
+a_n\omega_n\in E_n$ for which
\[
a_i=\langle x,\alpha^\vee_i \rangle > 0,\qquad i=1,2,\dots
,n.
\]
For any f\/ixed root $\alpha$, the domain $D^e_+=D_+\cup r_\alpha D_+$
(where each point is taken only once)
can be taken as a closure of the fundamental domain
of the even Weyl group $W_e$. The fundamental domain of $W_e$ is the set
$D_+^e$ without its boundary.

Let us describe a fundamental domain for the group $W^{\rm
af\/f}$. Since $W\subset W^{\rm af\/f}$, it can be chosen as a subset
of the dominant Weyl chamber for $W$.

We have seen that the element $r_0\in W^{\rm af\/f}$ is a ref\/lection
with respect to the hyperplane $X_{n-1}+\xi^\vee/2$, orthogonal to
the root $\xi$ and containing the point $\xi^\vee/2$. This
hyperplane is given by the equation~\eqref{hyperpl}. This equation
shows that the hyperplane $X_{n-1}+\xi^\vee/2$ intersects the axes,
determined by the vectors $\omega_i$, in the points $\omega_i/q_i$,
$i=1,2,\dots ,n$, where $q_i$ are such as in~\eqref{hyperpl}. We
create the simplex with $n+1$ vertices in the points
 \begin{equation}\label{sympl}
0,\ \frac{\omega_1}{q_1},\dots , \frac{\omega_n}{q_n} .
 \end{equation}
By \eqref{hyperpl}, this simplex consists of all points $y$ of the
dominant Weyl chamber for which $\langle y,\xi \rangle \le 1$.
Clearly, the interior $F$ of this simplex belongs to the dominant
Weyl chamber. The following theorem is true (see, for example,
\cite{KP06}):

\begin{theorem}\label{theorem1}
The set $F$ is a fundamental domain for the
affine Weyl group $W^{\rm af\/f}$.
\end{theorem}

For the rank 2 cases the fundamental domain is the interior of
the  simplex with the following vertices:
 \begin{alignat}{2} A_2
&:&\quad& \{ 0,\ \omega_1,\ \omega_2\} ,
\notag\\
C_2 &:&\quad& \{ 0,\ \omega_1,\ \omega_2\} , \notag
\\
G_2 &:&\quad&\{ 0,\ \tfrac{\omega_1}2,\ \omega_2\} .\notag
\end{alignat}

A fundamental domain of the even af\/f\/ine group $W_e^{\rm af\/f}$
can be taken in such a way that it is contained in the
fundamental domain $D_+^e$ of $W_e$. Namely, the set
$\bar F\cup r_\alpha \bar F$ (where each point is taken
only once) without its boundary satisf\/ies this condition and is a
$W_e^{\rm af\/f}$-fundamental domain.

\section[$W_e$-orbits]{$\boldsymbol{W_e}$-orbits}\label{s-orbi}

\subsection[Definition]{Def\/inition}\label{orbi}
As we have seen, the $(n-1)$-dimensional linear subspaces
$X_\alpha$ of $E_n$, orthogonal to positive roots $\alpha$
and containing the origin,  divide the space $E_n$
into connected parts, which are called Weyl chambers. A
number of such chambers is equal to an order of the corresponding
Weyl group $W$. Elements of the Weyl group permute these chambers.
A single chamber $D_+$ such that $\l\alpha_i, x \r \ge 0$,
$x\in D_+$, $i=1,2,\dots ,n$, is the dominant Weyl chamber.
We f\/ix a root $\alpha$
and create a set $D_+^e=D_+\cup r_\alpha D_+$, where $r_\alpha$ is
the ref\/lection corresponding to the root $\alpha$.
The sets, received from $D^e_+$ by action by
elements of $W_e$ are called {\it even Weyl chambers}. Clearly,
they depend on choosing of the root $\alpha$. However, dif\/ferent
choices of $\alpha$ (and therefore, of even Weyl chambers)
does not change a set of $W_e$-orbits, which are considered
below.

The cone of dominant integral weights $P_+$ belongs to the
dominant Weyl chamber $D_+$. By~$P^e_+$ we denote the set
$P^e_+ \cup r_\alpha P^e_+$ (where each point is taken only once).
Then $P^e_+\subset D^e_+$.

Let $y$ be an arbitrary dominant element of the Euclidean space
$E_n$. We act upon
$y$ by all elements of the Weyl group $W$. As a result, we obtain the
set $wy$, $w\in W/W_y$ (where $W_y$ is the subgroup of elements
of $W$ leaving $y$ invariant), which is called a {\it Weyl group
orbit} or a~$W$-orbit of the point $y$. A $W$-orbit of a~point $y\in D_+$ is
denoted by $O(y)$ or $O_W(y)$. A size of an orbit~$O(y)$ is a
number $|O(y)|$ of its elements. Each Weyl chamber contains only one
point of a~f\/ixed orbit $Q(y)$.

Now we act upon element
$y\in E_n$ by all elements of the even Weyl group $W_e$. As a result, we obtain a
set of elements $wy$, $w\in W_e$ (each point is taken only once),
which is called an {\it even Weyl group
orbit} or a $W_e$-orbit of the point $y$. A $W_e$-orbit of a point $y\in D_+^e$ is
denoted by~$O_e(y)$ or $O_{W_e}(y)$. Each even Weyl chamber contains only one
point of a f\/ixed orbit $Q_e(y)$.

$W_e$-orbits $O_e(y)$ do not depend on a choice of a root
$\alpha$ with respect to which we construct the even dominant
Weyl chamber $D^e_+=D_+\cup r_\alpha D_+$.

There are two types of $W_e$-orbits: orbits $O_e(y)$ which contain a point
of the dominant Weyl chamber $D_+$ (we call them $W_e$-{\it orbits of the
fist type}) and orbits which do not contain such
a~point (we call them $W_e$-{\it orbits of the second type}).
It is easy to see that each $W$-orbit $O(y)$ with
strictly dominant $y$ consists of two $W_e$-orbits,
one of them is of the f\/irst type and the second of the second type.
Namely,
\[
 O(y)=O_e(y)\bigcup O_e(r_\alpha y),
\]
where the $W_e$-orbit $O_e(r_\alpha y)$ does not depend on a choice
of the root $\alpha$, that is, $O_e(r_\alpha y)$ are the same for
all positive roots $\alpha$ of a given root system.
Moreover, the orbit $O_e(r_\alpha y)$ is obtained from the
orbit $O_e(y)$ by acting upon each point of $O_e(y)$ by
ref\/lection $r_\alpha$, $O_e(r_\alpha y)=r_\alpha O_e(y)$.

If $y$ belongs to some wall of the dominant Weyl chamber, then
the $W_e$-orbit $O_e(y)$ coincides with the $W$-orbit $O(y)$.
 \medskip

\noindent {\bf Example.} {\it The case $A_1$}. The Weyl group $W$ of
$A_1$ consists of two elements 1 and $r_\alpha$, where $\alpha$ is a
unique positive root of $A_1$. The element $r_\alpha$ is a
ref\/lection and, therefore, $\det r_\alpha=-1$. Thus, the subgroup
$W_e$ in this case contains only one element 1. This means that each
point of the real line is a $W_e$-orbit for $A_1$. In particular,
$O_e(y)$, $y>0$, belongs to orbits, corresponding to dominant
elements. The other orbits (except for the orbit $O_e(0)$) are
obtained by acting by the ref\/lection $r_\alpha$.

\subsection[$W_e$-orbits of $A_2$, $C_2$, $G_2$]{$\boldsymbol{W_e}$-orbits of $\boldsymbol{A_2}$, $\boldsymbol{C_2}$, $\boldsymbol{G_2}$}\label{orbits}

In this subsection we give $W_e$-orbits for the rank two cases.
Orbits will be given by coordinates in the $\omega$-basis. Points of the orbits
will be denoted as $(a\; b)$, where $a$ and $b$ are
$\omega$-coordinates.

If $a> 0$ and $b> 0$, then the $W_e$-orbits $O_e(a\; b)$ and
$r_\alpha O(a\; b)\equiv O_e(-a\; a+b)$ of $A_2$ contain points
\begin{align}
A_2 :\quad
           &O_e(a\ b)\ni(a\ b),\ ({-}a{-}b\ a),\ (b\ {-}a{-}b),\\
           &O_e({-}a\ a{+}b)\ni ({-}a\ a{+}b),\ (a{+}b\ {-}b),\  ({-}b\ {-}a).
\end{align}
The other  $W_e$-orbits of $A_2$ are
\begin{align}
A_2 :\quad
           &O_e(a\ 0)\ni (a\ 0),\ ({-}a\ a),\ (0\ {-}a),\\
           &O_e(0\ b)\ni (0\ b),\ (b\ {-}b),\ ({-}b\ 0).
\end{align}

In the cases of $C_2$ and $G_2$ (where the second simple
root is the longer one for $C_2$ and the shorter one for $G_2$)
for $a> 0$ and $b>0$ we have
\begin{align}
C_2 :\quad
   &O_e(a\ b) \ni (a\ b),\  (a{+}2b\ {-}a{-}b),\ ({-}a\ {-}b),\
           ({-}a{-}2b\ a{+}b),   \\
&O_e({-}a\ a{+}b) \ni ({-}a\ a{+}b),\ (a{+}2b\ {-}b),\
          (a\ {-}a{-}b),\ ({-}a{-}2b\ b), \\
G_2 :\quad
  &O_e(a\ b) \ni\pm(a\ b),\ \pm(2a{+}b\ {-}3a{-}b),\ \pm({-}a{-}b\ 3a{+}2b),\\
    &O_e({-}a\ 3a{+}b) \ni  \pm({-}a\ 3a{+}b),\ \pm(a{+}b\ {-}b),\
    \pm({-}2a{-}b\  3a{+}2b) ,
\end{align}
where $\pm (c\ d)$ means two points $(c \ d)$ and $(-c \ -d)$.

We also give $W_e$-orbits for which one of the numbers $a$, $b$ vanish:
\begin{align}
C_2  :\quad
  &O_e(a\ 0)\ni\pm(a\ 0),\ \pm({-}a\ a),\\
  & O_e(0\ b)\ni\pm(0\ b),\ \pm(2b\ {-}b), \\
G_2 :\quad
  &O_e(a\ 0)\ni\pm(a\ 0),\ \pm({-}a\ 3a),\ \pm(2a\ {-}3a),\\
  &O_e(0\ b)\ni\pm(0\ b),\ \pm(b\ {-}b),\ \pm({-}b\ 2b).
  \end{align}

As we see, for each point $(c\ d)$ of an orbit $O_e$ of $C_2$ or
$G_2$ there exists in the orbit the point $({-}c\ {-}d)$.


\subsection[The case of $A_n$]{The case of $\boldsymbol{A_n}$}\label{An}

In the cases of Coxeter--Dynkin diagrams $A_{n-1}$, $B_n$, $C_n$, $D_n$, root
and weight lattices, even Weyl groups and orbits are described in
a simple way if to use the orthogonal coordinate system in~$E_n$. In
particular, this coordinate system is useful under practical work
with orbits.

In the case $A_n$ it is convenient to describe root and weight
lattices, even Weyl group and orbit functions in the
subspace of the Euclidean space $E_{n+1}$, given by the equation
\[
x_1+x_2+\cdots +x_{n+1}=0,
\]
where $x_1,x_2,\dots ,x_{n+1}$ are orthogonal coordinates of a
point $x\in E_{n+1}$. The unit vectors in directions of these
coordinates are denoted by ${\bf e}_j$, respectively. Clearly, ${\bf
e}_i\bot {\bf e}_j$, $i\ne j$. The set of roots of $A_n$ is given by
the vectors
\[
\alpha_{ij}={\bf e}_i-{\bf e}_j, \qquad i\ne j.
\]
The roots $\alpha_{ij}={\bf e}_i-{\bf e}_j$, $i< j$,
are positive and the roots
\[
\alpha_i\equiv \alpha_{i,i+1}={\bf e}_i-{\bf e}_{i+1},\qquad
i=1,2,\dots ,n,
\]
constitute the system of simple roots.

If $x=\sum\limits_{i=1}^{n+1} x_i{\bf e}_i$, $x_1+x_2+\cdots +x_{n+1}=0$, is a
point of $E_{n+1}$, then this point belongs to the dominant Weyl
chamber $D_+$ if and only if
\[
x_1\ge x_2\ge \cdots \ge x_{n+1}.
\]
Indeed, if this condition is fulf\/illed, then $\langle
x,\alpha_i\rangle =x_i-x_{i+1}\geq 0$, $i=1,2,\dots, n$, and $x$ is
dominant. Conversely, if $x$ is dominant, then $\langle
x,\alpha_i\rangle \geq 0$ and this condition is fulf\/illed. {\it The
point $x$ is strictly dominant if and only if}
\[
x_1> x_2> \cdots > x_{n+1}.
\]

If $\alpha=\alpha_{12}$, the even dominant chamber
$D^e_+=D_+\cup r_\alpha D_+$ consists of
points $x$ such that
\[
x_1\ge x_2\ge x_3\ge \cdots \ge x_{n+1}\qquad {\rm or}\qquad
x_2\ge x_1\ge x_3\ge \cdots \ge x_{n+1},
\]
that is, such that
\[
x_2, x_2\ge x_3\ge\cdots \ge x_{n+1}.
\]

If $\lambda =\sum\limits_{i=1}^n \lambda_i \omega_i$, then
the $\omega$-coordinates $\lambda_i$ are connected
with the orthogonal coordinates $x_j$ of $\lambda=\sum\limits_{i=1}^{n+1}
x_i{\bf e}_i$ by the formulas
 \begin{gather*}
 x_1  = \frac{n}{n+1} \lambda_1
  + \frac{n-1}{n+1}\lambda_2 + \frac{n-2}{n+1}\lambda_3
 + \cdots  + \frac{2}{n+1}\lambda_{n-1} + \frac{1}{n+1}\lambda_n,\\
 x_2  = -\frac{1}{n+1} \lambda_1
  + \frac{n-1}{n+1}\lambda_2 + \frac{n-2}{n+1}\lambda_3
 + \cdots  + \frac{2}{n+1}\lambda_{n-1} + \frac{1}{n+1}\lambda_n,\\
 x_3  = -\frac{1}{n+1} \lambda_1
  - \frac{2}{n+1}\lambda_2 + \frac{n-2}{n+1}\lambda_3
 + \cdots  + \frac{2}{n+1}\lambda_{n-1} + \frac{1}{n+1}\lambda_n,\\
    \cdots      \cdots    \cdots    \cdots \cdots    \cdots    \cdots      \cdots    \cdots    \cdots \cdots    \cdots    \cdots      \cdots    \cdots    \cdots  \cdots    \cdots    \cdots      \cdots    \cdots    \cdots \\
 x_{n}  = -\frac{1}{n+1} \lambda_1
  - \frac{2}{n+1}\lambda_2 - \frac{3}{n+1}\lambda_3
 - \cdots  - \frac{n-1}{n+1}\lambda_{n-1} + \frac{1}{n+1}\lambda_n,\\
 x_{n+1}  = -\frac{1}{n+1} \lambda_1
  - \frac{2}{n+1}\lambda_2 - \frac{3}{n+1}\lambda_3
 - \cdots
  - \frac{n-1}{n+1}\lambda_{n-1} - \frac{n}{n+1}\lambda_n.
 \end{gather*}
 The inverse formulas are
  \begin{equation}\label{***}
 \lambda_i=x_i-x_{i+1},\qquad i=1,2,\dots ,n.
 \end{equation}

By means of the formula
 \begin{equation}\label{refl}
 r_\alpha \lambda=\lambda -\frac{2\l \lambda,\alpha\r}{\l \alpha,
 \alpha \r}\alpha \qquad
  \end{equation}
for the ref\/lection with respect to the hyperplane, orthogonal to a
root $\alpha$, we can f\/ind how elements of the Weyl group $W(A_n)$
act upon points $\lambda\in E_{n+1}$. We conclude that
the Weyl group $W(A_n)$ consists of all
permutations of the orthogonal coordinates $x_1,x_2,\dots ,x_{n+1}$
of a~point~$\lambda$, that is, $W(A_n)$ coincides with the symmetric
group $S_{n+1}$. Even permutations of $W(A_n)$ constitute the even
Weyl group $W_e(A_n)$. It is the alternating subgroup $S^e_{n+1}$
of the group $S_{n+1}$. This subgroup is simple. Transformations of
$S^e_{n+1}$ are elements of the rotation group $SO(n+1)$.

Sometimes (for example, if we wish for coordinates to be integers or
non-negative integers),
it is convenient to introduce orthogonal coordinates $y_1,y_2,\dots,
y_{n+1}$ for $A_n$ in such a way that
\[
y_1+y_2+\cdots +y_{n+1}=m,
\]
where $m$ is some f\/ixed real number. They are obtained from the
previous orthogonal coordinates by adding the same number $m/(n+1)$
to each coordinate. Then, as one can see from \eqref{***},
$\omega$-coordinates $\lambda_i=y_i-y_{i+1}$ and the group $W$ and $W_e$
do not change. Sometimes, it is natural to use orthogonal coordinates
$y_1,y_2,\dots,y_{n+1}$ for which all $y_i$ are non-negative.

$W_e$-orbits $O_e(\lambda)$ for strictly dominant $\lambda$
can be constructed by means of signed $W$-orbits.
Signed $W$-orbit $O^\pm (\lambda)$ were introduced in \cite{KP07} .
The signed orbit $O^\pm(\lambda)$,
$\lambda=(x_1,x_2,\dots,x_{n+1})$, $x_1> x_2> \cdots > x_{n+1}$,
consists of all points
\[
(x_{i_1},x_{i_2},\dots,x_{i_{n+1}})^{{\rm sgn}\, (\det w)}
\]
obtained from $(x_1,x_2,\dots,x_{n+1})$ by permutations $w\in
W\equiv S_{n+1}$ (instead of ${\rm sgn}\, (\det w)$ we sometimes write
simply $\det w$).

The signed $W$-orbit $O^\pm (\lambda)$ splits
into two $W_e$-orbits: one of them coincides with $O_e(\lambda)$
and the second with $O_e(r_\alpha \lambda)$, where $\alpha$ is
a positive root of $A_n$.
The $W_e$-orbit $O_e (\lambda)$ contains the points of
the $W$-orbit $O^\pm (\lambda)$, which have the sign plus and
$O_e(r_\alpha \lambda)$ contains the points with the sign
minus.

If $\lambda$ is dominant but not strictly dominant, then
the $W_e$-orbit $O_e (\lambda)$ coincides with the $W$-orbit
$O(\lambda)$. Description of $W$-orbits of $A_n$ in orthogonal
coordinates see in \cite[Subsection 3.1]{KP06}.

\subsection[The case of $B_n$]{The case of $\boldsymbol{B_n}$}\label{Bn}

Orthogonal coordinates of a point $x\in E_n$ are denoted by
$x_1,x_2,\dots ,x_n$. We denote by ${\bf e}_i$ the corresponding
unit vectors. Then the set of roots of $B_n$ is given by the vectors
\[
\alpha_{\pm i,\pm j}=\pm {\bf e}_i\pm {\bf e}_j, \qquad i\ne j,
\qquad  \alpha_{\pm i}=\pm {\bf e}_i,\qquad i=1,2,\dots ,n
\]
(all combinations of signs must be taken). The roots
$\alpha_{i,\pm j}={\bf e}_i\pm {\bf e}_j$, $i< j$,
$\alpha_{i}={\bf e}_i$, $i=1,2,\dots ,n$,
are positive and $n$ roots
\[
\alpha_i:={\bf e}_i-{\bf e}_{i+1},\qquad i=1,2,\dots ,n-1, \qquad
\alpha_n={\bf e}_n
\]
constitute the system of simple roots.

A point $\lambda=\sum\limits_{i=1}^{n} x_i{\bf e}_i\in E_{n}$
belongs to the dominant Weyl chamber $D_+$ if and only if
\[
x_1\ge x_2\ge \cdots \ge x_{n}\ge 0.
\]
Moreover, {\it this point is strictly dominant if and only if}
\[
x_1> x_2> \cdots > x_{n}> 0.
\]

The even dominant chamber can be taken consisting of
points $x$ such that
\[
x_1\ge x_2\ge x_3\ge\cdots \ge x_{n}\ge 0\qquad {\rm or}\qquad
x_2\ge x_1\ge x_2\ge\cdots \ge x_{n}\ge 0,
\]
that is, such that
\[
x_1, x_2\ge x_3\ge \cdots \ge x_{n}\ge 0.
\]

If $\lambda =\sum\limits_{i=1}^n \lambda_i \omega_i$, then the
$\omega$-coordinates $\lambda_i$ are connected
with the orthogonal coordinates $x_j$ of
$\lambda=\sum\limits_{i=1}^{n} x_i{\bf e}_i$ by the formulas
 \begin{gather*}
 x_1 =  \lambda_1 + \lambda_2 + \cdots  + \lambda_{n-1} + \tfrac 12
 \lambda_n, \\
 x_2  =  \phantom{\lambda_1 +{}} \lambda_2 + \cdots  + \lambda_{n-1} + \tfrac 12
 \lambda_n, \\
   \cdots   \cdots   \cdots    \cdots    \cdots \cdots  \cdots   \cdots   \cdots    \cdots    \cdots \\
 x_{n}=\phantom{\lambda_1 +\lambda_2 + \cdots  + \lambda_{n-1} +{}} \tfrac 12  \lambda_n,
 \end{gather*}
 The inverse formulas are
 \[
 \lambda_i=x_i-x_{i+1},\qquad i=1,2,\dots ,n-1,\qquad
 \lambda_n=2x_n.
 \]
It is easy to see that if $\lambda\in P_+$, then the coordinates
$x_1,x_1,\dots ,x_n$ are all integers or all half-integers.

The Weyl group $W(B_n)$ of $B_n$ consists of all permutations of the orthogonal
coordinates $x_1,x_2,\dots ,x_{n}$ of a point $\lambda$ with
possible sign alternations of any number of them.
Moreover,
$\det w$ is equal to $\pm 1$ depending on whether $w$ is a product of
even or odd number of ref\/lections and alternations of signs. A sign
of $\det w$ can be determined as follows. We represent $w$ as a~product
$w=\epsilon s$, where $s$ is a permutation of $(x_1,x_2,\dots,x_n)$
and $\epsilon$ is an alternation of signs of coordinates. Then $\det
w=(\det s)\epsilon_{i_1}\epsilon_{i_2}\cdots \epsilon_{i_n}$, where
$\det s$ is def\/ined as in the previous subsection and $\epsilon_{i_j}=-1$
in the case of change of a sign of $i_j$-th coordinate and  $\epsilon_{i_j}=1$
otherwise. This show haw to determine the even
Weyl group $W_e(B_n)$ as a subgroup of $W(B_n)$.

$W_e(B_n)$-orbits $O_e(\lambda)$ with strictly dominant $\lambda$
can be constructed by means of signed orbits $O^\pm (\lambda)$.
The signed orbit $O^\pm(\lambda)$, $\lambda=(x_1,x_2,\dots,x_{n})$,
$x_1> x_2> \cdots > x_{n}> 0$, consists of all points
  \begin{equation}\label{B-altern}
(\pm x_{i_1}, \pm x_{i_2},\dots,\pm x_{i_{n}})^{\det w}
  \end{equation}
(each combination of signs is possible) obtained from $(x_1,x_2,
\dots,x_{n})$ by permutations and alternations of signs which
constitute an element $w$ of the Weyl group $W(B_n)$.
The signed orbit $O^\pm(\lambda)$  splits into two $W_e$-orbits:
one of them $O_e(\lambda)$ consists of all points of
$O^\pm(\lambda)$, which have the sign plus, and the second one
consists of all points with the sign minus.

If a dominant element $\lambda$ is not strictly dominant, then
the $W_e$-orbit $O_e(\lambda)$ coincides with the $W$-orbit
$O(\lambda)$.

\subsection[The case of $C_n$]{The case of $\boldsymbol{C_n}$}\label{Cn}

In the orthogonal coordinate system of the Euclidean space
$E_{n}$ the set of roots of $C_n$ is given by the vectors
\[
\alpha_{\pm i,\pm j}=\pm {\bf e}_i\pm {\bf e}_j, \qquad i\ne j,
\qquad  \alpha_{\pm i}=\pm 2{\bf e}_i,\qquad i=1,2,\dots ,n,
\]
where ${\bf e}_i$ is the unit vector in the direction of $i$-th
coordinate $x_i$ (all combinations of signs must be taken). The
roots
$\alpha_{i,\pm j}={\bf e}_i\pm {\bf e}_j$, $i< j$, and
$\alpha_{i}=2{\bf e}_i$, $i=1,2,\dots ,n$,
are positive and $n$ roots
\[
\alpha_i:={\bf e}_i-{\bf e}_{i+1},\qquad i=1,2,\dots ,n-1, \qquad
\alpha_n=2{\bf e}_n
\]
constitute the system of simple roots.

A point $\lambda=\sum\limits_{i=1}^{n} x_i{\bf
e}_i\in E_n$ belongs to the dominant Weyl chamber $D_+$ if and only
if
\[
x_1\ge x_2\ge \cdots \ge x_{n}\ge 0.
\]
This point is strictly dominant if and only if
\[
x_1> x_2> \cdots > x_{n}> 0.
\]

The even dominant chamber can be taken consisting of
points $x$ such that
\[
x_1\ge x_2\ge x_3\ge\cdots \ge x_{n}\ge 0\qquad {\rm or}\qquad
x_2\ge x_1\ge x_3\ge \cdots \ge x_{n}\ge 0,
\]
that is, such that
\[
x_1, x_2\ge x_3\ge \cdots \ge x_{n}\ge 0.
\]

If $\lambda =\sum\limits_{i=1}^n \lambda_i \omega_i$, then
the $\omega$-coordinates $\lambda_i$ are connected
with the orthogonal coordinates $x_j$ of $\lambda=\sum\limits_{i=1}^{n} x_i{\bf
e}_i$ by the formulas
 \begin{gather*}
 x_1  = \lambda_1 + \lambda_2 + \cdots  + \lambda_{n-1} +
 \lambda_n,\\
 x_2 = \phantom{\lambda_1 +{}}\lambda_2 + \cdots  + \lambda_{n-1} +  \lambda_n,\\
\cdots  \cdots   \cdots   \cdots \cdots  \cdots   \cdots   \cdots     \cdots    \cdots \\
 x_{n} = \phantom{\lambda_1 +\lambda_2 + \cdots  + \lambda_{n-1} +{}} \lambda_n .
 \end{gather*}
 The inverse formulas are
 \[
 \lambda_i=x_i-x_{i+1},\qquad i=1,2,\dots ,n-1,\qquad
 \lambda_n=x_n.
 \]
If $\lambda\in P_+$, then all coordinates $x_i$ are integers.

The Weyl group $W(C_n)$ of $C_n$ consists of all permutations of the
orthogonal coordinates $x_1,x_2,\dots ,x_{n}$ of a point
$\lambda$ with sign alternations of some of them, that is, this
Weyl group acts on orthogonal coordinates exactly in the same way as the
Weyl group $W(B_n)$ does. Moreover,
$\det w$ is equal to $\pm 1$ depending on whether $w$ consists of
even or odd numbers of ref\/lections and alternations of signs. Since
$W(C_n)=W(B_n)$, then a sign of $\det w$ is determined as in
the case~$B_n$.

The signed orbit $O^\pm(\lambda)$, $\lambda=(x_1,x_2,\dots,x_{n})$,
$x_1> x_2> \dots > x_{n}> 0$, consists of all points
\[
(\pm x_{i_1}, \pm x_{i_2},\dots,\pm x_{i_{n+1}})^{\det w}
\]
(each combination of signs is possible) obtained from $(x_1,x_2,
\dots,x_{n})$ by permutations and alternations of signs which
constitute an element $w$ of the Weyl group $W(C_n)$,
that is, in the orthogonal coordinates signed orbits for $C_n$
coincide with signed orbits of $B_n$. This determine how to separate the
subgroup $W_e(C_n)$ in the group $W(C_n)$.

The signed orbit $O^\pm(\lambda)$, $\lambda=(x_1,x_2,\dots,x_{n})$,
$x_1> x_2> \cdots > x_{n}> 0$, splits into two $W_e$-orbits:
one of them $O_e(\lambda)$ consists of all points of
$O^\pm(\lambda)$, which have the sign plus, and the second one
consists of all points with the sign minus.

If a dominant element $\lambda$ is not strictly dominant, then
the $W_e$-orbit $O_e(\lambda)$ coincides with the $W$-orbit
$O(\lambda)$.

\subsection[The case of $D_n$]{The case of $\boldsymbol{D_n}$}\label{Dn}

In the orthogonal coordinate system of the Euclidean space
$E_{n}$ the set of roots of $D_n$ is given by the vectors
\[
\alpha_{\pm i,\pm j}=\pm {\bf e}_i\pm {\bf e}_j, \qquad i\ne j,
\]
where ${\bf e}_i$ is the unit vector in the direction of $i$-th
coordinate (all combinations of signs must be taken). The roots
$\alpha_{i,\pm j}={\bf e}_i\pm {\bf e}_j$, $i< j$,
are positive and $n$ roots
\[
\alpha_i:={\bf e}_i-{\bf e}_{i+1},\qquad i=1,2,\dots ,n-1, \qquad
\alpha_n={\bf e}_{n-1}+ {\bf e}_n
\]
constitute the system of simple roots.

If $\lambda=\sum\limits_{i=1}^{n} x_i{\bf e}_i$,
then this point belongs to the dominant Weyl
chamber $D_+$ if and only if
\[
x_1\ge x_2\ge \cdots \ge x_{n-1}\ge |x_n|.
\]
{\it This point is strictly dominant if and only if}
\[
x_1> x_2> \cdots > x_{n-1}> |x_n|
\]
(in particular, $x_n$ can take the value 0).

The even dominant chamber can be taken consisting of
points $x$ such that
\[
x_1\ge x_2\ge x_3\ge\cdots \ge |x_{n}|\qquad {\rm or}\qquad
x_2\ge x_1\ge x_3\ge \cdots \ge |x_{n}|,
\]
that is, such that
\[
x_1, x_2\ge x_3\ge \cdots \ge x_{n-1}\ge |x_{n}|.
\]

If $\lambda =\sum\limits_{i=1}^n \lambda_i \omega_i$, then
the $\omega$-coordinates $\lambda_i$ are connected
with the orthogonal coordinates $x_j$ of $\lambda=\sum\limits_{i=1}^{n} x_i{\bf
e}_i$ by the formulas
  \begin{gather*}
 x_1 = \lambda_1 + \lambda_2 + \cdots  + \lambda_{n-2} + \tfrac 12
 (\lambda_{n-1}  + \lambda_n),\\
 x_2 =   \phantom{\lambda_1 +{}}    \lambda_2 + \cdots  + \lambda_{n-2} + \tfrac 12
 (\lambda_{n-1} +   \lambda_n),\\
\cdots  \cdots   \cdots   \cdots   \cdots    \cdots  \cdots \cdots  \cdots   \cdots   \cdots   \cdots    \cdots  \cdots\\
 x_{n-1} = \phantom{\lambda_1 +\lambda_2 + \cdots  + \lambda_{n-2}\, }
 \tfrac 12 (\lambda_{n-1} +  \lambda_n),\\
 x_{n} = \phantom{\lambda_1 +\lambda_2 + \cdots  + \lambda_{n-2} +{}}
 \tfrac 12 (\lambda_{n-1} -  \lambda_n),
  \end{gather*}
 The inverse formulas are{\samepage
 \[
 \lambda_i=x_i-x_{i+1},\qquad i=1,2,\dots ,n-2,\qquad
 \lambda_{n-1}=x_{n-1}+x_n,\qquad \lambda_{n}=x_{n-1}-x_n .
 \]
If $\lambda\in P_+$, then the coordinates $x_1,x_2,\dots,x_n$ are
all integers or all half-integers.}

The Weyl
group $W(D_n)$ of $D_n$ consists of all permutations of the orthogonal
coordinates $x_1,x_2,\dots ,x_{n}$ of a point $\lambda$ with sign
alternations of even number of them.
Moreover, $\det w$ is equal to $\pm 1$
and a sign of $\det w$ is determined as follows. The element $w\in
W(D_n)$ can be represented as a product $w=\tau s$, where $s$ is a
permutation from $S_n$ and $\tau$ is an alternation of even number
of coordinates. Then $\det w=\det s$.
Indeed, since an alternation of signs of two coordinates $x_i$ and $x_j$ is a
product of two ref\/lections $r_\alpha$ with $\alpha={\bf e}_i + {\bf e}_j$
and with $\alpha={\bf e}_i - {\bf e}_j$, a sign of the determinant
of this alternation is plus. (Note that $|W(D_n)|=\frac 12 |W(B_n)|$.)

Now we may state that the even Weyl group $W_e(D_n)$ consists of
products $\tau s$, where $s$ runs over even permutations of $S_n$
and $\tau$ runs over alternations of even numbers of
coordinates.

The signed orbit $O^\pm(\lambda)$, $\lambda=(x_1,x_2,\dots,x_{n})$,
$x_1> x_2> \cdots > x_{n}> 0$, for $D_n$ consists of all points
\[
(\pm x_{i_1}, \pm x_{i_2},\dots,\pm x_{i_{n}})^{\det w}
\]
obtained from $(x_1,x_2, \dots,x_{n})$ by permutations and
alternations of even number of signs which constitute an element $w$
of the Weyl group $W(D_n)$.
This signed orbit splits into two $W_e$-orbits:
one of them $O_e(\lambda)$ consists of all points of
$O^\pm(\lambda)$, which have the sign plus, and the second one
consists of all points  with the sign minus.

If a dominant element $\lambda$ is not strictly dominant, then
the $W_e(D_n)$-orbit $O_e(\lambda)$ coincides with the $W(D_n)$-orbit
$O(\lambda)$.

\subsection[Fundamental domains of $W^{\rm aff}_e$]{Fundamental domains of $\boldsymbol{W^{\rm af\/f}_e}$}

Using the explicit formula for the antisymmetric orbit function
$\varphi_\rho(x)$, where $\rho$ is a half of positive roots, we have
derived in \cite{KP07} explicit forms of the fundamental domains of
$W^{\rm af\/f}$ for the cases $A_n$, $B_n$, $C_n$, $D_n$. They easily
determine fundamental domains $F_e$ for the corresponding even
af\/f\/ine Weyl groups $W^{\rm af\/f}_e$.
 \medskip

(a) The fundamental domain $F_e(A_n)$ of the even af\/f\/ine Weyl group
$W^{\rm af\/f}_e(A_n)$ is contained in the domain of real points
$x=(x_1,x_2,\dots$, $x_{n+1})$ such that
\[
x_1, x_2>x_3>\cdots>x_{n+1},\qquad x_1+x_2+\cdots +x_{n+1}=0.
\]
Moreover, a point $x$ of this domain belongs to $F_e(A_n)$ if and only if
$x_1+|x_{n+1}|<1$ and $x_1>x_2$, or $x_2+|x_{n+1}|<1$ and
$x_2>x_1$.
\medskip

(b) The fundamental domain $F_e(B_n)$ of $W^{\rm af\/f}_e(B_n)$
is contained in the domain of points
$x=(x_1,x_2,\dots, x_{n})$ such that
\[
1>x_1, x_2>x_3>\cdots>x_{n}>0.
\]
Moreover, a point $x$ of this domain belongs to $F_e(B_n)$ if and only if
$x_1+x_2<1$.
\medskip

(c) The fundamental domain $F_e(C_n)$ of $W^{\rm af\/f}_e(C_n)$
consists of all points
$x=(x_1,x_2,\dots,x_{n})$ such that
\[
\frac12 >x_1, x_2>x_3>\cdots>x_{n}>0.
\]

(d) The fundamental domain $F_e(D_n)$ of $W^{\rm af\/f}_e(D_n)$
is contained in the domain of points
$x=(x_1,x_2,\dots, x_{n})$  such that
\[
1> x_1,x_2>x_3>\cdots>x_{n-1}>|x_n|.
\]
Moreover, a point $x$ of this domain belongs to $F_e(D_n)$ if and only if
$x_1+x_2<1$.

\subsection[$W_e$-orbits of $A_3$]{$\boldsymbol{W_e}$-orbits of $\boldsymbol{A_3}$}\label{Orb2}

$W_e$-orbits for $A_3$, $B_3$ and $C_3$ can be calculated by using
the orthogonal coordinates on the corresponding Euclidean
space, described above, and the description of action of
the Weyl groups $W_e(A_3)$, $W_e(B_3)$ and $W_e(C_3)$
in the orthogonal coordinate systems. Below we give results of such
calculations. Points $\lambda$ of $W_e$-orbits are given in the
$\omega$-coordinates in the form $(a\,b\,c)$, where
$\lambda=a\omega_1+b\omega_2+c\omega_3$.

If $a>0$, $b>0$, $c>0$, then $W_e$-orbits $O_e(a\ b\ c)$ and
$O_e(a{+}b\ {-}b\ b{+}c)\equiv r_\alpha O_e(a\ b\ c)$ of
$A_3$ contain the points
\begin{gather*}
O_e(a\ b\ c)\ni (a\ b\ c),\  (a{+}b\ c\ {-}b{-}c),\ (a{+}b{+}c\ {-}b{-}c\ b),
\\
\quad\qquad\qquad ({-}a\ a{+}b{+}c\ {-}c),\ (b\ {-}a{-}b\ a{+}b{+}c),\
({-}a{-}b\ a\ b{+}c),
\\
O_e(a{+}b\ {-}b\ b{+}c)\ni (a{+}b\ {-}b\ b{+}c),\ (a\ b{+}c\ {-}c),\
(a{+}b{+}c\ {-}c\ {-}b),
\\
\quad\qquad\qquad ({-}a\ a{+}b\ c),\  (b{+}c\ {-}a{-}b{-}c\ a{+}b),\
 ({-}b\ {-}a\ a{+}b{+}c)
\end{gather*}
and the points, contragredient to these points, where the
contragredient of the point $(a'\ b'\ c')$ is $({-}c'\ {-}b'\
{-}a')$.

There exist also the $W_e$-orbits
 \begin{gather*}
O_e(a\ b\ 0)\ni (a\ b\ 0), (a{+}b\ {-}b\ b), (a{+}b\ 0\ {-}b),
   ({-}a\ a{+}b\ 0), ({-}a{-}b\ a\ b),
 \\   \qquad\qquad
(b\ {-}a{-}b\ a{+}b)\ {\rm and\  contragredient\  points};
 \\
O_e(a\ 0\ c)\ni (a\ 0\ c), (a\ c\ {-}c), (a{+}c\ {-}c\ 0),
   ({-}a\ a\ c), (0\ {-}a\ a{+}c),
 \\   \qquad\qquad
({-}a\ a{+}c\ {-}c)\ {\rm and\  contragredient\  points};
 \\
O_e(0\ b\ c)\ni (0\ b\ c), (b\ {-}b\ b{+}c), (0\ b{+}c\ {-}c),
   (b{+}c\ {-}b{-}c\ b), ({-}b\ 0\ b{+}c),
 \\   \qquad\qquad
(b\ c\ {-}b{-}c)\ {\rm and\  contragredient\  points};
 \\
O_e(a\ 0\ 0)\ni (a\ 0\ 0), ({-}a\ a\ 0), (0\ 0\ {-}a),
   (0\ {-}a\ a);
 \\
O_e(0\ b\ 0)\ni (0\ b\ 0), (b\ {-}b\ b), (b\ 0\ {-}b), ({-}b\ b\ {-}b),
    (0\ {-}b\ 0), ({-}b\ 0\ b);
 \\
O_e(0\ 0\ c)\ni (0\ 0\ c), (0\ c\ {-}c), (c\ {-}c\ 0),
   ({-}c\ 0\ 0).
 \end{gather*}

\subsection[$W_e$-orbits of $B_3$]{$\boldsymbol{W_e}$-orbits of $\boldsymbol{B_3}$} \label{Orb3}

As in the previous case, points $\lambda$ of $W_e$-orbits are given
by the $\omega$-coordinates $(a\ b\ c)$, where
$\lambda=a\omega_1+b\omega_2+c\omega_3$. The $W_e$-orbits $O_e(a\
b\ c)$ and $O_e(a{+}b\ {-}b\ 2b{+}c)\equiv r_\alpha O_e(a\
b\ c)$, $a>0$, $b>0$, $c>0$, of $B_3$
contain the points
\begin{gather*}
O_e(a\ b\ c)\ni  (a\ b\ c),\  (b\ {-}a{-}b\ 2a{+}2b{+}c),\
({-}a{-}b\ a\ 2b{+}c),\   (a{+}b{+}c\ {-}b{-}c\ 2b{+}c),
\\
\qquad\qquad ({-}a\ a{+}b{+}c\ {-}c),\  ({-}b{-}c\ {-}a\ 2a{+}2b{+}c),\
({-}a{-}b{-}c\ {-}b\ 2b{+}c),
\\
\qquad\qquad (a{+}2b{+}c\ {-}a{-}b{-}c\ c),\
({-}b\ a{+}2b{+}c\ {-}2a{-}2b{-}c),\ ({-}a{-}2b{-}c\ b{+}c\ {-}c),
\\
\qquad\qquad (b{+}c\ a{+}b\ {-}2a{-}2b{-}c),\
   (a{+}b\ {-}a{-}2b{-}c\ 2b{+}c),\
\\
O_e(a{+}b\ {-}b\ 2b{+}c)\ni   (a{+}b\ {-}b\ 2b{+}c),\
 ({-}a\ a{+}b\ c),\
({-}b\ {-}a\ 2a{+}2b{+}c),\ (a\ b{+}c\ {-}c),
\\
\qquad\qquad  (b{+}c\ {-}a{-}b{-}c\ 2a{+}2b{+}c),\
   ({-}a{-}b{-}c\ a\ 2b{+}c),\
({-}a{-}2b{-}c\ b\ c),
\\
\qquad\qquad  (b\ a{+}b{+}c\ {-}2a{-}2b{-}c),\
(a{+}b{+}c\ {-}a{-}2b{-}c\ 2b{+}c),\
({-}a{-}b\ {-}b{-}c\ 2b{+}c),
\\
\qquad\qquad  (a{+}2b{+}c\ {-}a{-}b\ {-}c),\
 ({-}b{-}c\ a{+}2b{+}c\ {-}2a{-}2b{-}c)
\end{gather*}
and also all these points taken with opposite signs of coordinates.

In the case $B_3$ there exist also the $W_e$-orbits $O_e(a\ b\ 0)$,
$O_e(a\ 0\ c)$, $O_e(0\ b\ c)$, which are of the form
 \begin{gather*}
O_e(a\ b\ 0)\ni \pm (a\ b\ 0), \pm (a{+}b\ {-}b\ 2b), \pm ({-}a\ a{+}b\ 0),
\pm (b\ {-}a{-}b\ 2a{+}2b),
\\   \qquad\qquad
\pm ({-}a{-}b\ a\ 2b), \pm ({-}b\ {-}a\ 2a{+}2b),\pm ({-}a{-}2b\ b\ 0), \pm
({-}a{-}b\ {-}b\ 2b),
\\    \qquad\qquad
\pm (a{+}2b\ {-}a{-}b\ 0), \pm (b\ a{+}b\ {-}2a{-}2b), \pm
(a{+}b\ {-}a{-}2b\ 2b), \pm ({-}b\ a{+}2b\ {-}2a{-}2b);
\\
O_e(a\ 0\ c)\ni \pm (a\ 0\ c), \pm ({-}a\ a\ c), \pm (0\ {-}a\ 2a{+}c),
\pm (a\ c\ {-}c),
\\    \qquad\qquad
\pm (a{+}c\ {-}c\ c), \pm ({-}a\ a{+}c\ {-}c),\pm (c\ {-}a{-}c\ 2a{+}c), \pm
({-}a{-}c\ a\ c),
\\      \qquad\qquad
\pm ({-}c\ {-}a\ 2a{+}c), \pm ({-}a{-}c\ 0\ c), \pm (a{+}c\ {-}a{-}c\ c), \pm (0\
a{+}c\ {-}2a{-}c);
\\
 O_e(0\ b\ c)\ni \pm (0\ b\ c), \pm (b\ {-}b\ 2b{+}c), \pm ({-}b\ 0\
2b{+}c), \pm (0\ b{+}c\ {-}c),
\\      \qquad\qquad
\pm (b{+}c\ {-}b{-}c\ 2b{+}c), \pm ({-}b{-}c\ 0\ 2b{+}c),\pm ({-}2b{-}c\ b\ c), \pm
({-}b{-}c\ {-}b\ 2b{+}c),
\\      \qquad\qquad
\pm (2b{+}c\ {-}b{-}c\ c), \pm (b\ b{+}c\ {-}2b{-}c), \pm (b{+}c\ {-}2b{-}c\
2b{+}c), \pm ({-}b\ 2b{+}c\ {-}2b{-}c).
\end{gather*}
The $W_e$-orbits $O_e(a\ 0\ 0)$, $O_e(0\ b\ 0)$ and $O_e(0\ 0\ c)$ consist of
the points
\begin{gather*}
O_e(a\ 0\ 0)\ni \pm (a\ 0\ 0), \pm (a\ {-}a\ 0), \pm (0\ a\ {-}2a);
\\
O_e(0\ b\ 0)\ni \pm (0\ b\ 0), \pm (b\ {-}b\ 2b), \pm ({-}b\ 0\ 2b), \pm
({-}2b\ b\ 0),  \pm ({-}b\ {-}b\ 2b), \pm (b\ {-}2b\ 2b);
\\
O_e(0\ 0\ c)\ni \pm (0\ 0\ c), \pm (c\ {-}c\ c), \pm (0\ c\ {-}c), \pm
({-}c\ 0\ c) .
\end{gather*}

\subsection[$W_e$-orbits of $C_3$]{$\boldsymbol{W_e}$-orbits of $\boldsymbol{C_3}$}\label{Orb4}

As in the previous cases, points $\lambda$ of $W_e$-orbits are given
by the $\omega$-coordinates $(a\ b\ c)$, where
$\lambda=a\omega_1+b\omega_2+c\omega_3$. The $W_e$-orbits $O_e(a\
b\ c)$ and $O_e(a{+}b\ {-}b\ b{+}c)\equiv r_\alpha O_e(a\
b\ c)$, $a>0$, $b>0$, $c>0$, of $C_3$ contain the points
\begin{gather*}
O_e(a\ b\ c)\ni  (a\ b\ c),\  (b\ {-}a{-}b\ a{+}b{+}c),\
({-}a{-}b\ a\ b{+}c),\  (a{+}b{+}2c\ {-}b{-}2c\ b{+}c),
\\
\qquad\qquad ({-}a\ a{+}b{+}2c\ {-}c),\
({-}b{-}2c\ {-}a\ a{+}b{+}c),\ ({-}a{-}b{-}2c\ {-}b\ b{+}c),
\\
\qquad\qquad (a{+}2b{+}2c\ {-}a{-}b{-}2c\ c),\
({-}b\ a{+}2b{+}2c\ {-}a{-}b{-}c),\ ({-}a{-}2b{-}2c\ b{+}2c\ {-}c),\
\\
\qquad\qquad (b{+}2c\ a{+}b\ {-}a{-}b{-}c),\
(a{+}b\ {-}a{-}2b{-}2c\ b{+}c),
\\
O_e(a{+}b\ {-}b\ b{+}c)\ni  (a{+}b\ {-}b\ b{+}c),\  ({-}a\ a{+}b\ c),\
({-}b\ {-}a\ a{+}b{+}c),\ (a\ b{+}2c\ {-}c),
\\
\qquad\qquad  (b{+}2c\ {-}a{-}b{-}2c\ a{+}b{+}c),\
({-}a{-}b{-}2c\ a\ b{+}c),\
 ({-}a{-}2b{-}2c\ b\ c),
\\
\qquad\qquad  (b\ a{+}b{+}2c\ {-}a{-}b{-}c),\
(a{+}b{+}2c\ {-}a{-}2b{-}2c\ b{+}c),\
({-}a{-}b\ {-}b{-}2c\ b{+}c),\
\\
\qquad\qquad (a{+}2b{+}2c\ {-}a{-}b\ {-}c),\
 ({-}b{-}2c\ a{+}2b{+}2c\ {-}a{-}b{-}c)
\end{gather*}
and also all these points taken with opposite signs of coordinates.

For the $W_e$-orbits $O_e(a\ b\ 0)$, $O_e(a\ 0\ c)$ and $O_e(0\ b\ c)$ we have
 \begin{gather*}
O_e(a\ b\ 0)\ni \pm (a\ b\ 0), \pm (a{+}b\ {-}b\ b), \pm ({-}a\ a{+}b\ 0),
\pm (b\ {-}a{-}b\ a{+}b),
\\   \qquad\qquad
\pm ({-}a{-}b\ a\ b), \pm ({-}b\ {-}a\ a{+}b),\pm ({-}a{-}2b\ b\ 0),
\pm ({-}a{-}b\ {-}b\ b),
\\     \qquad\qquad
\pm (a{+}2b\ {-}a{-}b\ 0), \pm (b\ a{+}b\ {-}a{-}b),
\pm (a{+}b\ {-}a{-}2b\ b), \pm ({-}b\ a{+}2b\ {-}a{-}b);
\\
O_e(a\ 0\ c)\ni \pm (a\ 0\ c), \pm ({-}a\ a\ c), \pm (0\ {-}a\ a{+}c), \pm
(a\ 2c\ {-}c),
\\      \qquad\qquad
\pm (a{+}2c\ {-}2c\ c), \pm (a{+}2c\ {-}a{-}2c\ c),
\pm (0\ a{+}2c\ {-}a{-}c), \pm ({-}a\ a{+}2c\ {-}c),
\\        \qquad\qquad
\pm (2c\ {-}a{-}2c\ a{+}c), \pm ({-}a{-}2c\ a\ c),
\pm ({-}2c\ {-}a\ a{+}c), \pm  ({-}a{-}2c\ 0\ c);
\\
 O_e(0\ b\ c)\ni \pm (0\ b\ c), \pm (b\ {-}b\ b{+}c), \pm ({-}b\ 0\
b{+}c), \pm (0\ b{+}2c\ {-}c),
\\       \qquad\qquad
\pm (b{+}2c\ {-}b{-}2c\ b{+}c), \pm ({-}b{-}2c\ 0\ b{+}c),
\pm ({-}2b{-}2c\ b\ c), \pm ({-}b{-}2c\ -b\ b{+}c),
\\         \qquad\qquad
\pm (2b{+}2c\ {-}b{-}2c\ c), \pm (b\ b{+}2c\ {-}b{-}c),
\pm (b{+}2c\ {-}2b{-}2c\ b{+}c), \pm ({-}b\ 2b{+}2c\ {-}b{-}c).
 \end{gather*}
The $W_e$-orbits $O_e(a\ 0\ 0)$, $O_e(0\ b\ 0)$ and $O_e(0\ 0\ c)$ consist of
the points
 \begin{gather*}
O_e(a\ 0\ 0)\ni \pm (a\ 0\ 0), \pm (a\ {-}a\ 0), \pm (0\ a\ {-}a);
\\
O_e(0\ b\ 0)\ni \pm (0\ b\ 0), \pm (b\ {-}b\ b), \pm (b\ 0\ {-}b), \pm
(2b\ {-}b\ 0),  \pm ({-}b\ {-}b\ b), \pm (b\ {-}2b\ b);
\\
O_e(0\ 0\ c)\ni \pm (0\ 0\ c), \pm (0\ 2c\ {-}c), \pm (2c\ {-}2c\ c),
\pm (2c\ 0\ {-}c) .
 \end{gather*}

\section[$E$-orbit functions]{$\boldsymbol{E}$-orbit functions}\label{Orb}

\subsection[Definition]{Def\/inition}\label{Orb-1}

$E$-orbit functions are obtained from
the exponential functions $e^{2\pi{\rm i}\langle \lambda,x\rangle}$, $x\in
E_n$, with f\/ixed $\lambda=(\lambda_1,\lambda_2,\dots ,\lambda_n)$ by the procedure of
symmetrization by means of the even Weyl group $W_e$.
$E$-orbit functions are closely related to symmetric and
antisymmetric $W$-orbit functions. For this reason, we f\/irst
def\/ine the last functions.

Let $W$ be a Weyl group of transformations of the Euclidean space
$E_n$. To each element $\lambda\in E_n$ from the dominant Weyl
chamber (that is, $\langle \lambda, \alpha_i\rangle \ge 0$
for all simple roots $\alpha_i$)
there corresponds a {\it symmetric orbit function}
$\phi_\lambda$ on $E_n$, which is given by the formula
 \begin{equation}\label{orb-s}
\phi_\lambda(x)=\sum_{\mu\in O(\lambda)} e^{2\pi{\rm i}\langle
\mu,x\rangle}, \qquad x\in E_n,
 \end{equation}
where $O(\lambda)$ is the $W$-orbit of the element $\lambda$. The
number of summands is equal to the size $|O(\lambda)|$ of the orbit
$O(\lambda)$ and we have $\phi_\lambda(0)=|O(\lambda)|$.
Sometimes (see, for example, \cite{Pat-Z-1} and \cite{Pat-Z-2}), it
is convenient to use a modif\/ied def\/inition of orbit functions:
\begin{equation}\label{orb-2}
\hat \phi_\lambda(x)=|W_\lambda| \phi_\lambda (x),
\end{equation}
where $W_\lambda$ is a subgroup in $W$ whose elements leave $\lambda$
f\/ixed. Then for all orbit functions $\hat\phi_\lambda$ we have $\hat
\phi_\lambda (0)=|W|$. The functions $\hat\phi_\lambda(x)$ can be
def\/ined as
\[
\hat\phi_\lambda(x)=\sum_{w\in W} e^{2\pi{\rm i}\langle
\mu,x\rangle}.
\]

Antisymmetric orbit functions are def\/ined
(see \cite{P-SIG-05} and \cite{PZ-06}) for dominant elements
$\lambda$, which do not belong to a wall of the dominant Weyl chamber (that
is, for strictly dominant elements~$\lambda$). The {\it antisymmetric
orbit function}, corresponding to such an element, is def\/ined as
 \begin{equation}\label{orb-a}
\varphi_\lambda(x)=\sum_{w\in W} (\det w) e^{2\pi{\rm i}\langle
w\lambda,x\rangle}, \qquad x\in E_n.
 \end{equation}
A number of summands in \eqref{orb-a} is equal to the size $|W|$ of
the Weyl group $W$. We have $\varphi_\lambda(0)=0$.

$E$-orbit functions are def\/ined for each element $\lambda$ of the domain
$D_+^e=D_+ \cup r_\alpha D_+$, where $D_+$ is the set of dominant
elements of $E_n$ and $\alpha$ is a root of the root system (we assume
that each point is taken only once). The $E$-orbit function $E_\lambda(x)$,
$\lambda\in D^e_+$, is given by the formula
 \begin{equation}\label{orb-ex}
E_\lambda(x)=\sum_{\mu\in O_e(\lambda)} e^{2\pi{\rm i}\langle
\mu,x\rangle}, \qquad x\in E_n,
 \end{equation}
where $O_e(\lambda)$ is the $W_e$-orbit of the point $\lambda$.

Sometimes, it is convenient to use normalized $E$-orbit functions
def\/ined as
 \begin{equation}\label{orb-ex-1}
\hat E_\lambda(x)=\sum_{w\in W_e} e^{2\pi{\rm i}\langle
w\lambda,x\rangle}.
 \end{equation}
We have $\hat E_\lambda(x)=|W^\lambda_e| E_\lambda(x)$, where
$W^\lambda_e$ is a subgroup of $W_e$ whose elements leave
$\lambda$ invariant.
\medskip

\noindent
{\bf Example.} {\it $E$-orbit functions for $A_1$}. In this
case, there exists only one simple (positive) root~$\alpha$. We have
$\langle \alpha,\alpha \rangle =2$. Then the relation
$2\langle \omega,\alpha \rangle / \langle \alpha,\alpha \rangle =1$
means that $\langle \omega,\alpha \rangle =1$. Therefore,
$\omega=\alpha/2$ and $\langle \omega,\omega \rangle =1/2$.
We identify
points $x$ of $E_1\equiv {\mathbb R}$ with $\theta \omega$. The
Weyl group $W(A_1)$ consists of two elements 1 and $r_\alpha$ and
$\det r_\alpha=-1$. The even Weyl group $W_e(A_1)$ consists of
one element~1. For this reason, $W_e$-orbit functions
$\varphi_\lambda(x)$,
$\lambda=m\omega$, $m\in {\mathbb R}$, in this case
coincide with exponential functions:
\[
E_\lambda(x)=e^{2\pi{\rm i}\langle m\omega,\theta\omega
\rangle}=e^{\pi{\rm i}m\theta}.
\]
Note that for the symmetric and antisymmetric orbit functions
$\phi_\lambda(x)$ and $\varphi_\lambda(x)$ we have
\begin{gather*}
\phi_\lambda(x)=2\cos (\pi m\theta), \qquad
\varphi_\lambda(x)=2{\rm i}\sin (\pi m\theta).
\end{gather*}
Therefore, $\phi_\lambda(x)=E_\lambda(x)+E_{-\lambda}(x)$ and
$\varphi_\lambda(x)=E_\lambda(x)-E_{-\lambda}(x)$.

\subsection[$E$-orbit functions of $A_2$]{$\boldsymbol{E}$-orbit functions of $\boldsymbol{A_2}$}\label{Orb-2}

Put $\lambda=a\omega_1+b\omega_2\equiv (a\ b)$ with $a>0$, $b>0$. Then
for $\varphi_\lambda(x)\equiv \varphi_{(a\ b)}(x)$ we receive from
\eqref{orb-a} that
\begin{gather*}
  E_{(a\ b)}(x)
   =    e^{2\pi i\l(a\ b), x\r} + e^{2\pi i\l(b\ -a-b), x\r}
       + e^{2\pi i\l(-a-b\ a), x\r},  \notag\\
  E_{(-a\ a+b)}(x)
   =    e^{2\pi i\l(-a\ a+b), x\r}+ e^{2\pi i\l(a+b\ -b), x\r}
       + e^{2\pi i\l(-b\ -a), x\r} . \notag
\end{gather*}
Using the representation $x=\psi_1\alpha_1+\psi_2\alpha_2$,
one obtains
\begin{gather}
  E_{(a\ b)}(x)
    =   e^{2\pi i(a\psi_1+b\psi_2)} + e^{2\pi i(b\psi_1-(a+b)\psi_2)}
      + e^{2\pi i((-a-b)\psi_1+a\psi_2)},\\
  E_{(-a\ a+b)}(x)
    =    e^{2\pi i(-a\psi_1+(a+b)\psi_2)}+ e^{2\pi i((a+b)\psi_1-b\psi_2)}
        + e^{2\pi i(-b\psi_1-a\psi_2)} .
\end{gather}
An expression for $ E_{(a\ b)}(x)$ depends on a
choice of coordinate systems for $\lambda$ and $x$. Setting
$x=\theta_1\omega_1+\theta_2\omega_2$ and $\lambda$ as before, we
get
\begin{gather*}
  E_{(a\ b)}(x)
    =   e^{\tfrac{2\pi i}3((2a+b)\theta_1+(a+2b)\theta_2)}
              + e^{-\tfrac{2\pi i}3((a-b)\theta_1+(2a+b)\theta_2)}
    + e^{-\tfrac{2\pi i}3((a+2b)\theta_1+(-a+b)\theta_2)},    \notag\\
  E_{(-a\ a+b)}(x)
    =    e^{\tfrac{2\pi i}3((-a+b)\theta_1+(a+2b)\theta_2)}
        + e^{\tfrac{2\pi i}3((2a+b)\theta_1+(a- b)\theta_2)}
        + e^{-\tfrac{2\pi i}3((a+2b)\theta_1+(2a+b)\theta_2)}.\notag
\end{gather*}

Similarly one f\/inds that $ E_{(a\ 0)}(x)$ and $ E_{(0\ b)}(x)$ are of
the form
\begin{gather}
 E_{(a\ 0)}(x)  =e^{\tfrac{2\pi i}3a(2\theta_1+ \theta_2)}
              +e^{\tfrac{2\pi i}3a(-\theta_1+ \theta_2)}
              +e^{\tfrac{2\pi i}3a(-\theta_1-2\theta_2)},\\
 E_{(0\ b)}(x)  =e^{\tfrac{2\pi i}3b( \theta_1+2\theta_2)}
              +e^{\tfrac{2\pi i}3b( \theta_1- \theta_2)}
              +e^{\tfrac{2\pi i}3b(-2\theta_1-\theta_2)}.
\end{gather}
Note that the pairs $ E_{(a\ b)}(x)+ E_{(b\ a)}(x)$ are
always real functions.

\subsection[$E$-orbit functions of $C_2$ and $G_2$]{$\boldsymbol{E}$-orbit functions of $\boldsymbol{C_2}$ and $\boldsymbol{G_2}$}

Putting again $\lambda=a\omega_1+b\omega_2=(a\,b)$,
$x=\theta_1\omega_1+\theta_2\omega_2$ and using the matrices $S$
from \eqref{matr}, which are of the form
\[
S(C_2)=\frac12\begin{pmatrix}1&1\\1&2\end{pmatrix} ,\qquad
S(G_2)=\frac16\begin{pmatrix}6&3\\3&2\end{pmatrix} ,
\]
we f\/ind the orbit functions $E_{(a\ b)}(x)$ for $C_2$ and $G_2$
with $a>0$ and $b>0$:
\begin{gather}
C_2 :\quad  E_{(a\ b)}(x)
  =2\cos\pi ((a+b)\theta_1+(a+2b)\theta_2)
   +2\cos\pi(b\theta_1-a\theta_2), \\
\phantom{C_2 :\quad{}}{} E_{(-a\ a+b)}(x)
  =2\cos\pi (b\theta_1+(a+2b)\theta_2)
+2\cos\pi ((a+b)\theta_1+a\theta_2) ,  \\
G_2 :\quad  E_{(a\ b)}(x)
  =2\cos\pi (2a+b)\theta_1+(a+\tfrac23b)\theta_2)\notag\\
\phantom{G_2 :\quad E_{(a\ b)}(x)=}{} +2\cos\pi((a+b)\theta_1+\tfrac13 b\theta_2)
       +2\cos\pi(a\theta_1+(a+\tfrac13b)\theta_2),\\
\phantom{G_2 :\quad{}}{}  E_{(-a\ 3a+b)}(x)
  =2\cos\pi ((a+b)\theta_1+(a+\tfrac23 b)\theta_2)\notag\\
\phantom{G_2 :\quad E_{(-a\ 3a+b)}(x)=}{}+2\cos\pi((2a+b)\theta_1+(a+\tfrac13 b)\theta_2)
           +2\cos\pi(a\theta_1-\tfrac13b\theta_2).
\end{gather}
When one of the numbers $a$ and $b$ vanishes, then we have
\begin{align}
C_2 : &&  E_{(a\ 0)}(x)
  &=2\cos\pi a(\theta_1+\theta_2) +2\cos\pi a\theta_2,&\\
      &&E_{(0\ b)}(x)
  &=2\cos\pi b(\theta_1+2\theta_2)
     +2\cos\pi b\theta_1, & \\
G_2 : && \ E_{(a\ 0)}(x)
  &=2\cos\pi a(2\theta_1+\theta_2)+2\cos\pi a(\theta_1+\theta_2)
       +2\cos\pi a\theta_1,&\\
      && \ E_{(0\ b)}(x)
  &=2\cos\pi b(\theta_1+\tfrac23\theta_2)
            +2\cos\pi b(\theta_1+\tfrac13\theta_2)
            +2\cos\pi \tfrac13b\theta_2.&
\end{align}
As we see, $E$-functions for $C_2$ and $G_2$ are real.

\subsection[$E$-orbit functions of $A_n$]{$\boldsymbol{E}$-orbit functions of $\boldsymbol{A_n}$}

It is dif\/f\/icult to write down an explicit form of $E$-orbit functions
for $A_n$, $B_n$, $C_n$ and $D_n$ in coordinates with respect to the
$\omega$- or $\alpha$-bases. For this reason, for these cases we use
the orthogonal coordinate systems, described in Section~\ref{s-orbi}.

Let $\lambda=(m_1,m_2,\dots ,m_{n+1})$ be a strictly dominant
element for $A_n$ in orthogonal coordinates described in Subsection~\ref{An}. Then $m_1> m_2> \cdots > m_{n+1}$. The Weyl group in this case
coincides with the symmetric group $S_{n+1}$ and the even Weyl group
coincides with the alternating subgroup~$S^e_{n+1}$ of $S_{n+1}$.
The $W_e$-orbit
$O_e(\lambda)$ consists of points $(w\lambda)$, $w\in
W_e$. Represen\-ting points $x\in E_{n+1}$ in the
orthogonal coordinate system, $x=(x_1,x_2,\dots ,x_{n+1})$, and
using formula~\eqref{orb-a} we f\/ind that
\begin{gather}\label{orb-f}
 E_\lambda(x)  =
 \sum _{w\in S^e_{n+1}}  e^{2\pi{\rm i}\langle w(m_1,
 \dots ,m_{n+1}),(x_1,\dots,x_{n+1})\rangle}
 = \sum _{w\in S^e_{n+1}}  e^{2\pi{\rm i}((w\lambda)_1x_1+
 \cdots +(w\lambda)_{n+1}x_{n+1})} ,
\end{gather}
where $(w\lambda)_{1},(w\lambda)_{2},\dots ,(w\lambda)_{n+1}$ are
the orthogonal coordinates of the point $w\lambda$.

The second type of $E$-orbit functions correspond to elements
$\lambda=(m_1,m_2,\dots ,m_{n+1})$, for which
$m_2> m_1> \cdots > m_{n+1}$. For this case the $E$-orbit functions
are given by the same formula~\eqref{orb-f}.

If $\lambda=(m_1,m_2,\dots ,m_{n+1})$ is dominant but not strictly dominant
(that is, some of $m_i$ are coinciding), then the corresponding
$E$-orbit function is equal to the symmetric orbit function
$\phi_{(m_{1},m_2,\dots ,m_{n+1})}(x)$ and, thus, we have
\begin{align}
 E_\lambda(x) & =
 \sum _{w\in S_{n+1}/S_\lambda} e^{2\pi{\rm i}\langle w(m_1,
 \dots ,m_{n+1}),(x_1,\dots,x_{n+1})\rangle} \notag \\
  & =
 \sum _{w\in S_{n+1}/S_\lambda} e^{2\pi{\rm i}((w\lambda)_1x_1+
 \cdots +(w\lambda)_{n+1}x_{n+1})} ,\label{orb-f-m}
\end{align}
where $S_\lambda$ is a subgroup of element of $S_{n+1}$
leaving $\lambda$ invariant.

Note that the element $-(m_{n+1},m_n,\dots ,m_1)$ is strictly dominant
if the element $(m_1,m_2,\dots$, $m_{n+1})$ is strictly dominant.
In the Weyl group $W(A_n)$ there exists an element $w_0$ such that
\[
w_0(m_{1},m_2,\dots ,m_{n+1})= (m_{n+1},m_n,\dots ,m_1).
\]
Moreover, we have{\samepage
\begin{gather*}
\det w_0=1\qquad {\rm for}\qquad A_{4k-1}\qquad {\rm and}\qquad
A_{4k},
\\
\det w_0=-1\qquad {\rm for}\qquad A_{4k+1}\qquad {\rm and}\qquad
A_{4k+2}.
\end{gather*}
This means that $w_0\in W_e$ for $A_{4k-1}$ and $A_{4k}$ and
$w_0\not\in W_e$ for $A_{4k+1}$ and $A_{4k+2}$.}

It follows from here that for $A_{4k-1}$ and $A_{4k}$
in the expressions for the orbit functions
$E_{(m_{1},m_2,\dots ,m_{n+1})}(x)$ and
$E_{-(m_{n+1},m_n,\dots ,m_{1})}(x)$ there are summands
\begin{equation}\label{w_0}
e^{2\pi{\rm i}\langle w_0\lambda ,x\rangle}= e^{2\pi{\rm
i}(m_{n+1}x_1+
 \cdots +m_{1}x_{n+1})}\qquad
{\rm and}\qquad e^{-2\pi{\rm i}(m_{n+1}x_1+
 \cdots +m_{1}x_{n+1})} ,
\end{equation}
respectively, which are complex conjugate to each other.

Similarly, for $A_{4k-1}$ and $A_{4k}$, in the
expressions \eqref{orb-f} for the functions
$E_{(m_{1},m_2,\dots ,m_{n+1})}(x)$ and
$E_{-(m_{n+1},m_n,\dots ,m_{1})}(x)$ all other summands are
pairwise complex conjugate. Therefore,
\begin{equation}\label{compl}
E_{(m_{1},m_2,\dots
,m_{n+1})}(x)=\overline{E_{-(m_{n+1},m_n,\dots ,m_{1})}(x)}
\end{equation}
for $n=4k-1,4k$.
If to use for $\lambda$ the coordinates $\lambda_i=\langle
\lambda,\alpha^\vee_i\rangle$ in the $\omega$-basis instead of the
orthogonal coordinates $m_j$, then this equation can be written as
\[
E_{(\lambda_{1},\dots
,\lambda_{n})}(x)=\overline{E_{(\lambda_{n},\dots
,\lambda_{1})}(x)}.
\]

If $n=4k+1$ or $n=4k+2$, then the $E$-orbit function
$E_{-(m_{n+1},m_n,\dots ,m_{1})}(x)$ belongs to the second type
of $E$-orbit functions. In this case the orbit functions
$E_{(m_{1},m_2,\dots ,m_{n+1})}(x)$ and
$E_{-(m_{n+1},m_n,\dots ,m_{1})}(x)$ have no common summands.
In this case we have
\[
E_{-r_\alpha(m_{n+1},m_n,\dots ,m_{1})}(x)=
\overline{E_{(m_{1},m_2,\dots ,m_{n+1})}(x)},
\]
where $\alpha$ is a positive root of our root system.

According to \eqref{compl}, if
\begin{equation}\label{m-A}
(m_1,m_2,\dots ,m_{n+1})=-(m_{n+1},m_n,\dots ,m_{1})
\end{equation}
(that is, the element $\lambda$ has in the $\omega$-basis the
coordinates $(\lambda_1,\lambda_2,\dots ,\lambda_2,\lambda_1)$),
then {\it the $E$-orbit function $E_\lambda$ is real for
$n=4k-1,4k$.} This orbit function can be
represented as a sum of cosines of angles.

It is know from Proposition 2 in \cite{KP06} that
in the orthogonal coordinates
antisymmetric orbit functions $\varphi_{(m_{1},m_2,\dots ,m_{n+1})}(x)$,
$m_1>m_2>\cdots >m_{n+1}$,
of $A_n$ can be represented as determinants of certain matrices:
\begin{alignat}{2}  \label{det-A}
\varphi_{(m_{1},m_2,\dots ,m_{n+1})}(x)=& \det \left( e^{2\pi{\rm
i}m_ix_j}\right)_{i,j=1}^{n+1}
\notag\\
\equiv & \det \left(
 \begin{array}{cccc}
 e^{2\pi{\rm i}m_1x_1}& e^{2\pi{\rm i}m_1x_2}&\cdots & e^{2\pi{\rm
i}m_1x_{n+1}}\\
  e^{2\pi{\rm i}m_2x_1}& e^{2\pi{\rm i}m_2x_2}&\cdots & e^{2\pi{\rm
i}m_2x_{n+1}}\\
  \cdots & \cdots &
\cdots & \cdots \\\
 e^{2\pi{\rm i}m_{n+1}x_1}& e^{2\pi{\rm i}m_{n+1}x_2}&\cdots & e^{2\pi{\rm
i}m_{n+1}x_{n+1}} \end{array} \right) .
 \end{alignat}
It follows from this formula that the corresponding $E$-orbit functions
$E_{(m_{1},m_2,\dots ,m_{n+1})}(x)$ and
$E_{(m_{2},m_1,m_3,\dots ,m_{n+1})}(x)$ can be represented as
\begin{gather}\label{e-a-n-1}
E_{(m_{1},m_2,m_3,\dots ,m_{n+1})}(x)=\left[ \det \left( e^{2\pi{\rm
i}m_ix_j}\right)_{i,j=1}^{n+1}\right]^+,
\\
\label{e-a-n-2}
E_{(m_{2},m_1,m_3,\dots ,m_{n+1})}(x)=\left[ \det \left( e^{2\pi{\rm
i}m_ix_j}\right)_{i,j=1}^{n+1}\right]^-,
\end{gather}
where $\left[ \det C\right]^+$ and $\left[ \det C\right]^-$
mean a parts of the expression for the determinant of $C$
containing all terms with sigh plus and with sign minus, respectively.

\subsection[$E$-orbit functions of $B_n$]{$\boldsymbol{E}$-orbit functions of $\boldsymbol{B_n}$}

Let $\lambda=(m_1,m_2,\dots ,m_{n})$ be a strictly dominant element
for $B_n$ in orthogonal coordinates described in Subsection~\ref{Bn}.
Then $m_1> m_2> \cdots > m_{n}> 0$. The Weyl group $W(B_n)$ consists
of permutations of the coordinates $m_i$ with sign alternations of
some of them. The even Weyl group $W_e(B_n)$ consists of those
elements of $w\in W(B_n)$ for which $\det w=1$.
Representing points $x\in E_{n}$ also in the
orthogonal coordinate system, $x=(x_1,x_2,\dots ,x_{n})$, and using
formula \eqref{orb-a} we f\/ind that the antisymmetric orbit function
of $W(B_n)$, corresponding to element $\lambda$, coincides with
\begin{align}\label{orb-B}
 \varphi_\lambda(x) & =
\sum _{\varepsilon_i=\pm 1}\sum _{w\in S_{n}} (\det w)\varepsilon_1
\varepsilon_2\cdots \varepsilon_n e^{2\pi{\rm i}\langle
w(\varepsilon_1m_1, \dots ,\varepsilon_nm_{n}),
(x_1,\dots,x_{n})\rangle} \notag \\
  & =\sum _{\varepsilon_i=\pm 1}\sum _{w\in S_{n}}
(\det w)\varepsilon_1 \varepsilon_2\cdots \varepsilon_n e^{2\pi{\rm
i}((w(\varepsilon \lambda))_1x_1+
 \cdots +(w(\varepsilon\lambda))_{n}x_{n})} ,
\end{align}
where $(w(\varepsilon \lambda))_1,
 \dots ,(w(\varepsilon\lambda))_{n}$ are the orthogonal
 coordinates of the
 points $w(\varepsilon \lambda)$ if
$\varepsilon \lambda=(\varepsilon_1m_1$, $\dots
,\varepsilon_nm_{n})$.

In order to obtain the corresponding $E$-orbit function $E_\lambda(x)$ we
have to take in the expres\-sion~\eqref{orb-B} only those terms,
for which $(\det w)\varepsilon_1 \varepsilon_2\cdots \varepsilon_n=1$.
It is easy to see that
\begin{align}\label{orb-B-e}
 E_\lambda(x) &
   ={\sum _{\varepsilon_i=\pm 1}}'\sum _{w\in S^e_{n}}
 e^{2\pi{\rm i}((w(\varepsilon \lambda))_1x_1+
 \cdots +(w(\varepsilon\lambda))_{n}x_{n})} \notag \\
  &\quad +{\sum _{\varepsilon_i=\pm 1}}''\sum _{w\in S_n\backslash S^e_{n}}
 e^{2\pi{\rm i}((w(\varepsilon \lambda))_1x_1+
 \cdots +(w(\varepsilon\lambda))_{n}x_{n})} ,
\end{align}
where ${\sum\limits_{\varepsilon_i=\pm 1}}'$ means the sum over $\varepsilon_i$
such that $\varepsilon_1 \varepsilon_2\cdots \varepsilon_n=1$ and
${\sum \limits_{\varepsilon_i=\pm 1}}''$ means the sum over $\varepsilon_i$
such that $\varepsilon_1 \varepsilon_2\cdots \varepsilon_n=-1$. The
notation $S_n\backslash S^e_{n}$ means a complement of $S_n$ to
$S^e_{n}$, where, as before, $S_n^e$ is the alternating subgroup
of $S_n$.

For the $E$-orbit function $E_{r_\alpha\lambda}(x)$ we
respectively have
\begin{align}\label{orb-B-ee}
 E_{r_\alpha\lambda}(x) &
   ={\sum _{\varepsilon_i=\pm 1}}''\sum _{w\in S^e_{n}}
 e^{2\pi{\rm i}((w(\varepsilon \lambda))_1x_1+
 \cdots +(w(\varepsilon\lambda))_{n}x_{n})} \notag \\
  &\quad +{\sum _{\varepsilon_i=\pm 1}}'\sum _{w\in S_n\backslash S^e_{n}}
 e^{2\pi{\rm i}((w(\varepsilon \lambda))_1x_1+
 \cdots +(w(\varepsilon\lambda))_{n}x_{n})} ,
\end{align}
where ${\sum\limits _{\varepsilon_i=\pm 1}}'$ and ${\sum\limits_{\varepsilon_i=\pm 1}}''$
are such as in \eqref{orb-B-e}.

In $W(B_n)$ there exists an element $w_-$ which change signs of all
coordinates $m_i$. Then $\det w_-=1$ if $n=2k$ and
$\det w_-=-1$ if $n=2k+1$.
Therefore, for each summand $e^{2\pi{\rm
i}((w(\varepsilon \lambda))_1x_1+  \cdots +(w(\varepsilon
\lambda))_{n}x_{n})}$ in the expressions \eqref{orb-B-e} for the
$E$-orbit function $E_{(m_{1},m_2,\dots ,m_{n})}(x)$
there exists exactly one summand complex conjugate to it,
 $e^{-2\pi{\rm i}(((w(\varepsilon \lambda))_1x_1+  \cdots
+(w(\varepsilon \lambda))_{n}x_{n})}$, if $n=2k$.
This means that {\it $E$-orbit functions of $B_n$ are real
if $n=2k$.} These orbit functions can be represented as sums
of cosines of the corresponding angles.
If $n=2k+1$, then the expression $e^{-2\pi{\rm i}(((w(\varepsilon
\lambda))_1x_1+  \cdots +(w(\varepsilon \lambda))_{n}x_{n})}$
is not contained in the $E$-orbit function
$E_{(m_{1},m_2,\dots ,m_{n})}(x)$. Therefore, in this case
this expression belongs to the $E$-orbit function of
the second type, that is, to the $E$-orbit function
$E_{(m_{2},m_1,m_3,\dots ,m_{n})}(x)$. We conclude that
when $n=2k+1$, then the $E$-orbit functions $E_\lambda(x)$ and
$E_{r_\alpha\lambda}(x)$ are pairwise complex conjugate
to each other.

If $\lambda$ is dominant but not strictly dominant, then
the $E$-orbit function $E_\lambda(x)$ coincides with the
symmetric orbit function $ \phi_\lambda(x)$ and we have
\begin{align}\label{orb-B-dom}
 E_\lambda(x) & =
\sum _{\varepsilon_i=\pm 1}\sum _{w\in S_{n}/S_\lambda}
e^{2\pi{\rm i}\langle w(\varepsilon_1m_1, \dots
,\varepsilon_nm_{n}),
(x_1,\dots,x_{n})\rangle} \notag \\
  & =\sum _{\varepsilon_i=\pm 1}\sum _{w\in S_{n}/S_\lambda}
 e^{2\pi{\rm i}((w(\varepsilon \lambda))_1x_1+
 \cdots +(w(\varepsilon\lambda))_{n}x_{n})} .
\end{align}
The summation here is over those
$\varepsilon_i=\pm 1$ for which $m_i\ne 0$.

\subsection[$E$-orbit functions of $C_n$]{$\boldsymbol{E}$-orbit functions of $\boldsymbol{C_n}$}

Let $\lambda=(m_1,m_2,\dots ,m_{n})$ be a strictly dominant element
for $C_n$ in the orthogonal coordinates described in Subsection~\ref{Cn}.
Then $m_1> m_2> \cdots > m_{n}> 0$.
Representing points $x\in E_{n}$ also in the orthogonal
coordinate system, $x=(x_1,x_2,\dots ,x_{n})$, we f\/ind that
the antisymmetric orbit function
of $W(C_n)$, corresponding to $\lambda$, coincides with
\begin{align}\label{orb-C}
 \varphi_\lambda(x) & =
\sum _{\varepsilon_i=\pm 1}\sum _{w\in S_{n}} (\det w)
\varepsilon_1\varepsilon_2\dots \varepsilon_n e^{2\pi{\rm i}\langle
w(\varepsilon_1m_1, \dots ,\varepsilon_nm_{n}),
(x_1,\dots,x_{n})\rangle} \notag \\
  & =\sum _{\varepsilon_i=\pm 1}\sum _{w\in S_{n}} (\det w)
\varepsilon_1\varepsilon_2\cdots \varepsilon_n
 e^{2\pi{\rm i}((w(\varepsilon \lambda))_1x_1+
 \cdots +(w(\varepsilon\lambda))_{n}x_{n})} ,
\end{align}
where, as above, $(w(\varepsilon \lambda))_1, \dots ,
(w(\varepsilon\lambda))_{n}$ are the orthogonal coordinates of the
points $w(\varepsilon \lambda)$, where
$\varepsilon \lambda=(\varepsilon_1m_1$, $\dots
,\varepsilon_nm_{n})$.

In order to obtain the corresponding $E$-orbit function $E_\lambda(x)$ we
have to take in the expression~\eqref{orb-C} only those terms,
for which $(\det w)\varepsilon_1 \varepsilon_2\cdots \varepsilon_n=1$.
It is easy to see that
\begin{align}\label{orb-C-e}
 E_\lambda(x) &
   ={\sum _{\varepsilon_i=\pm 1}}'\sum _{w\in S^e_{n}}
 e^{2\pi{\rm i}((w(\varepsilon \lambda))_1x_1+
 \cdots +(w(\varepsilon\lambda))_{n}x_{n})} \notag \\
  &\quad +{\sum _{\varepsilon_i=\pm 1}}''\sum _{w\in S_n\backslash S^e_{n}}
 e^{2\pi{\rm i}((w(\varepsilon \lambda))_1x_1+
 \cdots +(w(\varepsilon\lambda))_{n}x_{n})} ,
\end{align}
where ${\sum\limits_{\varepsilon_i=\pm 1}}'$ means the sum over $\varepsilon_i$
such that $\varepsilon_1 \varepsilon_2\cdots \varepsilon_n=1$ and
${\sum\limits_{\varepsilon_i=\pm 1}}''$ means the sum over $\varepsilon_i$
such that $\varepsilon_1 \varepsilon_2\cdots \varepsilon_n=-1$. The
notation $S_n\backslash S^e_{n}$ means a complement of $S_n$ to
$S^e_{n}$. For the $E$-orbit functions $E_{r_\alpha\lambda}(x)$
we have the expression
\begin{align}\label{orb-C-ee}
 E_\lambda(x) &
   ={\sum _{\varepsilon_i=\pm 1}}''\sum _{w\in S^e_{n}}
 e^{2\pi{\rm i}((w(\varepsilon \lambda))_1x_1+
 \cdots +(w(\varepsilon\lambda))_{n}x_{n})} \notag \\
  &\quad +{\sum _{\varepsilon_i=\pm 1}}'\sum _{w\in S_n\backslash S^e_{n}}
 e^{2\pi{\rm i}((w(\varepsilon \lambda))_1x_1+
 \cdots +(w(\varepsilon\lambda))_{n}x_{n})} ,
\end{align}
where ${\sum\limits _{\varepsilon_i=\pm 1}}'$ and
${\sum\limits_{\varepsilon_i=\pm 1}}''$ are such as in \eqref{orb-C-e}.

In $W(C_n)$ there exists an element $w_-$ which change signs of all
coordinates $m_i$. Then $\det w_-=1$ if $n=2k$ and
$\det w_-=-1$ if $n=2k+1$.
Therefore, for each summand $e^{2\pi{\rm
i}((w(\varepsilon \lambda))_1x_1+  \cdots +(w(\varepsilon
\lambda))_{n}x_{n})}$ in the expressions \eqref{orb-C-e} for the
$E$-orbit function $E_{(m_{1},m_2,\dots ,m_{n})}(x)$
there exists exactly one summand complex conjugate to it,
 $e^{-2\pi{\rm i}(((w(\varepsilon \lambda))_1x_1+  \cdots
+(w(\varepsilon \lambda))_{n}x_{n})}$, if $n=2k$.
This means that {\it $E$-orbit functions of $C_n$ are real
if $n=2k$.} This orbit function of $C_n$ can be represented as a sum
of cosines of the corresponding angles.
If $n=2k+1$, then the expression $e^{-2\pi{\rm i}(((w(\varepsilon
\lambda))_1x_1+  \cdots +(w(\varepsilon \lambda))_{n}x_{n})}$
is contained in the $E$-orbit function
$E_{r_\alpha(m_{1},m_2,\dots ,m_{n})}(x)$. We conclude that
when $n=2k+1$, then the $E$-orbit functions $E_\lambda(x)$ and
$E_{r_\alpha\lambda}(x)$ are pairwise complex conjugate
to each other.

If $\lambda$ is dominant but not strictly dominant, then
the $E$-orbit function $E_\lambda(x)$ coincides with the
symmetric orbit function $ \phi_\lambda(x)$ and we have
\begin{align}\label{orb-C-dom}
 E_\lambda(x) & =
\sum _{\varepsilon_i=\pm 1}\sum _{w\in S_{n}/S_\lambda}
e^{2\pi{\rm i}\langle w(\varepsilon_1m_1, \dots
,\varepsilon_nm_{n}),
(x_1,\dots,x_{n})\rangle} \notag \\
  & =\sum _{\varepsilon_i=\pm 1}\sum _{w\in S_{n}/S_\lambda}
 e^{2\pi{\rm i}((w(\varepsilon \lambda))_1x_1+
 \cdots +(w(\varepsilon\lambda))_{n}x_{n})} ,
\end{align}
where the summation is over those
$\varepsilon_i=\pm 1$ for which $m_i\ne 0$.

Note that in the orthogonal coordinates expressions for the $E$-orbit functions
$E_{(m_{1},m_2,\dots ,m_{n})}(x)$ of $C_n$ coincide with
the expressions for the corresponding $E$-orbit functions
$E_{(m_{1},m_2,\dots ,m_{n})}(x)$ of~$B_n$.
However, $\alpha$-coordinates of the element $(m_1,m_2,\dots,m_n)$
for $C_n$ do not coincide with $\alpha$-coordinates of the
element $(m_1,m_2,\dots,m_n)$ for $B_n$, that is, in $\alpha$-coordinates
expressions for the corresponding $E$-orbit functions of $B_n$ and $C_n$
are dif\/ferent.

\subsection[$E$-orbit functions of $D_n$]{$\boldsymbol{E}$-orbit functions of $\boldsymbol{D_n}$}
Let $\lambda=(m_1,m_2,\dots ,m_{n})$ be a strictly dominant element
for $D_n$ in the orthogonal coordinates described in Subsection~\ref{Dn}.
Then $m_1> m_2> \cdots > m_{n-1}> |m_n|$. The Weyl group $W(D_n)$
consists of permutations of the coordinates with sign alternations
of even number of them. The even Weyl group $W_e(D_n)$
consists of even permutations of the coordinates with sign alternations
of even number of them.

Representing points $x\in E_{n}$ also in the
orthogonal coordinate system, $x=(x_1,x_2,\dots ,x_{n})$, and using
formula \eqref{orb-a} we f\/ind that the antisymmetric orbit function
$\varphi_\lambda(x)$ of $D_n$ is given by the formula
\begin{align}\label{orb-D}
 \varphi_\lambda(x) & =
{\sum_{\varepsilon_i=\pm 1}}'\; \sum _{w\in S_{n}} (\det w)
  e^{2\pi{\rm i}\langle
w(\varepsilon_1m_1, \dots ,\varepsilon_nm_{n}),
(x_1,\dots,x_{n})\rangle} \notag \\
  & ={\sum_{\varepsilon_i =\pm 1}}'\; \sum _{w\in S_{n}}  (\det
w)   e^{2\pi{\rm i}((w(\varepsilon\lambda))_1x_1+
 \cdots +(w(\varepsilon\lambda))_nx_{n})} ,
\end{align}
where $(w(\varepsilon \lambda))_1, \dots ,
(w(\varepsilon\lambda))_{n}$ are the orthogonal
coordinates of the points
$w(\varepsilon \lambda)$ and the prime at the sum symbol means that
the summation is over values of $\varepsilon_i$ with even number
of sign minus. We have taken into account that an alternation of
coordinates without any permutation does not change a
determinant.

Therefore, if $\lambda$ is strictly dominant, then the
$E$-orbit function $E_\lambda(x)$ is of the form
\begin{align}\label{orb-D-even}
 E_\lambda(x)  =
  & {\sum_{\varepsilon_i =\pm 1}}'\; \sum _{w\in S^e_{n}}
   e^{2\pi{\rm i}((w(\varepsilon\lambda))_1x_1+
 \cdots +(w(\varepsilon\lambda))_nx_{n})} ,
\end{align}
where $S_n^e$ is the even part of the symmetric group $S_n$ and the
f\/irst sum is such as in \eqref{orb-D}.
For the corresponding $E$-orbit functions $E_{r_\alpha\lambda}(x)$
we have
\begin{align}\label{orb-D-even-0}
 E_{r_\alpha\lambda}(x)  =
  & {\sum_{\varepsilon_i =\pm 1}}'\; \sum _{w\in S_{n}\backslash S^e_n}
   e^{2\pi{\rm i}((w(\varepsilon\lambda))_1x_1+
 \cdots +(w(\varepsilon\lambda))_nx_{n})} ,
\end{align}
where the f\/irst sum is such as in \eqref{orb-D-even}.

Let $m_n\ne 0$. Then in the expressions \eqref{orb-D-even}
for the $E$-orbit function
$E_{(m_{1},m_2,\dots ,m_{n})}(x)$ of~$D_{n=2k}$ for each
summand $ e^{2\pi{\rm i}((w(\varepsilon\lambda))_1x_1+
 \cdots +(w(\varepsilon\lambda))_nx_{n})}$
there exists exactly one summand complex
conjugate to it. This means that
these {\it $E$-orbit functions of $D_{2k}$ are real.}
Each orbit function of~$D_{2k}$ can be
represented as a sum of cosines of the corresponding angles.

It is also proved by using \eqref{orb-D-even} that for $m_n\ne 0$
{\it the $E$-orbit
functions $E_{(m_{1},\dots, m_{2k},m_{2k+1})}(x)$ and
$E_{(m_{1},\dots ,m_{2k},-m_{2k+1})}(x)$ of $D_{2k+1}$ are
complex conjugate.}

If $m_n=0$, then it follows from \eqref{orb-D-even} that
$E$-orbit functions $E_\lambda(x)$ of $D_n$ are real and
can be represented as a sum of cosines of certain angles.

For dominant, but not strictly dominant, elements $\lambda$ the
$E$-orbit functions coincide with the corresponding symmetric orbit functions
and we have
\begin{align}\label{orb-D-dom}
 E_\lambda(x) & =
{\sum_{\varepsilon_i=\pm 1}}'\; \sum _{w\in S_{n}/S_\lambda}
e^{2\pi{\rm i}\langle w(\varepsilon_1m_1, \dots
,\varepsilon_nm_{n}),
(x_1,\dots,x_{n})\rangle} \notag \\
  & ={\sum_{\varepsilon_i =\pm 1}}'\; \sum _{w\in S_{n}/S_\lambda}
 e^{2\pi{\rm i}((w(\varepsilon\lambda))_1x_1+
 \cdots +(w(\varepsilon\lambda))_nx_{n})} ,
\end{align}
where $S_\lambda$ is the subgroup of $S_n$ consisting of elements
leaving $\lambda$ invariant.

Note that in the expressions \eqref{orb-D-dom} for the orbit function
$E_{(m_{1},m_2,\dots ,m_{n})}(x)$ of $D_{2k}$ for each summand
$ e^{2\pi{\rm i}((w(\varepsilon\lambda))_1x_1+
 \cdots +(w(\varepsilon\lambda))_nx_{n})}$
there exists exactly one summand complex conjugate to it. This
means that {\it all orbit functions of $D_{2k}$ are real.}
Each orbit function of $D_{2k}$
can be represented as a sum of cosines of the corresponding angles.

It also follows from \eqref{orb-D-dom} that {\it for $D_{2k+1}$ the
$E$-orbit function $E_{(m_{1},m_2,\dots ,m_{n})}(x)$ is real if
and only if the condition $m_{2k+1}=0$ is fulfilled. The $E$-orbit
functions $E_{(m_{1},\dots, m_{2k},m_{2k+1})}(x)$ and
$E_{(m_{1},\dots ,m_{2k},-m_{2k+1})}(x)$ of $D_{2k+1}$ are
complex conjugate.}

\section[Properties of $E$-orbit functions]{Properties of $\boldsymbol{E}$-orbit functions}\label{sec5}

\subsection{Invariance with respect to the even Weyl group}
Since the scalar product $\langle\cdot ,\cdot \rangle$ in $E_n$ is
invariant with respect to the Weyl group $W$, that is,
\[
\langle wx,wy\rangle =\langle x,y\rangle, \qquad
 w\in W,\qquad x,y\in E_n,
\]
it is invariant with respect to the even Weyl group $W_e$, which
is a subgroup of $W$.
It follows from here that $E$-orbit functions $E_\lambda$ for strictly
dominant elements $\lambda$
are invariant with respect to $W_e$, that is,
\[
E_\lambda (w'x)=E_\lambda (x), \qquad w'\in W_e.
\]
Indeed, this relation is equivalent to the relation
$\hat E_\lambda (w'x)=\hat E_\lambda (x)$ for the functions
\eqref{orb-ex-1}. For $\hat E_\lambda(x)$ we have
\begin{equation*}
\hat E_\lambda (w'x)= \sum_{w\in W_e}  e^{2\pi{\rm i}\langle
w\lambda,w'x\rangle} = \sum_{w\in W_e} e^{2\pi{\rm i}\langle
{w'}^{-1}w\lambda,x\rangle}
=  \sum_{w\in W_e} (\det w) e^{2\pi{\rm i}\langle
w\lambda,x\rangle} =\hat E_\lambda (x)
\end{equation*}
since ${w'}^{-1}w$ runs over the whole group $W_e$ when $w$ runs over
$W_e$.

In the same way it is shown that the corresponding $E$-orbit functions
$E_{r_\alpha\lambda} (x)$, where $\lambda$ is as above and $\alpha$ is a
root, is invariant with respect to $W_e$.

If $\lambda$ is dominant, but not strictly dominant, then
$E_{\lambda} (x)$ coincides with the corresponding symmetric orbit function,
which is invariant with respect to the Weyl group $W$, and therefore with respect to
$W_e$.

\subsection[Invariance with respect to the even affine Weyl group]{Invariance with respect to the even af\/f\/ine Weyl group}

If $\lambda$ belongs to the set of dominant integral elements $P_+$ or to
$r_\alpha P_+$, then $E_{\lambda} (x)$ is invariant with respect to
the even af\/f\/ine group $W_e^{\rm af\/f}$. Recall that
$W_e^{\rm af\/f}$ is a semidirect product of the even Weyl group $W_e$
and the group $\hat Q$ of translations by elements of the
dual root system $Q^\vee$. In order to prove this invariance of
$E$-orbit functions $E_\lambda(x)$
we need only to prove their invariance with respect to
the subgroup $\hat Q$.
For this we note that for $\mu\in P$ and for $\nu\in Q^\vee$ we have
\[
\langle\mu,x+\nu\rangle=\langle\mu,x\rangle +\langle\mu,\nu\rangle
=\langle\mu,x\rangle + {\rm integer}.
\]
Hence,
\begin{equation}\label{even-invar}
E_\lambda(x+\nu)=|W_\lambda|^{-1} \sum_{w\in W_e} e^{2\pi{\rm i}\langle
w\lambda,x+\nu\rangle}=|W_\lambda|^{-1}
\sum_{w\in W_e}  e^{2\pi{\rm i}\langle
\lambda,x\rangle}=E_\lambda(x),
\end{equation}
where $W_\lambda$ is the subgroup of $W_e$ consisting of elements $w$
such that $w\lambda=\lambda$. Thus, $E_\lambda(x)$ is invariant
with respect to the even af\/f\/ine Weyl group $W_e^{\rm af\/f}$.

If $\lambda\not\in P$, then $E_\lambda$ is not invariant with
respect to $W_e^{\rm af\/f}$. It is invariant only under action by
elements of the even Weyl group $W_e$.

Due to the invariance of $E$-orbit functions
$E_\lambda$, $\lambda\in P_+$, with respect to the group
$W_e^{\rm af\/f}$, it is enough to consider them only on the even fundamental
domain $F_e$ of $W_e^{\rm af\/f}$ (see Subsection~\ref{Fund}). Values of
$E_\lambda$ on other points of $E_n$ are determined by using
the action of $W_e^{\rm af\/f}$ on $F_e$ or taking a limit.

\subsection{Relation to symmetric and antisymmetric orbit functions}

$E$-orbit functions are closely related to symmetric and antisymmetric
orbit functions $\phi_\lambda(x)$ and~$\varphi_\lambda(x)$.
We have seen that if $\lambda$ is lying on a wall of the dominant
Weyl chamber, then
\begin{equation}\label{even-symm}
E_\lambda(x)=\phi_\lambda(x).
\end{equation}

Since the Weyl group $W$ can be represented as a union of the even
Weyl group $W_e$ and of the set $r_\alpha W_e$,
\[
 W=W_e\bigcup r_\alpha W_e,
\]
then it follows from the def\/initions of symmetric and antisymmetric
orbit functions that for strictly dominant $\lambda$ we have
\begin{gather}\label{even-symm-1}
\phi_\lambda(x)=E_\lambda(x)+E_{r_\alpha\lambda}(x),
\\
\label{even-symm-2}
\varphi_\lambda(x)=E_\lambda(x)-E_{r_\alpha\lambda}(x).
\end{gather}
It follows from here that for such $\alpha$ one gets
\begin{gather}\label{even-symm-3}
E_\lambda(x)=\frac12 \left( \phi_\lambda(x) +\varphi_\lambda(x)\right),
\\
\label{even-symm-4}
E_{r_\alpha\lambda}(x)=\frac12 \left( \phi_\lambda(x) -\varphi_\lambda(x)\right).
\end{gather}
It is directly derived from \eqref{even-symm-1}--\eqref{even-symm-4}
that the relations
\begin{gather}\label{even-symm-5}
\phi^2_\lambda(x)-\varphi^2_\lambda(x)=4 E_\lambda(x)E_{r_\alpha\lambda}(x),
\\
\label{even-symm-6}
\phi^2_\lambda(x)+\varphi^2_\lambda(x)=2(
E^2_\lambda(x)+E^2_{r_\alpha\lambda}(x))
\end{gather}
are true, where $\lambda$ is strictly dominant.

\subsection{Continuity}
An $E$-orbit function $E_\lambda(x)$ is a f\/inite sum of
exponential functions. Therefore, it is continuous and has continuous
derivatives of all orders on the Euclidean space $E_n$.

Antisymmetric orbit functions $\varphi_\lambda$ vanish on the
boundary of the fundamental domain $F(W)$ of the Weyl group $W$. The
normal derivative $\partial {\bf n}$ of symmetric orbit functions
$\phi_\lambda$ to the boundary $\partial F(W)$ of the fundamental
domain $F(W)$ equals zero. The reason of these properties is
(anti)symmetry with respect to the ref\/lections with respect to walls
of the dominant Weyl chamber. These ref\/lections do not belong to
$W_e$ and, therefore, $E$-orbit functions do not have these
properties.

\subsection{Scaling symmetry}
Let $O_e(\lambda)$ be an $E$-orbit of $\lambda$, $\lambda\in D_+$.
Since $w(c\lambda)=cw(\lambda)$ for any $c\in
{\mathbb R}$ and for any $w\in W_e$, then the orbit $O_e(c\lambda)$ is an
orbit consisting of the points $cw\lambda$, $w\in W_e$.
Let $E_\lambda=\sum\limits_{w\in W_e} e^{2\pi{\rm
i}w\lambda}$ be the $E$-orbit function for $\lambda\in
D_+$. Then
\[
E_{c\lambda}(x)=| W_\lambda|^{-1} \sum_{w\in W_e} e^{2\pi{\rm i}\langle
cw\lambda ,x\rangle} =| W_\lambda|^{-1} \sum_{w\in W_e} e^{2\pi{\rm i}\langle
w\lambda ,cx\rangle} = E_{\lambda}(cx).
\]
The equality $E_{c\lambda}(x)=E_{\lambda}(cx)$ expresses
the {\it scaling symmetry of $E$-orbit functions.}

\subsection{Duality}
Due to the invariance of the scalar product $\langle \cdot, \cdot
\rangle$ with respect to the even Weyl group $W_e$, $\langle w\mu ,wy
\rangle= \langle \mu ,y \rangle$,  for $E$-orbit functions $\hat E_\lambda(x)$
(see Subsection~\ref{Orb-1}) we have
 \[
\hat E_\lambda(x)=\sum_{w\in W_e} e^{2\pi {\rm i}\langle
\lambda,w^{-1}x \rangle} =\sum_{w\in W_e} e^{2\pi {\rm
i}\langle \lambda,wx \rangle}= \hat E_x(\lambda).
 \]
It is easy to see that this relation is also true for $E$-orbit
functions $E_\lambda(x)$.
The relation $ E_\lambda(x)=E_x(\lambda)$ expresses the
{\it duality} of $E$-orbit functions.

$E$-orbit functions have also the following property
\begin{equation}\label{r-alpha}
 E_{r_\alpha\lambda}(x)=E_{\lambda}(r_\alpha x),
\end{equation}
where $\alpha$ is a root of the corresponding root system.
This relation follows from the fact that $r_\alpha W_e=W_er_\alpha$
and from the equalities
\[
\hat E_{r_\alpha\lambda}(x)= \sum_{w\in W_e} e^{2\pi{\rm i}\langle
w r_\alpha \lambda ,x\rangle}=  \sum_{w\in W_e} e^{2\pi{\rm i}\langle
 r_\alpha w\lambda ,x\rangle}= \sum_{w\in W_e} e^{2\pi{\rm i}\langle
w  \lambda ,r_\alpha x\rangle}=\hat E_{\lambda}(r_\alpha x),
\]
where $\lambda$ is strictly dominant.

\subsection{Orthogonality}
If values of $\lambda$ are integral points lying inside of the
fundamental domain $F_e$ of the even af\/f\/ine Weyl group $W_e^{\rm af\/f}$,
then the corresponding $E$-orbit functions are orthogonal on
the closure $\overline{F_e}$ of the fundamental domain
$F_e$ with respect to the Euclidean measure:
 \begin{equation}\label{ortog}
|\overline{F_e}|^{-1}\int_{\overline{F_e}} E_\lambda(x)
\overline{E_{\lambda'}(x)}dx=
|W_e|\delta_{\lambda\lambda'} ,
 \end{equation}
where the overbar over $E_{\lambda'}(x)$
means complex conjugation. This relation directly
follows from the orthogonality of the exponential functions
$e^{2\pi{\rm i}\langle \mu,x \rangle}$ (entering into the def\/inition
of $E$-orbit functions) for dif\/ferent weights $\mu$ and from the fact
that a given element $\nu\in P$ belongs to precisely one $E$-orbit
function. In \eqref{ortog}, $|\overline{F_e}|$ means an area of the
domain $\overline{F_e}$.

Sometimes, it is dif\/f\/icult to f\/ind the area $|\overline{F_e}|$. In this case it is
useful the following form of the formula \eqref{ortog}:
\[
\int_{\sf T} E_\lambda(x)\overline{E_{\lambda'}(x)}dx=
|W_e|\delta_{\lambda\lambda'} ,
 \]
where ${\sf T}$ is the torus in $E_n$ described in Subsection~9.1 of
\cite{KP07}.
If to assume that an area of ${\sf T}$ is equal to 1, $|{\sf T}|=1$,
then $|\overline{F_e}|=|W_e|^{-1}$ and formula \eqref{ortog} takes the form
\begin{equation}\label{ortog-tor}
\int_{\overline{F_e}} E_\lambda(x)\overline{E_{\lambda'}(x)}dx=
\delta_{\lambda\lambda'} .
 \end{equation}

If $\lambda$ is an integral point which lies on a wall of the even dominant
Weyl chamber $D_+^e$, then instead of \eqref{ortog-tor} we have the relation
\begin{equation}\label{ortog-tor-0}
\int_{\overline{F_e}} E_\lambda(x)\overline{E_{\lambda'}(x)}dx=
|W_\lambda|^{-1}\delta_{\lambda\lambda'} ,
 \end{equation}
where $W_\lambda$ is the subgroup of $W_e$ consisting of elements
$w\in W_e$ such that $w\lambda=\lambda$.

\subsection[Solutions of the Laplace equation: the cases
$A_n$, $B_n$, $C_n$ and $D_n$]{Solutions of the Laplace equation: the cases
$\boldsymbol{A_n}$, $\boldsymbol{B_n}$, $\boldsymbol{C_n}$ and $\boldsymbol{D_n}$}

We use orthogonal coordinates $x_1,x_2,\dots
,x_{n+1}$ in the case of $A_n$ and the
orthogonal coordinates $x_1,x_2,\dots ,x_n$ in the cases
$B_n$, $C_n$ and $D_n$ (see section 3).
The Laplace operator on $E_r$ in the orthogonal coordinates
has the form
\[
\Delta=\frac{\partial^2}{\partial
x^2_1}+\frac{\partial^2}{\partial x^2_2}+\cdots
+\frac{\partial^2}{\partial x^2_r} ,
\]
where $r=n+1$ for $A_n$ and $r=n$ for $B_n$, $C_n$ and
$D_n$. Let us consider the case $B_n$. We take a~summand from
the expression \eqref{orb-B} for
the $E$-orbit function $E_\lambda(x)$ of $B_n$ and
act upon it by the operator $\Delta$. We get
\begin{gather*}
\Delta  e^{2\pi{\rm i}((w(\varepsilon \lambda))_1x_1+
 \cdots + (w(\varepsilon \lambda))_nx_{n})}
\\
\qquad\qquad  =-4\pi^2[(\varepsilon_1 m_1)^2+
 \cdots +(\varepsilon_n m_{n})^2]e^{2\pi{\rm i}((w(\varepsilon \lambda))_1x_1+
 \cdots + (w(\varepsilon \lambda))_nx_{n})}
\\
\qquad\qquad  =-4\pi^2(m_1^2+\cdots +m_n^2)e^{2\pi{\rm i}((w(\varepsilon
\lambda))_1x_1+
 \cdots + (w(\varepsilon \lambda))_nx_{n})}
\\
\qquad\qquad  =-4\pi^2 \langle \lambda,\lambda \rangle\, e^{2\pi{\rm
i}((w(\varepsilon \lambda))_1x_1+
 \cdots + (w(\varepsilon \lambda))_nx_{n})},
\end{gather*}
where $\lambda=(m_1,m_2,\dots ,m_n)$ is the element
of $D_+^e(B_n)$, determining
$E_\lambda(x)$, in the orthogonal coordinates and $w\in S^e_n\equiv
S_n/S_2$.
Since this action of $\Delta$ does not depend on a summand from~\eqref{orb-B}, we have
\begin{equation}\label{Lap}
\Delta E_\lambda(x)= -4\pi^2\langle \lambda,\lambda \rangle
E_\lambda(x).
 \end{equation}
For $A_n$, $C_n$ and $D_n$ this formula also holds and the
corresponding proofs are the same. Remark that in the case $A_n$
the scalar product $\langle \lambda,\lambda \rangle$ is equal to
\[
\langle \lambda,\lambda \rangle =m_1^2+m_2^2+\cdots +m_{n+1}^2.
\]

Thus, $E$-orbit functions are eigenfunctions of the
Laplace operator on the Euclidean space~$E_r$.

\subsection[The Laplace operator in $\omega$-basis]{The Laplace operator in $\boldsymbol{\omega}$-basis}

We may parametrize points of $E-n$ by coordinates in the
$\omega$-basis: $x=\theta_1\omega_1+\cdots+ \theta_n\omega_n$.
Denoting by $\p_k$ partial derivative with respect to $\theta_k$,
we have the Laplace operator $\Delta$ in the form
\begin{equation}\label{operator}
\Delta = \frac12\sum_{i,j=1}^n\l\alpha_j ,\alpha_j\r ^{-1}
       M_{ij}\p_i\p_j,
\end{equation}
where $(M_{ij})$ is the corresponding Cartan matrix. One can see
that it is indeed the Laplace operator as follows. The matrix
$(S_{ij})=(\l\alpha_j ,\alpha_j\r ^{-1}M_{ij})$ is symmetric with
respect to transposition and its determinant is positive.
Hence it can be diagonalized, so that $\Delta$ becomes a sum of
second derivatives with no mixed derivative terms.

We write
down explicit form of the Laplace operators in coordinates in the
$\omega$-basis for ranks~2 and~3. For rank two the operator $\Delta$
is of the form
\begin{gather}
A_2: \quad (\p_1^2-\p_1\p_2+\p_2^2)E_\lambda(x)
     =-\tfrac{4\pi^2}3(a^2+ab+b^2)E_\lambda(x),
\\
C_2:\quad  (2\p_1^2-2\p_1\p_2+\p_2^2)E_\lambda(x)
     =-2\pi^2(a^2+4ab+4b^2)E_\lambda(x),
\\
G_2 :\quad (\p_1^2-3\p_1\p_2+3\p_2^2)E_\lambda(x)
     =-\tfrac{4\pi^2}3(3a^2+3ab+b^2)E_\lambda(x).
   \end{gather}
Here, $\lambda=(a\;b)$ and $x=(\theta_1\;\theta_2)$.

In the semisimple case $A_1\times A_1$ one has
$M=2\left(\begin{smallmatrix}1&0\\0&1\end{smallmatrix}\right)$,
therefore $\Delta=2\p_1^2+2\p_2^2$, and $E_\lambda(x)$ is the
product of two $E$-orbit functions, one from each $A_1$.

There are three 3-dimensional cases, namely $A_3$,
$B_3$, and $C_3$. For $A_3$, $B_3$, and $C_3$
the result can be represented by the formulas
 \begin{gather} A_3:\quad\Delta=\p_1^2+\p_2^2+\p_3^2-\p_1\p_2-\p_2\p_3,
\notag\\
B_3:\quad\Delta=\p_1^2+\p_2^2+2\p_3^2-\p_1\p_2-2\p_2\p_3,
\notag
\\
C_3:\quad\Delta=2 \p_1^2+2 \p_2^2+2\p_3^2-2 \p_1\p_2-2\p_2\p_3.
\end{gather}
In all these case we have
\begin{equation}\label{Laplac}
\Delta E_\lambda(x)= -4\pi^2\langle \lambda,\lambda \rangle
E_\lambda(x).
 \end{equation}

\subsection[$E$-orbit functions as eigenfunctions of
other operators]{$\boldsymbol{E}$-orbit functions as eigenfunctions of
other operators}

$E$-orbit functions are eigenfunctions of many other
operators. Let us consider examples of such operators.

With each $y\in E_n$ we associate the shift operator $T_y$ which
acts on the exponential functions $e^{2\pi{\rm i}\langle \lambda,x
\rangle}$ as
\[
T_y e^{2\pi{\rm i}\langle \lambda,x \rangle}=e^{2\pi{\rm i}\langle
\lambda,x+y \rangle}=e^{2\pi{\rm i}\langle \lambda,y
\rangle}e^{2\pi{\rm i}\langle \lambda,x \rangle}.
\]
An action of elements of the even Weyl group $W_e$ on
functions, given on $E_n$, is given as $wf(x)=f(wx)$. For each $y\in
E_n$ we def\/ine an operator acting on orbit functions by the
formula
\[
D_y=\sum_{w\in W_e} (\det w) wT_y.
\]
Then
\begin{alignat}{3}
D_y \hat E_\lambda(x)&=D_y\sum _{w\in W}
e^{2\pi{\rm i}\langle w\lambda,x \rangle}
=\sum_{w'\in W_e} \sum _{w\in W_e}  e^{2\pi{\rm
i}\langle w\lambda,y \rangle} e^{2\pi{\rm i}\langle w\lambda,w'x
\rangle}
\notag \\
&=\sum _{w\in W_e}  e^{2\pi{\rm i}\langle w\lambda,y \rangle}
\sum _{w'\in W_e}  e^{2\pi{\rm i}\langle w\lambda,w'x
\rangle}
=\sum _{w\in W_e}  e^{2\pi{\rm i}\langle w\lambda,y \rangle}
\sum _{w'\in W_e}  e^{2\pi{\rm i}\langle {w'}^{-1}w\lambda,x
\rangle}
\notag \\
&= \sum _{w\in W_e} e^{2\pi{\rm i}\langle w\lambda,y
\rangle} \hat E_\lambda(x) =\hat E_\lambda(y)\hat E_\lambda(x),
\notag
\end{alignat}
that is, $\hat E_\lambda(x)$ (and therefore $E_\lambda(x)$)
is an eigenfunction of the operator
$D_y$ with eigen\-va\-lue~$\hat E_\lambda(y)$.

It is shown similarly that in the cases of $A_n$, $B_n$, $C_n$,
$D_n$ the functions $E_\lambda(x)$ are
eigenfunctions of the operators
\[
\sum_{w\in W_e} w\frac{\partial^2}{\partial x_i^2}, \qquad
 i=1,2,\dots ,r,
\]
where $x_1,x_2,\dots,x_r$ are orthogonal coordinates of the point
$x$, $r=n+1$ for $A_n$ and $r=n$ for other cases. In fact, these
operators are multiple to the Laplace operator $\Delta$.

\section[Decomposition of products of $E$-orbit functions]{Decomposition of products of $\boldsymbol{E}$-orbit functions}
\label{operat}

In this section we show how to decompose products of
$E$-orbit functions into sums of $E$-orbit
functions. Such operations are fulf\/illed by means of the
corresponding decompositions of $W_e$-orbits.

\subsection{Products of orbit functions}
\label{Prod-f}
Invariance of $E$-orbit functions $E_\lambda$ with respect to
the even Weyl group $W_e$
leads to the following statement:

\begin{proposition}\label{proposition1} A product of $E$-orbit functions
expands into a sum of $W_e$-orbit functions:
 \begin{equation}\label{dec-1}
 E_\lambda E_\mu =\sum _\nu n_\nu E_\nu,
 \end{equation}
where $n_\nu$ are non-negative integers, which
shows how many times the orbit function
$E_\nu$ is contained in the product $ E_\lambda E_\mu$.
\end{proposition}

\begin{proof} For $w\in W_e$ we have
\[
 E_\lambda(wx) E_\mu(wx)= E_\lambda(x)
 E_\mu(x).
\]
Therefore, the product $E_\lambda(x) E_\mu(x)$ is a f\/inite
sum of exponential functions, which
is invariant
with respect to $W_e$. Hence, it can be expanded into $E$-orbit
functions. Representing the product $E_\lambda(x) E_\mu(x)$
as a sum of exponential functions we see that these exponential
functions enter into this sum with positive integral
coef\/f\/icients. This means that coef\/f\/icients $n_\nu$ in
\eqref{dec-1} are non-negative integers. The proposition is
proved.
\end{proof}

Under termwise multiplication of $E$-orbit functions
$E_\lambda(x)$ and $E_\mu(x)$ we multiply exponential functions,
\[
e^{2\pi {\rm i}\langle \nu,x  \rangle}
e^{2\pi {\rm i}\langle \nu',x  \rangle}=
e^{2\pi {\rm i}\langle \nu+\nu',x  \rangle}.
\]
This reduces the procedure of multiplication of $E$-orbit
functions $E_\lambda(x)$ and $E_\mu(x)$  to the procedure of
multiplication of the corresponding $W_e$-orbits
$O_e(\lambda)$ and $O_e(\mu)$.
A product of these orbits is def\/ined as follows.

A product $O_e(\lambda)\otimes O_e(\lambda')$ of two $W_e$-orbits
$O_e(\lambda)$ and $O_e(\lambda')$ is the set of all points of the
form $\lambda_1+\lambda_2$, where $\lambda_1\in O_e(\lambda)$ and
$\lambda_2\in O_e(\lambda')$. Since a set of points
$\lambda_1+\lambda_2$, $\lambda_1\in O_e(\lambda)$, $\lambda_2\in
O_e(\lambda')$, is invariant with respect to action of the
corresponding even Weyl group $W_e$, each product of $W_e$-orbits is decomposable
into a sum of $W_e$-orbits. If $\lambda=0$, then $O_e(\lambda)\otimes
O_e(\lambda')=O_e(\lambda')$. If $\lambda'=0$, then $O_e(\lambda)\otimes
O_e(\lambda')=O_e(\lambda)$. In what follows we assume that
$\lambda\ne 0$ and $\lambda'\ne 0$.
Decomposition of products of two $W_e$-orbits into separate
$W_e$-orbits is not a simple task.

\subsection[Decomposition of products of $W_e$-orbits]{Decomposition of products of $\boldsymbol{W_e}$-orbits}\label{Decomp1}

If $O_e(\lambda)$ and $O_e(\lambda')$ are two $W_e$-orbits such that
$\lambda$ and $\lambda'$ are dominant and lie on walls of the
dominant Weyl chamber, then these orbits in fact coincide with the
corresponding $W$-orbits $O(\lambda)$ and $O(\lambda')$,
respectively. In this case we can apply to this product a procedure
of decomposition of a product of $W$-orbits (see Section~4 in~\cite{KP06}). Decomposing this product into $W$-orbits, we make
further the following. If a resulting $W$-orbit $O(\mu)$ is such
that $\mu$ lies on a wall of the dominant Weyl chamber, then
$O(\mu)$ is $W_e$-orbit, and $O_e(\mu)$ is contained in the product
$O_e(\lambda)\otimes O_e(\lambda')$ of $W_e$-orbits with a
multiplicity equal to a multiplicity of $W$-orbit $O(\mu)$ in the
product $O(\mu)\otimes O(\mu')$. If a resulting $W$-orbit $O(\mu)$
is such that $\mu$ does not lie on a wall of the dominant Weyl
chamber, then this $W$-orbit consists of two $W_e$-orbit $O_e(\mu)$
and $O(r_\alpha \mu)$. Moreover, multiplicities of these
$W_e$-orbits are determined as in the previous case. Thus, in our
case a decomposition of $W_e$-orbits is completely determined by a
decomposition of the corresponding $W$-orbits.

Let now the $W_e$-orbits $O_e(\lambda)$ and $O_e(\lambda')$ be such that
$\lambda$ is on a wall of the dominant Weyl chamber and $\lambda'$
is a strictly dominant element. Then $O_e(\lambda)$ can be considered
as $W$-orbit $O(\lambda)$. Instead of the $W_e$-orbit $O_e(\lambda')$
we consider the signed $W$-orbit $O^{\pm}(\lambda')$. (This signed orbit
consists of two $W_e$-orbits $O_e(\lambda)$ and $O_e(r_\alpha\lambda)$,
and points of $O_e(r_\alpha\lambda)$ are taken with the sign minus.)
Our problem of decomposition of the product
$O_e(\lambda)\otimes O_e(\lambda')$
is reduced to decomposition of the product
$O(\lambda)\otimes O^{\pm}(\lambda')$ of a $W$-orbit and a signed $W$-orbit.
Decompositions of such products are studied in Section~7 in \cite{KP07}.
The product $O(\lambda)\otimes O^{\pm}(\lambda')$ decomposes into
signed orbits, which are taken with sign plus or sign minus,
that is,
\[
O(\lambda)\otimes O^{\pm}(\lambda')=\bigcup_\mu n_\mu O^{\pm}(\mu),
\]
where $n_\mu$ are positive or negative integers.
Now a result for decomposition of the product
$O_e(\lambda)\otimes O_e(\lambda')$ can be formulated as follows.
If a sign orbit $O^{\pm}(\mu)$ is contained in the decomposition of
$O(\lambda)\otimes O^{\pm}(\lambda')$ with positive coef\/f\/icient
$n_\mu$, then
the product $O_e(\lambda)\otimes O_e(\lambda')$ contains the
$W_e$-orbit $O_e(\mu)$ with multiplicity $n_\mu$.
If an orbit $O^{\pm}(\mu)$ is contained in decomposition of
$O(\lambda)\otimes O^{\pm}(\lambda')$ with negative coef\/f\/icient
$n_\mu$, then
the product $O_e(\lambda)\otimes O_e(\lambda')$ contains the
$W_e$-orbit $O_e(r_\alpha \mu)$ with multiplicity $-n_\mu$.
This procedure determines in the decomposition
of $O_e(\lambda)\otimes O_e(\lambda')$ all $W_e$-orbits
$O_e(\mu)$ and $O_e(r_\alpha \mu)$ for which $\mu$ is strictly
positive and does not give $W_e$-orbits $O_e(\mu)$ with
$\mu$ lying on a wall of the dominant Weyl chamber.

In order to f\/ind in the decomposition
$O_e(\lambda)\otimes O_e(\lambda')$ the $W_e$-orbits $O_e(\mu)$ with
$\mu$ lying on a~wall of the dominant Weyl chamber, along with
the product $O(\lambda)\otimes O^{\pm}(\lambda')$ we have to
consider also the product $O(\lambda)\otimes O(\lambda')$ of
$W$-orbit functions (which are not signed orbits).

Below we consider the case of products $O_e(\lambda)\otimes O_e(\lambda')$
of $W_e$-orbits, when $\lambda$ and $\lambda'$ do not lie on
walls of even dominant Weyl chamber. For simplicity we consider the case
when $\lambda$ and $\lambda'$ are strictly dominant.

Let $O_e(\lambda)=\{ w\lambda | w\in W_e\}$ and $O_e(\mu)=\{
w\mu | w\in W_e\}$ be two $W_e$-orbits with strictly dominant
$\lambda$ and $\mu$. Then
 \begin{gather}
O_e(\lambda)\otimes O_e(\mu)= \{ w\lambda +w'\mu | w, w'\in
W_e\}\nonumber\\
 \label{decom}  \qquad
=\{ w\lambda{+}w_1\mu | w{\in}
W_e\}\cup \{ w\lambda{+}w_2\mu | w{\in} W_e\}\cup
\cdots \cup \{ w\lambda{+}w_s\mu | w{\in} W_e\},
 \end{gather}
where $w_1,w_2,\dots ,w_s$ is the set of elements of $W_e$.
Since a product of $W_e$-orbits is invariant with respect to $W_e$, for
decomposition of the product $O_e(\lambda)\otimes O_e(\mu)$ into
separate $W_e$-orbits it is necessary to separate
from each term of the right hand side of~\eqref{decom} elements,
which belong to the even Weyl chamber $D^e_+$. That is,
$O_e(\lambda)\otimes O_e(\mu)$ is a union of the $W_e$-orbits, determined by
points from
 \begin{equation} \label{decom1}
 D_e(\{ w\lambda+w_1\mu | w\in W_e\}),\ D(\{
w\lambda+w_2\mu | w\in W_e\}), \dots ,
D_e(\{ w\lambda+w_s\mu | w\in W_e\}),
 \end{equation}
where $D_e(\{ w\lambda+w_i\mu | w\in W_e\})$ means the set
of elements in the collection $\{ w\lambda+w_i\mu | w\in W_e\}$
belonging to $D_+^e$.

\begin{proposition}\label{proposition2}
Let $\lambda$ and $\mu$ be elements of
$D_+^e$, which do not lie on walls of $D_+^e$.
The product $O_e(\lambda)\otimes O_e(\mu)$
consists only of $W_e$-orbits of the form $O_e(|w\lambda+\mu|)$, $w\in
W_e$, where $|w\lambda+\mu|$ is an element of $D^e_+$ in the
$W_e$-orbit containing $w\lambda+\mu$. Moreover, each such $W_e$-orbit
$O_e(|w\lambda+\mu|)$, $w\in W_e$, belongs to the product
$O_e(\lambda)\otimes O_e(\mu)$.
\end{proposition}

\begin{proof} For each dominant element $w\lambda+w_i\mu$ from
\eqref{decom1} there exists an element $w''\in W_e$ such that $w''(
w\lambda+w_i\mu)=w'\lambda+\mu$. It means that $w\lambda+w_i\mu$
is of the form $|w'\lambda+\mu|$, $w'\in W_e$. Conversely,
take any element $w\lambda+\mu$, $w\in W_e$. It belongs to
the product $O_e(\lambda)\otimes O_e(\mu)$. This means that
$|w\lambda+\mu|$ also belongs to this product. Therefore, the
orbit $O_e(|w\lambda+\mu|)$ is contained in $O_e(\lambda)\otimes
O_e(\mu)$. Proposition is proved.
\end{proof}

It follows from Proposition~\ref{proposition2} that for decomposition of the
product $O_e(\lambda)\otimes O_e(\mu)$ into separate orbits we have to
take all elements $w\lambda+\mu$, $w\in W_e$, and to f\/ind
the corresponding dominant elements $|w\lambda+\mu |$, $w\in
W_e$.

For $\lambda$ and $\mu$ from Proposition~\ref{proposition2} the product
$O_e(\lambda)\otimes O_e(\mu)$ contains $W_e$-orbits with multiplicities~1, that is,
 \begin{equation} \label{decom2}
O_e(\lambda)\otimes O_e(\mu)=\bigcup_{w\in W_e}
O_e(|w\lambda+\mu |).
 \end{equation}
If $\lambda$ and $\mu$ lie on walls of $D_+^e$, then
some orbits can be contained in the decomposition of
$O_e(\lambda)\otimes O_e(\mu)$ with a multiplicity. The most dif\/f\/icult
problem under consideration of products of orbits is to f\/ind these
multiplicities.

Formulas \eqref{decom} and \eqref{decom1} are related to decompositions
of products $O_e(\lambda)\otimes O_e(\mu)$, when $\lambda$ and $\mu$
do not lie on walls of the domain $D_+^e$. Now we assume that $\lambda$
or/and $\mu$ may lie on walls of $D_+^e$. Then formula
\eqref{decom} is replaced by
 \begin{gather}
O_e(\lambda)\otimes O_e(\mu)= \{ w\lambda +w'\mu | w\in W_e/W_\lambda,
 w'\in W_e/W_\mu\}\nonumber\\
 \label{decom-ad}  \qquad
=\{ w{+}w_1\mu | w{\in}
W_e/W_\lambda\}\cup \{ w\lambda{+}w_2\mu | w{\in} W_e/W_\lambda\}\cup
\cdots \cup \{ w\lambda{+}w_r\mu | w{\in} W_e/W_\lambda\},\!\!
 \end{gather}
where $W_\lambda$ is the subgroup of $W_e$ consisting of
elements leaving $\lambda$ invariant and
$w_1,w_2,\dots ,w_r$ is the set of elements of $W_e/W_\mu$.
In this case,
$O_e(\lambda)\otimes O_e(\mu)$ is a union of the $W_e$-orbits, determined by
points from
 \begin{gather}
 D_e(\{ w\lambda+w_1\mu | w\in W_e/W_\lambda\}),\ D(\{
w\lambda+w_2\mu | w\in W_e/W_\lambda\}), \ \dots ,\nonumber\\
\qquad \qquad
D_e(\{ w\lambda+w_r\mu | w\in W_e/W_\lambda\}),\label{decom1-ad}
 \end{gather}
where $D_e(\{ w\lambda+w_i\mu | w\in W_e\})$ has the same sense as
in~\eqref{decom1}.

\begin{proposition} \label{proposition3} Let $O_e(\lambda)$ and $O_e(\mu)$,
$\lambda,\mu\in D_+^e$, be $W_e$-orbits
such that $\lambda\ne 0$ and $\mu\ne 0$, and let elements
$w\lambda+\mu$, $w\in W_e/W_\lambda$, belong to $D_e^+$
and do not belong to walls of even Weyl dominant
chamber $D_+^e$. Then
 \begin{equation} \label{decom3}
O_e(\lambda)\otimes O_e(\mu)=\bigcup_{w\in W_e/W_\lambda}
O_e(w\lambda+\mu).
 \end{equation}
 \end{proposition}

\begin{proof} Under the conditions of the proposition the set of
elements $w\lambda+\mu$, $w\in W_e/W_\lambda$, is contained in the
f\/irst set of \eqref{decom1-ad} if $w_1$ coincides with the identical
transformation. Moreover, $W_{w\lambda+\mu}
\equiv \{ w'\in W_e; w'(w\lambda+\mu)=w\lambda+\mu\}=
\{ 1\}$ for all
elements $w\in W_e$. Then $\mu$ does not
lies on the boundary of $D_e^+$,
that is, $W_\mu=\{ 1\}$. Let us show that
the collection \eqref{decom1-ad} contains only one non-empty set $ D(\{
w\lambda+\mu | w\in W_e/W_\lambda\})$. Indeed, let
$w\lambda+w_i\mu$, $w_i\ne 1$, belongs to $D_e^+$. Then
 \[
w_i^{-1}(w\lambda+w_i\mu)=w_i^{-1}w\lambda+\mu \in D_+^e.
 \]
Since $W_{w_i^{-1}w\lambda+\mu}=\{ 1\}$, then
$w_i^{-1}(w\lambda+w_i\mu)$ is an intrinsic point of $D_+^e$.
Therefore, $w\lambda+w_i\mu\not\in D_+^e$.
This contradicts to the condition that $w\lambda+w_i\mu\in D_+^e$ and,
therefore, the collection \eqref{decom1-ad} contains only one non-empty
set. The set of orbits, corresponding to the points of $D(\{
w\lambda+\mu|w\in W_e/W_\lambda \} )$, coincides with the right hand
side of \eqref{decom3}. Proposition is proved.
\end{proof}

\begin{proposition} \label{proposition4} If $\lambda,\mu\in D_+^e$, then
$O_e(\lambda)\otimes O_e(\mu)$
consists only of $W_e$-orbits of the form $O_e(|w\lambda+\mu|)$, $w\in
W_e/W_\lambda$, where $|w\lambda+\mu|$ is an element of $D^e_+$ in the
$W_e$-orbit containing $w\lambda+\mu$. Moreover, each such $W_e$-orbit
$O_e(|w\lambda+\mu|)$ belongs to the product
$O_e(\lambda)\otimes O_e(\mu)$.
\end{proposition}

\noindent
{\bf Proof} is similar to that of Proposition \ref{proposition2} and we omit it.
 \medskip

Under conditions of Proposition \ref{proposition4} the relation
\eqref{decom2} in general is not true.
The simplest counterexample is when $\mu=0$.
Then according to this formula $O_e(\lambda)\otimes
O_e(\mu)=O_e(\lambda)\cup O_e(\lambda) \cup \cdots \cup O_e(\lambda)$
($|W_e/W_\lambda$ times). However, as we know, $O_e(\lambda)\otimes
O_e(\mu)=O_e(\lambda)$.

Proposition \ref{proposition3} states that instead of \eqref{decom2} we have
 \begin{equation} \label{decom-0}
O_e(\lambda)\otimes O_e(\mu) \subseteq \bigcup_{w\in W_e}
O_e(|w\lambda+\mu |).
 \end{equation}
Note that some orbits on the right hand side can coincide.

\begin{proposition} \label{proposition5} Let $O_e(\mu)$ and $O_e(\mu)$ be $W_e$-orbits
such that $\lambda\ne 0$ and $\mu\ne 0$, and let all elements
$w\lambda+\mu$, $w\in W_e$, belong to $D^e_+$. Then
 \[
O_e(\lambda)\otimes O_e(\mu)=\bigcup_{w\in W/W_\lambda}
n_{w\lambda+\mu} O_e(w\lambda+\mu),
 \]
where $n_{w\lambda+\mu}=|W_{w\lambda+\mu}|$.
\end{proposition}

\begin{proof} Since $\lambda\ne 0$, all elements $w\lambda+\mu$,
$w\in W_e/W_\lambda$, belong to $D_e^+$ if and only if $W_\mu=\{ 1\}$.
Then on the right hand side of \eqref{decom-ad} there are $|W_e|$
terms. If the element $w\lambda+\mu$ belongs to~$D_e^+$ and
does not lie on a wall, that
is, $W_{w\lambda+\mu}=\{ 1\}$, then this element is met only in
one term. This means that the multiplicity $n_{w\lambda+\mu}$
of $O_e(w\lambda+\mu)$ in the product $O_e(\lambda)\otimes O_e(\mu)$
is 1. If $w\lambda+\mu$
is placed on some wall of $D_+^e$, then it is met in
$n_{w\lambda+\mu}=|W_{w\lambda+\mu}|$ terms (since the elements
$w'w\lambda+w'\mu$, $w'\in W_{w\lambda+\mu}$, belong to pairwise
dif\/ferent terms in \eqref{decom-ad}). Therefore, there are
$n_{w\lambda+\mu}$ orbits $O_e(w\lambda+\mu)$ in the decomposition
of $O_e(\lambda)\otimes O_e(\mu)$. Proposition is proved.
\end{proof}

\begin{proposition} \label{proposition6} If $W_\mu=\{ 1\}$ and none of the points
$w\lambda+\mu$, $w\in W_e/W_\lambda$, lies on a wall of
some even Weyl chamber, then
 \[
O_e(\lambda)\otimes O_e(\mu)=\bigcup_{w\in W_e/W_\lambda} O_e(|w\lambda+\mu|).
 \]
\end{proposition}

\begin{proof}For the product $O_e(\lambda)\times O_e(\mu)$ the inclusion
 \begin{equation} \label{decom-00}
O_e(\lambda)\otimes O_e(\mu) \subseteq \bigcup_{w\in W_e}
O_e(|w\lambda+\mu |)
 \end{equation}
holds. Each orbit $O_e(|w\lambda+\mu|)$,
$w\in W_e/W_\lambda$,
has $|W_e|$ elements and is contained in $O_e(\lambda)\otimes O_e(\mu)$.
Therefore, numbers of elements in both sides of \eqref{decom-00}
coincide. This means that the inclusion~\eqref{decom-00} is in fact
an equality. Proposition is proved.
\end{proof}

We formulate a conjecture concerning decomposition of products of
$W_e$-orbits.

 \medskip

\noindent
{\bf Conjecture.} {\it Let $O_e(\lambda)$ and $O_e(\mu)$,
$\lambda\ne 0$, $\mu\ne 0$, be orbits, and let $\mu$ belong to $D_e^+$
and do not lie on a wall of $D_+^e$.
If for some $w\in W_e$ the element $w\lambda+\mu$
belongs to $D_e^+$ and does not lie on a wall, then a multiplicity
of $O_e(w\lambda+\mu )$ in $O_e(\lambda)\otimes O_e(\mu)$ is $1$.}

 \medskip

At the end of this subsection we formulate a method
for decomposition of products $O_e(\lambda)\otimes O_e(\mu)$, which
follows from the statement of Proposition \ref{proposition4}. On the f\/irst step we
shift all points of the orbit $O_e(\lambda)$ by $\mu$. As a result,
we obtain the set of points $w\lambda+\mu$, $w\in W_e$. On
the second step, we map elements of this set, which do not belong to
$D_e^+$, by elements of the even Weyl group $W_e$ to the chamber $D_e^+$.
On this step we obtain the set $|w\lambda+\mu|$, $w\in W_e$. Then
by Proposition \ref{proposition4}, $O_e(\lambda)\otimes O_e(\mu)$ consists of
the orbits $O_e(|w\lambda+\mu|)$. On the third step, we determine
multiplicities of these orbits, taking into account the above
propositions or making direct calculations.

\subsection[Decomposition of products for $A_2$]{Decomposition of products for $\boldsymbol{A_2}$}\label{Decomp2}

We give examples of decompositions of products of
$W_e$-orbits for the cases $A_2$ and $C_2$. Orbits for these cases are
placed on a plane. Therefore, decompositions can be done by
geometrical calculations on this plane. These cases can be easily
considered also by using for orbit points orthogonal coordinates from
Section~\ref{s-orbi}.  The corresponding even Weyl groups have a simple description
in these coordinates and this gives a possibility to make
calculations in a simple manner.

For the case of $A_2$ at $a\ne b$, $a>0$, $b>0$, and at $c>0$ we have
\begin{alignat*}{3}
&A_2:\quad &O_e(a\ b)\otimes O_e(c\ 0) &= O_e(a{+}c\ b)\cup
O_e(a{-}c\ b{+}c) &&\\
        & & &\qquad \cup O_e({-}a{-}b{+}c\ a)     \quad &&(a>c>b),\\
            &&O_e(a\ b)\otimes O_e(c\ 0) &= O_e(a{+}c\ b)\cup O_e(a{-}c\ b{+}c) \cup O_e(a\ b{-}c)     \quad &&(a>c, b>c),\\
            &&O_e(a\ b)\otimes O_e(c\ 0) &= O_e(a{+}c\ b)\cup
                  O({-}a{-}b{+}c\ a))       \cup O_e(0\ a+b) \quad &&(a=c>b),\\
            &&O_e(a\ b)\otimes O_e(c\ 0) &= O_e(a{+}c\ b)\cup
                  O_e(a\ b{-}c)\cup O_e(0\ a+b) \ \   && (b>a=c),\\
            &&O_e(a\ b)\otimes O_e(c\ 0) &= O_e(a{+}c\ b)\cup
                  O_e(a{-}c\ b{+}c)\cup O_e(a\ 0)  && (a<b=c),\\
            &&O_e(a\ b)\otimes O_e(c\ 0) &= O_e(a{+}c\ b)\cup O_e(a{-}c\
            b{+}c)&&\\
        & & &\qquad \cup O_e(c{-}a{-}b\ a)     \quad &&(c>a+b),\\
            &&O_e(a\ b)\otimes O_e(c\ 0) &= O_e(a{+}c\ b)\cup O^\pm(a{-}c\
            b{+}c)&&\\
        & & &\qquad \cup O^\pm({-}a{-}b{+}c\ b)     \quad &&(a+b>c>b),\\
            &&O_e(a\ b)\otimes O_e(c\ 0) &= O_e(a{+}c\ b)\cup O_e(a{-}c\
            b{+}c) \cup O_e(a\ b{-}c)     \quad &&(a+b>b>c),
\end{alignat*}
\begin{alignat*}{3}
&A_2:\quad &O_e({-}a\ a{+}b)\otimes O_e(c\ 0) &= O_e({-}a{-}c\ a{+}b{+}c)\cup
        O_e(c{-}a\ a{+}b) &&\\
        & & &\qquad \cup O_e(a{+}b{-}c\ c{-}b)     \quad &&(a>c>b),\\
            &&O_e({-}a\ a{+}b)\otimes O_e(c\ 0) &= O_e({-}a{-}c\ a{+}b{+}c)
              \cup O_e(c{-}a\ a{+}b)&&\\
        & & &\qquad \cup O_e({-}a\ a{+}b{-}c)     \quad &&(a>c, b>c),\\
            &&O_e({-}a\ a{+}b)\otimes O_e(c\ 0) &= O_e({-}a{-}c\ a{+}b{+}c)\cup
                 O_e(a{+}b{-}c\ c{-}b))&&\\
         & & &\qquad          \cup O_e(0\ a+b) \quad &&(a=c>b),\\
            &&O_e({-}a\ a{+}b)\otimes O_e(c\ 0) &= O_e({-}a{-}c\ a{+}b{+}c)\cup
                  O_e({-}a\ a{+}b{-}c)&&\\
          & & &\qquad         \cup O_e(0\ a+b)   && (b>a=c),\\
            &&O_e({-}a\ a{+}b)\otimes O_e(c\ 0) &= O_e({-}a{-}c\ a{+}b{+}c)\cup
                  O_e(c{-}a\ a{+}b)&&\\
           & & &\qquad        \cup O_e(a\ 0)  && (a<b=c),\\
            &&O_e({-}a\ a{+}b)\otimes O_e(c\ 0) &= O_e({-}a{-}c\ a{+}b{+}c)\cup
                      O_e(c{-}a\      a{+}b)&&\\
        & & &\qquad \cup O_e(a{+}b{-}c\ c{-}b)     \quad &&(c>a+b),\\
            &&O_e({-}a\ a{+}b)\otimes O_e(c\ 0) &= O_e({-}a{-}c\ a{+}b{+}c)\cup
                      O_e(c{-}a\   a{+}b)&&\\
        & & &\qquad \cup O_e(a{+}b{-}c\ c{-}a)     \quad &&(a+b>c>b),\\
            &&O_e({-}a\ a{+}b)\otimes O(c\ 0) &= O_e({-}a{-}c\ a{+}b{+}c)\cup
                      O_e(c{-}a\    a{+}b)&&\\
        & & &\qquad \cup O_e({-}a\ a{+}b{-}c)     \quad &&(a+b>b>c).
\end{alignat*}

Note that $(-a\ a{+}b)$ belongs to $D_+^e$ and does not
belong to $D_+$.
If $a=b$, then we get
\begin{alignat*}{3}
&\qquad &O_e(a\ a)\otimes O_e(c\ 0) &= O_e(a{+}c\ a)\cup
O_e(c{-}2a\ a)  \cup O_e(a{-}c\ a{+}c)     \quad &&(c>2a),\\
            &&O_e(a\ a)\otimes O_e(c\ 0) &= O_e(a{+}c\ a)\cup O_e(c{-}2a\ a) \cup O_e(a{-}c\ a+c)     \quad &&(2a>c>a),\\
            &&O_e(a\ a)\otimes O_e(c\ 0) &= O_e(a{+}c\ a)\cup O_e(a{-}c\ a{+}c) \cup O_e(a\ a{-}c)     \quad &&(a>c).
\end{alignat*}
The $W_e$-orbit $O_e(r_\alpha (a\ a))$ has the form $O_e(-a\ 2a)$.
The products of such orbits with the $W_e$-orbit $O_e(c\ 0)$
decompose into $W_e$-orbits as
\begin{alignat*}{3}
&\qquad &O_e({-}a\ 2a)\otimes O_e(c\ 0) &= O_e({-}a{-}c\ 2a+c)\cup
O_e(2a{-}c\ c-a) &&\\
        & & &\qquad \cup O_e(c{-}a\ 2a)     \quad &&(c>2a),\\
            &&O_e({-}a\ 2a)\otimes O_e(c\ 0) &=
            O_e({-}a{-}c\ 2a+c)\cup O_e(2a{-}c\ c{-}a)&&\\
        & & &\qquad \cup O_e(c{-}a\ 2a)     \quad &&(2a>c>a),\\
            &&O_e({-}a\ 2a)\otimes O_e(c\ 0) &=
            O_e({-}a{-}c\ 2a+c)\cup O_e(c{-}a\ 2a)&&\\
        & & &\qquad \cup O_e({-}a\ 2a{-}c)     \quad &&(a>c).
\end{alignat*}

We also give the following decompositions for $W_e$-orbits of $A_2$:
\begin{alignat*}{3}
&A_2:\quad  &O_e(a\ 0)\otimes O_e(b\ 0) &= O_e(a{+}b\ 0)\cup O_e({-}a{+}b\ a)
                  \cup O_e(a{-}b\ b)            &&    \quad (a<b),\\
            &&O_e(a\ 0)\otimes O_e(a\ 0) &= O_e(2a\ 0)\cup 2O_e(0\ a),&&\\
            &&O_e(a\ 0)\otimes O_e(0\ b) &= O_e(a\ b)\cup O_e({-}a\ a{+}b)
                    \cup O_e(0\ {-}a{+}b)      &&    \quad (a<b),\\
            &&O_e(a\ 0)\otimes O_e(0\ a) &= O_e(a\ a)\cup
            O_e({-}a\ 2a)  \cup 3O_e(0\ 0).&&
 \end{alignat*}

\subsection[Decomposition of products for $C_2$]{Decomposition of products for $\boldsymbol{C_2}$}\label{Decomp3}

For dominant elements $(a\ b)$ products $W_e$-orbits for $C_2$ are of the form
\begin{alignat*}{2}
&C_2 :\quad &O_e(a\ b)\otimes O_e(c\ 0) &= O_e(a{+}c\ b)\cup
           O_e(a{+}2b{-}c\ {-}a{-}b{+}c) \cup O_e(a{-}c\ b{+}c) \\
        & & &\qquad \cup O_e(c{-}2b{-}a\ a{+}b)  \qquad\qquad (a{+}b{-}c{<}b),\\
          & &O_e(a\ b)\otimes O_e(c\ 0) &= O_e(a{+}c\ b)\cup
           O_e(a{+}2b{-}c\ c{-}a{-}b) \cup O_e(a{-}c\ b{+}c) \\
        & & &\qquad \cup O_e(c{-}a{-}2b\ a{+}b)  \qquad\qquad (b{>}c{-}a{-}b,a{>}c),\\
        & &O_e(a\ b)\otimes O_e(c\ 0) &= O_e(a{+}c\ b)\cup
           O_e(a{+}2b{-}c\ c{-}a{-}b) \cup O_e(a{-}c\ b{+}c) \\
        & & &\qquad \cup O_e(c{-}a{-}2b\ a{+}b)  \qquad\qquad (b{>}c{-}a{-}b,c{>}a),\\
        & &O_e(a\ b)\otimes O_e(c\ 0) &= O_e(a{+}c\ b)\cup
           O_e(a{-}c\ b) \cup O_e(a{-}c\ b{+}c) \\
        & & &\qquad \cup O_e(c{-}a{-}2b\ a{+}b)  \qquad\qquad (a{+}b{>}b{+}c),\\
        & &O_e(a\ b)\otimes O_e(c\ 0) &= O_e(a{+}c\ b)\cup
           O_e(a{-}c\ b) \cup O_e(a{-}c\ b{+}c) \\
        & & &\qquad \cup O_e(c{-}a{-}2b\ a{+}b)  \qquad\qquad (b{+}c{>}a{+}b{>}c{-}b),\\
        & &O_e(a\ b)\otimes O_e(c\ 0) &= O_e(a{+}c\ b)\cup
           O_e(a{-}c\ b) \cup O_e(a{-}c\ b{+}c) \\
        & & &\qquad \cup O_e(c{-}a{-}2b\ a{+}b)  \qquad\qquad (a{+}b{<}c{-}b).
\end{alignat*}
The $W_e$-orbit $O_e(r_\alpha (a\ b))$ has the form $O_e(-a\ a+b)$.
The products of such orbits with the $W_e$-orbit $O_e(c\ 0)$
decompose into $W_e$-orbits as
\begin{alignat*}{2}
&C_2:\quad &O_e(-a\ a{+}b)\otimes O_e(c\ 0) &= O_e({-}a{-}c\ a{+}b{+}c)\cup
           O_e({-}a{-}2b{+}c\ b) \cup O_e(c{-}a\ a{+}b) \\
        & & &\qquad \cup O_e(a{+}2b{-}c\ b{+}c)  \qquad\qquad (a{+}b{-}c{<}b),\\
          & &O_e(-a\ a{+}b)\otimes O_e(c\ 0) &= O_e({-}a{-}c\ a{+}b{+}c)\cup
           O_e(c{-}a{-}2b\ b) \cup O_e(c{-}a\ a{+}b) \\
        & & &\qquad \cup O_e(a{+}2b{-}c\ c{-}b)  \qquad\qquad (b{>}c{-}a{-}b,a{>}c),\\
        & &O_e(-a\ a{+}b)\otimes O_e(c\ 0) &= O_e({-}a{-}c\ a{+}b{+}c)\cup
           O_e(c{-}a{-}2b\ b) \cup O_e(c{-}a\ a{+}b) \\
        & & &\qquad \cup O_e(a{+}2b{-}c\ c{-}b)  \qquad\qquad (b{>}c{-}a{-}b,c{>}a),\\
        & &O_e(-a\ a{+}b)\otimes O_e(c\ 0) &= O_e({-}a{-}c\ a{+}b{+}c)\cup
           O_e(c{-}a\ a{+}b{-}c) \cup O_e(c{-}a\ a{+}b) \\
        & & &\qquad \cup O_e(a{+}2b{-}c\ c{-}b)  \qquad\qquad (a{+}b{>}b{+}c),\\
        & &O_e(-a\ a{+}b)\otimes O_e(c\ 0) &= O_e({-}a{-}c\ a{+}b{+}c)\cup
           O_e(c{-}a\ a{+}b{-}c) \cup O_e(c{-}a\ a{+}b) \\
        & & &\qquad \cup O_e(a{+}2b{-}c\ c{-}b)  \qquad\qquad (b{+}c{>}a{+}b{>}c{-}b),\\
        & &O_e(-a\ a{+}b)\otimes O_e(c\ 0) &= O_e({-}a{-}c\ a{+}b{+}c)\cup
           O_e(c{-}a\ a{+}b{-}c) \cup O_e(c{-}a\ a{+}b) \\
        & & &\qquad \cup O_e(a{+}2b{-}c\ c{-}b)  \qquad\qquad (a{+}b{<}c{-}b).
\end{alignat*}

We also give the following decompositions of $E$-orbits of $C_2$:
\begin{alignat*}{3}
&C_2 :\quad && O_e(a\ 0){\otimes} O_e(b\ 0)  = O_e(a{+}b\ 0)\cup O_e(a{-}b\ 0)
                  \cup O_e(a{-}b\ b)\cup O_e(b{-}a\ a) \quad &&(a>b),\\
            &&&O_e(a\ 0){\otimes} O_e(a\ 0)   =  O_e(2a\ 0)\cup 2O_e(0\ 2a)
                                      \cup 4O_e(0\ 0),\quad &&\\
            &&&O_e(0\ a){\otimes} O_e(0\ b)   =  O_e(0\ a{+}b){\cup} O_e(2b\ a{-}b)
                 {\cup} O_e({-}2b\ a{+}b) {\cup} O_e(0\ a{-}b)\quad &&(a>b),\\
            &&&O_e(0\ a){\otimes} O_e(0\ a)   =  O_e(0\ 2a)\cup 2O_e(2a\ 0)
                                      \cup 4O_e(0\ 0),\quad &&\\
            &&&O_e(a\ 0){\otimes} O_e(0\ b)   =  O_e(a\ b){\cup} O_e({-}a\ a{+}b)
{\cup} O_e(a{-}2b\ b) {\cup} O_e(2b{-}a\ a{-}b)       \quad &&(a>2b),\\
            &&&O_e(a\ 0){\otimes} O_e(0\ b)   =  O_e(a\ b){\cup} O_e({-}a\ a{+}b)
        {\cup} O_e(2b{-}a\ a{-}b) {\cup} O_e(a{-}2b\ b)  \quad &&(2b{>}a{>}b),\\
            &&&O_e(a\ 0){\otimes} O_e(0\ b)   =  O_e(a\ b) \cup O_e({-}a\ a{+}b)
         \cup O(a\ b{-}a) \cup O({-}a\ b)        \quad &&(b>a),\\
            &&&O_e(a\ 0){\otimes} O_e(0\ a)   =   O_e(a\ a)\cup O_e({-}a\ 2a)
              \cup 2O_e(a\ 0),&&{}\\
            &&&O_e(a\ 0){\otimes} O_e(0\ 2a)   =  O_e(a\ 2a)\cup O_e({-}a\ 3a)
      \cup O_e(a\ a) \cup O_e({-}a\ 2a),   &&{}\\
            &&&O_e(2a\ 0){\otimes} O_e(0\ a)   =  O_e(2a\ a) \cup O_e({-}2a\ 3a)
              \cup 2O_e(0\ a).&&   \end{alignat*}

\subsection[Decomposition of products for $G_2$]{Decomposition of products for $\boldsymbol{G_2}$}\label{Decomp4}

We give some examples of decomposition of products of $W_e$-orbits
of $G_2$ using $\omega$-coordinates for elements of orbits:
\begin{alignat*}{2}
&G_2 :\quad &&O_e(a\ 0)\otimes O_e(b\ 0)  = O_e(a{+}b\ 0)\cup O(b{-}a\ 3a)\cup O(2a{+}b\ {-}3a)
                 \cup O(2a{-}b\ 3b{-}3a)\\
            &&  &   \qquad\qquad\qquad\qquad  \qquad \cup O(b{-}a\ 3a{-}3b)  \cup O_e(b{-}a\ 0)\qquad \qquad    (a<b<2a),\\
            &&&O_e(a\ 0)\otimes O_e(b\ 0)  = O_e(a{+}b\ 0)\cup O_e(b{-}a\ 3a)\cup O_e(2a{+}b\ {-}3a)
                    \cup O(b{-}2a\ 3a)\\
             &&  &  \qquad\qquad\qquad\qquad  \qquad \cup O_e(a{+}a\ {-}3a)  \cup O_e(b{-}a\ 0)\quad \qquad (b>2a),\\
            &&&O_e(a\ 0)\otimes O_e(a\ 0)  = O_e(2a\ 0)\cup 2O_e(0\ 3a)
                    \cup 2O_e(a\ 0)\cup 6O_e(0\ 0),\\
            &&&O_e(a\ 0)\otimes O_e(2a\ 0)  = O_e(3a\ 0)\cup O_e(a\ 0)
                     \cup O_e(a\ 3a)\cup O_e(4a\ {-}3a) \cup 2O_e(0\ 3a),\\
            &&&O_e(0\ a)\otimes O_e(0\ b)  = O_e(0\ a{+}b)\cup O_e(a\ b{-}a) \cup O_e(b\ a{-}b)
                 \cup O_e(b{-}a\ 2a{-}b)\\
            &&  &    \qquad\qquad\qquad\qquad  \qquad \cup O_e(a\ b{-}2a)  \cup O_e(0\ b{-}a)  \qquad\qquad (a<b<2a),\\
            &&&O_e(0\ a)\otimes O_e(0\ b)  = O_e(0\ a{+}b)\cup O_e(0\ b{-}a)
                   \cup O_e(a\ b{-}a) \cup O_e(b\ a{-}b)  \\
            &&  &    \qquad\qquad\qquad\qquad  \qquad  \cup O_e(a\ b{-}2a) \cup O_e(b{-}a\ 2a{-}b)  \qquad\qquad  (b>2a),\\
            &&&O_e(0\ a)\otimes O_e(0\ a)  = O_e(0\ 2a)\cup 2O_e(a\ 0)
                    \cup 2O_e(0\ a)\cup 6O_e(0\ 0),\\
            &&&O_e(0\ a)\otimes O_e(0\ 2a)  = O_e(0\ 3a)\cup 2O_e(a\ 0)
                     \cup O_e(a\ a) \cup O_e(2a\ {-}a) \cup O_e(0\ a) .
\end{alignat*}

\section[Decomposition of $W_e$-orbit functions into
$W_e'$-orbit functions]{Decomposition of $\boldsymbol{W_e}$-orbit functions into
$\boldsymbol{W_e'}$-orbit functions}\label{sec7}

For these decompositions it is enough to obtain the
corresponding decompositions for signed $W$-orbits and then to make a
corresponding separation of $W_e$-orbits. For this reason,
we shall deal mainly with signed orbits. Our reasoning here is
similar to that of Section~4 in \cite{KP06}.

\subsection{Introduction}\label{Introd}

Let $R$ be a root system with a Weyl group $W$ and let $R'$ be
another root system which is a~subset of the set $R$. Then the Weyl
group $W'$ for $R'$ can be considered as a subgroup of $W$.
Moreover, $W'_e$ is a subgroup of the even Weyl group $W_e$.

Let $O_e^W(\lambda)$ be a $W_e$-orbit. The set of points of
$O_e^W(\lambda)$ is invariant with respect to $W_e'$. This
means that the orbit $O_e^W(\lambda)$ consists of
$W_e'$-orbits.  In this section we deal with representing~$O_e^W(\lambda)$ as a union of $W_e'$-orbits. Properties of
such a representation depend on root systems~$R$ and~$R'$ (or on Weyl
groups $W$ and $W'$). We distinguish two cases:
\medskip

\noindent
{\bf Case 1.} {\it Root systems $R$ and $R'$ span vector spaces of
the same dimension.} In this case Weyl chambers for $W$ are smaller
than Weyl chambers for~$W'$. Moreover, each Weyl chamber for~$W'$
consists of $|W/W'|$ chambers for~$W$. Therefore, an even Weyl chamber
for $W'_e$ consists of $|W/W'|=|W_e/W'_e|$ even Weyl chambers of
$W_e$.
Let $D_e^+$ be an even dominant Weyl chamber for the root
system $R$. Then the even dominant Weyl chamber for~$W'_e$ consists of $W_e$-chambers  $w_iD_e^+$, $i=1,2,\dots ,k$,
$k=|W/W'|$, where $w_i$, $i=1,2,\dots ,k$, are representatives of
cosets in $W_e/W_e'$. If $\lambda$ does not lie on any wall of the
even dominant Weyl chamber $D_e^+$, then
  \begin{equation} \label{Decompo}
 O^W_e(\lambda)=\bigcup_{i=1}^k
O^{W'}_e(w_i\lambda),
 \end{equation}
where $O_e^{W'}$ are $W_e'$-orbits.

Representing $\lambda$ by
coordinates in $\omega$-basis it is necessary to take into account
that coordinates of the same point in $\omega$-bases related to the
root systems $R$ and $R'$ are dif\/ferent. There exist matrices
connecting coordinates in these dif\/ferent $\omega$-bases (see
\cite{MPS77}).

To the decomposition \eqref{Decompo} there corresponds the
following expansion for $E$-orbit functions:
\[
E^{(W)}_\lambda(x)=\sum_{i=1}^k E_{w_i\lambda}^{(W')}(x).
\]

\noindent
{\bf Case 2.} {\it Root systems $R$ and $R'$ span vector spaces of
different dimensions.} This case is more complicated. In order to
represent $O^W_e(\lambda)$ as a union of $W_e'$-orbits, it is
necessary to project points $\mu$ of $O^W_e(\lambda)$ to the
vector subspace $E_{n'}$ spanned by $R'$ and to select in the set of
these projected points dominant points with respect to the root
system $R'$. Note that under projection, dif\/ferent points of
$O^W_e(\lambda)$ can give the same point in $E_{n'}$. This leads
to appearing of coinciding $W_e'$-orbits in a representation of
$O^W_e(\lambda)$ as a union of $W_e'$-orbits.

Under expansion of an $E^W$-orbit
function $E^{(W)}_\lambda(x)$ into $E^{(W')}$-orbit
functions we have to consider $E^{(W)}_\lambda(x)$ on the subspace
$E_{n'}\subset E_n$ and to take into account the corresponding
decomposition of the orbit $O^W_e (\lambda)$. For this
reason, below in this section we consider decomposition of
$W_e$-orbits into $W_e'$-orbits. They uniquely determine the
corresponding expansions for $E$-orbit functions.

\subsection[Decomposition of  $W_e(A_n)$-orbits  into
$W_e(A_{n-1})$-orbits]{Decomposition of  $\boldsymbol{W_e(A_n)}$-orbits  into
$\boldsymbol{W_e(A_{n-1})}$-orbits} \label{W-A_n}

Below we shall consider decompositions for $W_e(A_n)$-orbits
$O_e(\lambda)$ and $O_e(r_\alpha\lambda)$ such that $\lambda$ is
strictly dominant.

If $\lambda$ is not strictly dominant, then $W_e(A_n)$-orbit
$O_e(\lambda)\equiv O(\lambda)$ coincides with the $W(A_n)$-orbit~$O(\lambda)$.
In this case, in order to decompose $W_e(A_n)$-orbit $O_e(\lambda)$ into
$W_e(A_{n-1})$-orbits we have to decompose the
$W(A_n)$-orbit $O(\lambda)$ into $W(A_{n-1})$-orbits
and then to split $W(A_{n-1})$-orbits into $W_e(A_{n-1})$-orbits.
Namely, if a $W(A_{n-1})$-orbit $O(\mu)$ is such that $\mu$
is not strictly dominant, then $O(\mu)$ is in fact
a $W_e(A_{n-1})$-orbit. If $\mu$
is strictly dominant, then $O(\mu)$ consists of two
$W_e(A_{n-1})$-orbits $O_e(\mu)$ and $O(r_\alpha\mu)$,
where $\alpha$ is a root of $A_{n-1}$. Thus, if
$\lambda$ is not strictly dominant, then decomposition of
$W_e(A_n)$-orbits  into $W_e(A_{n-1})$-orbits are reduced
to decomposition of $W(A_n)$-orbits  into $W(A_{n-1})$-orbits.
The last decomposition are studied in Subsection 4.5 in~\cite{KP06}.

So, let $\lambda$ be a strictly dominant element for $A_n$.
It is convenient to fulf\/il the decomposition of
$W_e(A_n)$-orbits
$O_e(\lambda)$ and $O_e(r_\alpha\lambda)$
by using a corresponding decomposition of a signed $W(A_n)$-orbit
$O^\pm(\lambda)$  into signed $W(A_{n-1})$-orbits
considered in Subsection 8.2 in~\cite{KP07}. For this we have to take
into account that a signed $W(A_n)$-orbit $O^\pm(\lambda)$ consists of
two $W_e(A_n)$-orbits. One of them consists of points with the sign
$+$ and the second with the sign $-$.

For such decomposition it is convenient to represent orbit elements
in orthogonal coordinates (see Section~\ref{s-orbi}).
Let $m_1,m_2,\dots,m_{n+1}$ be orthogonal coordinates of a strictly
dominant element~$\lambda$, that is,
\[
m_1>m_2>\cdots >m_n>m_{n+1}.
\]
The orthogonal coordinates $m_1,m_2,\dots,m_{n+1}$ satisfy the
conditions $m_1+m_2+\cdots +m_{n+1}=0$. However, we may add to all
coordinates $m_i$ the same real number, since under this procedure the
$\omega$-coordinates $\lambda_i=m_i-m_{i+1}$, $i=1,2,\dots, n$ do
not change (see Section~\ref{s-orbi}).

Let $O^\pm(\lambda)\equiv
O^\pm(m_1,m_2,\dots,m_{n+1})$ be a signed $W(A_n)$-orbit with dominant
element $\lambda=(m_1,m_2,\dots, m_{n+1})$.
This orbit consists of all points
  \begin{equation} \label{A_n-1}
w(m_1,m_2,\dots ,m_{n+1})=(m_{i_1},m_{i_2},\dots,m_{i_{n+1}}),\qquad w\in W({A_n}),
 \end{equation}
where $(i_1,i_2,\dots ,i_{n+1})$ is a permutation of the numbers
$1,2,\dots,n+1$, determined by $w$. The sign of $(\det w)$ is attached to
such a point.
Points of $O^\pm(\lambda)$ belong to the Euclidean
space~$E_{n+1}$. We restrict these points to the subspace~$E_n$, spanned by the simple roots $\alpha_1,\alpha_2,\dots,
\alpha_{n-1}$ of~$A_n$, which form a set of simple roots of
$A_{n-1}$. This restriction is reduced to removing the last
coordinate $m_{i_{n+1}}$ in points
$(m_{i_1},m_{i_2},\dots,m_{i_{n+1}})$ of the signed orbit
$O^\pm(\lambda)$ (see \eqref{A_n-1}). As a~result, we obtain the set
of points
  \begin{equation} \label{A_n-2}
(m_{i_1},m_{i_2},\dots,m_{i_{n}})
  \end{equation}
received from the points \eqref{A_n-1}. The point \eqref{A_n-2} is
dominant if and only if
\[
m_{i_1}\ge m_{i_2}\ge \cdots \ge m_{i_{n}}.
\]
It is easy to see that after restriction to $E_{n}$ (that is, under
removing the last coordinate) we obtain from the set of points
\eqref{A_n-1} the following set of dominant elements:
\[
(m_1,\dots,m_{i-1},\hat{m_i},m_{i+1},\dots, m_{n+1}),\qquad
i=1,2,\dots,n+1,
\]
where a hat over $m_i$ means that the coordinate $m_i$ must be
omitted.

Thus, the signed $W(A_n)$-orbit
$O^\pm (m_1,m_2,\dots,m_{n+1})$ consists of the following signed
\linebreak $W(A_{n-1})$-orbits:
\[
O^\pm (m_1,\dots,m_{i-1},\hat{m_i},m_{i+1},\dots, m_{n+1}),\qquad
\ i=1,2,\dots,n+1.
\]
Each of these signed orbits must be taken with a coef\/f\/icient $+1$
or $-1$. Moreover, a coef\/f\/icient at the orbit
$O^\pm (m_1,\dots,m_{i-1},\hat{m_i},m_{i+1},\dots, m_{n+1})$
is $1$, if after $\hat{m_i}$ in the point
\[
(m_1,\dots,m_{i-1},\hat{m_i},m_{i+1},\dots, m_{n+1})
\]
there exists an even number of coordinates, and $-1$ otherwise
(see Section~8.2 in~\cite{KP07}).
These statements can be written in the form
  \begin{equation} \label{An-100}
O^\pm_W(m_1,m_2,\dots,m_{n+1})= \bigcup_{i=1}^{n+1} (\det w(m_i))
O_{W'}^\pm (m_1,\dots,m_{i-1},\hat{m_i},m_{i+1},\dots, m_{n+1}),
 \end{equation}
where $w(m_i)$ is the permutation which send the coordinate $m_i$
to the end, not changing an order of other coordinates.

Now we have to split the left and the right hand sides of
\eqref{An-100} into two parts: the f\/irst part has to consist
of points with the sign plus and the second with the sign minus.
This splitting  for the left hand side leads to two $W_e(A_n)$-orbits
$O_e(\lambda)$ and $O_e(r_\alpha\lambda)$. Each signed
$W(A_{n-1})$-orbit on the right hand side splits into two
$W(A_{n-1})$-orbits, one contains points with the sign plus and
other with the sign minus; one of them is contained in $O_e(\lambda)$
and another in~$O_e(r_\alpha\lambda)$. If \mbox{$\det w(m_i)=1$}, then
the $W_e(A_{n-1})$-orbit of
$O_{W'}^\pm (m_1,\dots,m_{i-1},\hat{m_i},m_{i+1},\dots, m_{n+1})$ with
points having the sign plus belongs to~$O_e(\lambda)$.
The $W_e(A_{n-1})$-orbit with
points having the sign minus belongs to~$O_e(r_\alpha\lambda)$.
If $\det w(m_i)=-1$, then the $W_e(A_{n-1})$-orbit of
$O_{W'}^\pm (m_1,\dots,m_{i-1},\hat{m_i},m_{i+1}$, $\dots, m_{n+1})$ with
points having the sign plus belongs to~$O_e(r_\alpha\lambda)$
and the $W_e(A_{n-1})$-orbit with
points having the sign minus belongs to~$O_e(\lambda)$.
Fulf\/illing this splitting we obtain lists of
$W_e(A_{n-1})$-orbits, which are contained in
$W_e(A_{n})$-orbits $O_e(\lambda)$ and~$O_e(r_\alpha\lambda)$.

\subsection[Decomposition of $W_e(B_{n})$-orbits into
$W_e(B_{n-1})$-orbits and of $W_e(C_{n})$-orbits into
$W_e(C_{n-1})$-orbits]{Decomposition of $\boldsymbol{W_e(B_{n})}$-orbits into
$\boldsymbol{W_e(B_{n-1})}$-orbits\\ and of $\boldsymbol{W_e(C_{n})}$-orbits into
$\boldsymbol{W_e(C_{n-1})}$-orbits}\label{W-C_n}

Decomposition of $W_e(B_{n})$-orbits and decomposition of
$W_e(C_{n})$-orbits are fulf\/illed in the same way. For this
reason, we give a proof only for the case of
$W_e(C_{n})$-orbits. As in the previous case for fulf\/illing
the decompositions we use signed $W$-orbits.

A set of simple roots of $C_n$ consists of roots
$\alpha_1,\alpha_2,\dots,\alpha_n$. The roots
$\alpha_2,\dots,\alpha_n$ constitute a set of simple
roots of $C_{n-1}$. They span the subspace $E_{n-1}$.

We determine elements $\lambda$ of $E_n$ by using orthogonal
coordinates $m_1,m_2,\dots,m_n$. Then $\lambda$ is strictly
dominant if and only if
\[
m_1>m_2>\cdots >m_n>0.
\]
Then the signed $W(C_n)$-orbit $O^\pm(\lambda)$ consists of all points
  \begin{equation} \label{C_n-1}
w(m_1,m_2,\dots ,m_{n})=(\pm m_{i_1},\pm m_{i_2},\dots,\pm
m_{i_{n}}),\qquad w\in W({C_n}),
 \end{equation}
where $(i_1,i_2,\dots ,i_{n})$ is a permutation of the set
$1,2,\dots,n $, and all combinations of signs are possible.

Restriction of elements \eqref{C_n-1} to the vector subspace
$E_{n-1}$, def\/ined above, reduces to removing  the f\/irst coordinate
$\pm m_{i_1}$ in \eqref{C_n-1}. As a result, we obtain from the set
of points \eqref{C_n-1} the collection
\[
(\pm m_{i_2},\pm m_{i_3},\dots,\pm m_{i_{n}}),\qquad w\in W(C_n).
\]
Only the points $(m_{i_2}, m_{i_3},\dots,m_{i_{n-1}}, m_{i_{n}})$
with positive coordinates may be dominant. Moreover, such a point
is dominant if and only if
\[
 m_{i_2}> m_{i_3}>\cdots > m_{i_{n}}.
\]
Therefore, under restriction of points \eqref{C_n-1} to $E_{n-1}$ we
obtain the following strictly $W({C_{n-1}})$-dominant elements:
  \begin{equation} \label{C_n-2}
(m_1,m_2,\dots ,m_{i-1},\hat{m_i},m_{i+1},\dots ,m_{n}),\qquad
i=1,2,\dots,n,
  \end{equation}
where a hat over $m_i$ means that the coordinate $m_i$ must be
omitted. Moreover, the element \eqref{C_n-2} with f\/ixed $i$ can be
obtained from two elements in \eqref{C_n-1}, namely, from
$(m_1,m_2,\dots ,m_{i-1},\pm m_i$, $m_{i+1},\dots ,m_{n})$.
In the signed orbit $O^\pm (m_1,m_2,\dots,m_n)$ these two
elements have  opposite signs.

Thus, the signed $W({C_n})$-orbits $O^\pm(m_1,m_2,\dots,m_n)$
consists of the following signed $W({C_{n-1}})$-orbits:{\samepage
  \begin{gather}
O^\pm(m_1,m_2,\dots ,m_{i-1},\hat{m_i},m_{i+1},\dots ,m_{n})\nonumber\\
\qquad{}\equiv
O^\pm(m_1,\dots ,\hat{m_i},\dots ,m_{n}),\qquad
i=1,2,\dots,n.\label{C_n-2*}
  \end{gather}}
Each such signed $W({C_{n-1}})$-orbit is contained in
$O^\pm (m_1,m_2,\dots,m_n)$ twice with opposite signs.
Since the signed $W({C_{n}})$-orbit $O^\pm(m_1,m_2,\dots,m_n)$
consists of two $W_e(C_{n})$-orbits (one consists of points with
the sign + and the second with the sign $-$), then it follows
from these assertions that {\it the $W_e(C_{n})$-orbits
$O_e (m_1,m_2,\dots,m_n)$ and $O_e (r_\alpha(m_1,m_2,\dots,m_n))$
consist of the same set of $W_e(C_{n-1})$-orbits}, namely,
of $W_e(C_{n-1})$-orbits which are contained in the signed
$W(C_{n-1})$-orbits \eqref{C_n-2*}. This set consists of the
$W_e(C_{n-1})$-orbits $O^\pm(m_1,,\dots ,\hat{m_i},\dots ,m_{n})$
and $O^\pm(r_\beta(m_1,,\dots ,\hat{m_i},\dots ,m_{n}))$,
$i=1,2,\dots,n$, where $\beta$ is a root of $C_{n-1}$.

For $W_e(B_n)$-orbits we have similar assertions. A
$W_e(B_n)$-orbit $O_e(m_1,m_2,\dots,m_n)$, $m_1>m_2>\cdots
>m_n>0$, consists of $W_e(B_{n-1})$-orbits which are contained
in the signed $W^\pm(B_n)$-orbit
\[
O^\pm (m_1,m_2,\dots ,m_{i-1},\hat{m_i},m_{i+1},\dots ,m_{n}),\qquad
i=1,2,\dots,n,
\]
and each such $W_e(B_{n-1})$-orbit is contained in the decomposition
only once.

\subsection[Decomposition of $W_e(D_{n})$-orbits into
$W_e(D_{n-1})$-orbits]{Decomposition of $\boldsymbol{W_e(D_{n})}$-orbits into
$\boldsymbol{W_e(D_{n-1})}$-orbits}\label{W-D_n}

Assume that $\alpha_1,\alpha_2,\dots,\alpha_n$ is the set of simple
roots of $D_n$, $n>4$. Then $\alpha_2,\dots,\alpha_n$ are simple roots of
$D_{n-1}$. The last roots span the subspace $E_{n-1}$.

For elements $\lambda$ of $E_n$ we use orthogonal coordinates
$m_1,m_2,\dots,m_n$. Then $\lambda$ is strictly dominant if and
only if $m_1> m_2> \cdots > m_{n-1}> |m_n|$. We assume that
$\lambda$ satisf\/ies the condition
\[
m_1>m_2>\cdots >m_n>0.
\]
Then the signed $W(D_n)$-orbit $O^\pm(\lambda)$ consists of all points
  \begin{equation} \label{D_n-1}
w(m_1,m_2,\dots ,m_{n})=(\pm m_{i_1},\pm m_{i_2},\dots,\pm
m_{i_{n}}),\qquad w\in W_{D_n},
 \end{equation}
where $(i_1,i_2,\dots ,i_{n})$ is a permutation of the numbers
$1,2,\dots,n $ and there exists an even number of signs $-$.
Restriction of elements \eqref{D_n-1} to the subspace $E_{n-1}$
reduces to removing the f\/irst coordinate $\pm m_{i_1}$ in
\eqref{D_n-1}. As a result, we obtain from the set of points
\eqref{D_n-1} the collection
\[
(\pm m_{i_2},\pm m_{i_3},\dots,\pm m_{i_{n}}),\qquad w\in W({D_n}),
\]
where a number of signs $-$ may be either even or odd. Only
points of the form $(m_{i_2}, m_{i_3},\dots$, $m_{i_{n-1}},\pm
m_{i_{n}})$ may be dominant. Moreover, such a point is dominant if
and only if
\[
 m_{i_2}> m_{i_3}>\cdots >m_{i_{n-1}}> |m_{i_{n}}|.
\]
Therefore, under restriction of points \eqref{D_n-1} to $E_{n-1}$
we obtain the following $W({D_{n-1}})$-dominant elements:
  \begin{equation} \label{D_n-2}
(m_1,m_2,\dots ,m_{i-1},\hat{m_i},m_{i+1},\dots ,m_{n-1},\pm
m_{n}),\qquad i=1,2,\dots,n,
  \end{equation}
where a hat over $m_i$ means that the coordinate $m_i$ must be
omitted. Moreover, the element \eqref{D_n-2} with f\/ixed $i$ can be
obtained only from one element in \eqref{D_n-1}, namely, from
element $(m_1,m_2,\dots ,m_{i-1},\pm m_i,m_{i+1},\dots ,\pm
m_{n})$, where at $m_i$ and $m_n$ signs are coinciding.

Thus, the signed $W({D_n})$-orbit $O^\pm(m_1,m_2,\dots,m_n)$
with $m_1>m_2>\cdots >m_{n}>0$ consists of the following signed
$W({D_{n-1}})$-orbits:
  \begin{equation} \label{D_n-2*}
O^\pm(m_1,m_2,\dots ,m_{i-1},\hat{m_i},m_{i+1},\dots ,\pm m_{n}),\qquad i=1,2,\dots,n.
 \end{equation}
Each such signed $W({D_{n-1}})$-orbit is contained in
$O^\pm(m_1,m_2,\dots,m_n)$ only once $($with sign $+$ or sign $-$).
A sign of such an orbit depends on a number $i$ and does not
depend on a sign at
$m_n$. This sign is $+$ if after $\hat m_i$ in \eqref{D_n-2*}
there exists on even number of coordinates and $-1$ otherwise.

Now we split the signed $W({D_n})$-orbit $O^\pm(m_1,m_2,\dots,m_n)$
into two $W_e(D_n)$-orbits $O_e(m_1$, $m_2, \dots,m_n)$ and
$O_e(r_\alpha(m_1,m_2,\dots,m_n))$. Next, we split each of the signed
$W({D_{n-1}})$-orbits \eqref{D_n-2*} also into two
$W_e(D_{n-1})$-orbits
  \begin{gather} \label{D_n-2**}
O_e(m_1,m_2,\dots ,m_{i-1},\hat{m_i},m_{i+1},\dots ,\pm m_{n}),
\qquad i=1,2,\dots,n.
\\
 \label{D_n-2***}
O^e(r_\beta(m_1,m_2,\dots ,m_{i-1},\hat{m_i},m_{i+1},\dots ,\pm m_{n})),\qquad i=1,2,\dots,n.
 \end{gather}
It is necessary to split these $W_e(D_{n-1})$-orbits into two parts which
constitute the $W_e(D_{n})$-orbits
$O_e(m_1,m_2,\dots,m_n)$ and
$O_e(r_\alpha(m_1,m_2,\dots,m_n))$. This is done as follows.
If a f\/ixed signed orbit from \eqref{D_n-2*} is contained in the signed
$W({D_{n}})$-orbit $O^{\pm}(m_1,m_2,\dots,m_n)$ with the sign +,
then the corresponding $W_e(D_{n-1})$-orbit
\eqref{D_n-2**} is contained in the $W_e(D_{n})$-orbit
$O_e(m_1,m_2,\dots,m_n)$ and the
$W_e(D_{n-1})$-orbit \eqref{D_n-2***} is contained in
$O_e(r_\alpha(m_1,m_2,\dots,m_n))$. If
a~f\/ixed signed $W({D_{n-1}})$-orbit \eqref{D_n-2*} is contained
in $O^{\pm}(m_1,m_2,\dots,m_n)$ with the sign $-$,
then~the corresponding
$W_e(D_{n-1})$-orbit
\eqref{D_n-2**} is contained in the $W_e(D_{n})$-orbit
$O_e(r_\alpha(m_1,m_2$, $\dots,m_n))$ and the
$W_e(D_{n-1})$-orbit \eqref{D_n-2***} is contained in
$O_e(m_1,m_2,\dots,m_n)$. Therefore, the $W_e(D_{n})$-orbit
$O_e(m_1,m_2, \dots,m_n)$ consists of
$W_e(D_{n-1})$-orbit \eqref{D_n-2**} such that $n-i$ is an even
integer and of $W_e(D_{n-1})$-orbit \eqref{D_n-2***}
such that $n-i$ is odd. The
$W_e(D_{n})$-orbit
$O_e(r_\alpha(m_1,m_2$, $\dots,m_n))$ consists of
$W_e(D_{n-1})$-orbit \eqref{D_n-2**} such that $n-i$ is odd
and of $W_e(D_{n-1})$-orbit \eqref{D_n-2***}
such that $n-i$ is even.

It is shown similarly that {\it the $W_e(D_n)$-orbits
\[
O_e(m_1,\dots,m_{n-1},-m_n), \qquad
O_e(r_\alpha(m_1,\dots,m_{n-1},-m_n)),
\]
with $m_1>m_2>\dots >m_{n}>0$
consists of the same $W_e(D_{n-1})$-orbits as the
$W_e(D_n)$-orbits $O_e(m_1,\dots,m_{n-1},m_n)$ and
$O_e(r_\alpha(m_1,\dots,m_{n-1},m_n))$ with the same
numbers $m_1,\dots,m_{n-1},m_n$ do, respectively.}

\section[$E$-orbit function transforms]{$\boldsymbol{E}$-orbit function transforms}\label{sec8}

As in the case of symmetric and antisymmetric orbit functions, $E$-orbit
functions determine certain orbit function transforms which
generalize the Fourier transform (in the case of symmetric orbit
functions these transforms generalize the cosine transform and in the
case of antisymmetric orbit
functions these transforms generalize the sine transform)
\cite{KP06,KP07,MP06}.

As in the case of symmetric and antisymmetric orbit functions, $E$-orbit
functions determine three types of orbit function transforms: the
f\/irst one is related to the $E$-orbit functions
$E_\lambda(x)$ with integral $\lambda$, the second one is related
to $E_\lambda(x)$ with real values of coordinates of $\lambda$, and
the third one is the related discrete transform.

\subsection[Expansion in $E$-orbit functions on $F_e$]{Expansion in $\boldsymbol{E}$-orbit functions on $\boldsymbol{F_e}$}

The aim of this subsection is to obtain formulas for expansions of
functions on the closure of the fundamental domain $F_e$ of the
even af\/f\/ine Weyl group $W^{\rm af\/f}_e$ in $E$-orbit functions
$E_\lambda(x)$ with integral $\lambda$.

Let us start with the usual Fourier expansion of functions on $E_n$,
 \begin{equation}\label{decomp-e}
f(x)=\sum_{\lambda\in {\mathbb Z}^n} c_\lambda e^{2\pi \langle \lambda,x\rangle }.
 \end{equation}
with coef\/f\/icients
\begin{equation}\label{decomp-e-c}
c_\lambda=\int_{x\in {\sf T}} f(x) e^{-2\pi \langle \lambda,x\rangle }dx,
 \end{equation}
where ${\sf T}$ is a torus in $E_n$.

Let the function $f(x)$ be invariant with respect to the even Weyl
group $W_e$. It is easy to check that the coef\/f\/icients $c_\lambda$
are also $W_e$-invariant, $c_{w\lambda}=c_\lambda$, $w\in W_e$.
Replace in \eqref{decomp-e-c} $\lambda$ by~$w\lambda$, $w\in W_e$,
and sum up both side of \eqref{decomp-e-c} over $w\in W_e$. Then
instead of \eqref{decomp-e-c} we obtain
 \[
c_\lambda=|W_e|^{-1} \int_{\sf T} f(x) \hat E_\lambda(x)dx,
  \]
where $\hat E_\lambda(x) =\sum\limits_{w\in W_e} e^{2\pi{\rm i}\langle w\lambda,x
\rangle}$. (We have taken into account that both $f(x)$ and $E_\lambda(x)$ are
$W_e$-invariant.) This formula can be written as
 \begin{equation}\label{F-3-}
c_\lambda= \int_{\overline{F_e}} f(x) \hat E_\lambda(x)dx=
|W_\lambda| \int_{\overline{F_e}} f(x)  E_\lambda(x)dx,
  \end{equation}
where $W_\lambda$ is the subgroup of $W_e$ consisting of elements
leaving $\lambda$ invariant.

Similarly, starting from \eqref{decomp-e}, we obtain an inverse formula:
 \begin{equation}\label{F-4-}
 f (x)= \sum_{\lambda\in P^+_e} c_\lambda
 \overline{E_\lambda(x)} ,
  \end{equation}
where $P^+_e$ is the set of integral elements from $D^+_e$.
For the transforms \eqref{F-3-} and \eqref{F-4-} the Plancherel formula
 \[
 \int_{\overline{F_e}} |f(x)|^2 dx=\sum_{\lambda\in P^+_e}
 |W_\lambda|^{-1} |c_\lambda|^2
 \]
holds, which means that the Hilbert spaces with the appropriate scalar
products are isometric.

Formula \eqref{F-4-} is the symmetrized (by means of the group
$W_e$) Fourier transform of
the function $f(x)$. Formula \eqref{F-3-} gives an inverse
transform. These formulas  give the
{\it $E$-orbit function transforms} corresponding to $E$-orbit
functions $E_\lambda$, $\lambda\in P^+_e$.

Let ${\mathcal L}^2(F_e)$ denote the Hilbert space of functions on
the closure of the fundamental domain~$F_e$ of the group $W_e^{\rm af\/f}$
with the scalar product
\[
\langle f_1,f_2\rangle = \int_{\overline{F_e}} f_1(x)\overline{f_2(x)} dx.
\]
The formulas \eqref{F-3-} and \eqref{F-4-}
show that {\it the set of $E$-orbit functions $E_\lambda$,
$\lambda\in P^+_e$, form an orthogonal basis of ${\mathcal
L}^2(F_e)$.}

\subsection[$E$-orbit function transform on even dominant Weyl chamber]{$\boldsymbol{E}$-orbit function transform on even dominant Weyl chamber}

The expansion \eqref{F-4-} of functions on the
domain $\overline{F_e}$ is an expansion in the $E$-orbit functions~$E_\lambda (x)$ with integral elements $\lambda$.  The $E$-orbit
functions $E_\lambda (x)$ with $\lambda$ lying in the
even dominant Weyl chamber (and not obligatory integral) are not
invariant with respect to the corresponding even af\/f\/ine Weyl group
$W_e^{\rm af\/f}$. They are invariant only with respect to the even Weyl
group $W_e$. A~fundamental domain of $W_e$ coincides with the even dominant
Weyl chamber $D_e^+=D_+\cup r_\alpha D_+$. For this reason, the $E$-orbit functions
$E_\lambda (x)$, $\lambda\in D^+_e$, determine another orbit
function transform (a transform on $D_e^+$).

We began with the usual Fourier transforms on ${\mathbb R}^n$:
 \begin{gather}\label{F-1}
\tilde f (\lambda)=\int_{{\mathbb R}^n} f(x) e^{2\pi {\rm i}\langle
\lambda,x \rangle} dx,
\\
\label{F-2}
 f (x)=\int_{{\mathbb R}^n} \tilde f(\lambda) e^{-2\pi {\rm i}\langle
\lambda,x \rangle} d\lambda.
  \end{gather}
Let the function $f(x)$ be invariant with respect to the even Weyl
group $W_e$, that is, $f(wx)=(\det w)f(x)$, $w\in W_e$.
The function $\tilde f (\lambda)$ is also invariant with
respect to the even Weyl group~$W_e$. Replace in~\eqref{F-1} $\lambda$ by
$w\lambda$, $w\in W_e$, and sum up these
both side over $w\in W_e$. Then instead of~\eqref{F-1} we obtain
\[
\tilde f (\lambda)=|W_e|^{-1} \int_{\mathbb{R}^n} f(x)
\hat E_\lambda(x) dx,\qquad \lambda\in D^+_e,
\]
Therefore,
 \begin{equation}\label{F-3}
\tilde f (\lambda)= \int_{D_e^+} f(x) \hat E_\lambda(x) dx,\qquad
\lambda\in D^+_e,
  \end{equation}
where we have taken into account that $f(x)$
is invariant with respect to $W_e$.

Similarly, starting from \eqref{F-2}, we obtain the inverse formula:
 \begin{equation}\label{F-4}
 f (x)= \int_{D_e^+} \tilde f(\lambda)
 \overline{\hat E_\lambda(x)} d\lambda .
  \end{equation}
For the transforms \eqref{F-3} and \eqref{F-4} the Plancherel
formula
 \[
 \int_{D_e^+} |f(x)|^2 dx=
\int_{D_e^+} |\tilde f(\lambda) |^2  d\lambda
 \]
holds.

\section[Finite $E$-orbit function transforms]{Finite $\boldsymbol{E}$-orbit function transforms}\label{sec9}

\subsection{Introduction}\label{sec9.1}
It is possible to introduce  f\/inite
$E$-orbit function transforms, based on $E$-orbit functions (see~\cite{MP06}$\!$).
It is done in the same way as in the case of symmetric orbit
functions in \cite{KP06} by using the results of paper \cite{MP87}.
Finite $E$-orbit function transforms are generalizations
of the f\/inite (discrete) Fourier transforms, which are def\/ined as
follows.

Let us f\/ix a positive integer $N$ and consider the numbers
 \begin{equation}\label{f-F-1}
e_{mn}:=N^{-1/2}\exp (2\pi {\rm i}mn/N),\qquad m,n=1,2,\dots,N.
 \end{equation}
The matrix $(e_{mn})_{m,n=1}^N$ is unitary, that is,
 \begin{equation}\label{f-F-2}
\sum_k e_{mk}\overline{e_{nk}} =\delta_{mn},\qquad \sum_k
e_{km}\overline{e_{kn}} =\delta_{mn}.
 \end{equation}

Let $f(n)$ be a function of $n\in \{ 1,2,\dots ,N\}$. We may
consider the transform
 \begin{equation}\label{f-F-3}
\sum_{n=1}^N f(n)e_{mn}\equiv N^{-1/2} \sum_{n=1}^N f(n) \exp
(2\pi{\rm i}mn/N) =\tilde f (m).
 \end{equation}
Then due to unitarity of the matrix $(e_{mn})_{m,n=1}^N$, we
express $f(n)$ as a linear combination of conjugates of
the functions \eqref{f-F-1}:
 \begin{equation}\label{f-F-4}
f(n)= N^{-1/2} \sum_{m=1}^N {\tilde f}(m) \exp (-2\pi{\rm i}mn/N)
.
 \end{equation}
The function ${\tilde f}(m)$ is a {\it finite Fourier transform}
of $f(n)$. This transform is a linear map. The
formula \eqref{f-F-4} gives an inverse transform. The Plancherel
formula
\[
\sum_{m=1}^N |\tilde f(m)|^2=\sum_{n=1}^N | f(n)|^2
\]
holds for transforms \eqref{f-F-3} and \eqref{f-F-4}. This means
that the f\/inite Fourier transform conserves the norm introduced on
the space of functions on $\{ 1,2,\dots,N\}$.

The f\/inite Fourier transform on the $r$-dimensional linear space
$E_r$ is def\/ined similarly. We again f\/ix a positive integer $N$.
Let ${\bf m}=(m_1,m_2,\dots,m_r)$ be an $r$-tuple of integers such
that each $m_i$ runs over the integers $1,2,\dots,N$. Then the
f\/inite Fourier transform on $E_r$ is given by the kernel
\[
e_{\bf mn}:=e_{m_1n_1}e_{m_2n_2}\cdots e_{m_rn_r}=N^{-r/2} \exp
(2\pi{\rm i}{\bf m\cdot n}/N),
\]
where ${\bf m\cdot n}=m_1n_1+m_2n_2+\cdots +m_rn_r$. If $F({\bf
m})$ is a function of $r$-tuples ${\bf m}$, $m_i\in\{
1,2,\dots,N\}$, then the f\/inite Fourier transform of $F$ is given
by
\[
{\tilde F}({\bf n})=N^{-r/2}\sum_{\bf m}F({\bf m}) \exp (2\pi{\rm
i}{\bf m\cdot n}/N).
\]
The inverse transform is
\[
F({\bf m})=N^{-r/2}\sum_{\bf n}{\tilde F}({\bf n}) \exp (-2\pi{\rm
i}{\bf m\cdot n}/N).
\]
The corresponding Plancherel formula is of the form $\sum\limits_{\bf m}
|F({\bf m})|^2=\sum\limits_{\bf n} |\tilde F ({\bf n})|^2$.

\subsection[Grids on the fundamental domain $F_e$]{Grids on the fundamental domain $\boldsymbol{F_e}$}

In order to determine an analogue of the f\/inite Fourier transform,
based on $E$-orbit functions, we need an analogue of the
set
\[
\{ {\bf m}=\{ m_1,m_2,\dots,m_n\}\ |\  m_i\in \{1,2,\dots,N\}\},
\]
used for multidimensional f\/inite Fourier transform. Such a set has
to be invariant with respect to the even Weyl group $W_e$ (see \cite{MP87}).

We know that the coroot lattice $Q^\vee$ is a
discrete $W$-invariant subset of $E_n$.
Clearly, the set $\frac 1m Q^\vee$ is also $W$-invariant, where
$m$ is a f\/ixed positive integer. Then the set
\[
T_m={\textstyle \frac 1m} Q^\vee / Q^\vee
\]
is f\/inite and $W$-invariant. If $\alpha_1,\alpha_2,\dots
,\alpha_l$ is the set of simple root for the Weyl group $W$, then~$T_m$ can be identif\/ied with the set of elements
 \begin{equation}\label{lat-1}
m^{-1}\sum_{i=1}^l d_i\alpha_i^\vee,\qquad d_i=0,1,2,\dots, m-1.
 \end{equation}

We select from $T_m$ the set of elements which belongs to the closure
$\overline{F}_e$ of the fundamental domain $F_e$. These elements lie
in the collection $\frac 1m Q^\vee \cap \overline{F}_e$.

Let $\mu\in \frac 1m Q^\vee \cap \overline{F}_e$ be an element determining an
element of $T_m$ and let $M$ be the least positive integer such
that $M\mu \in P^\vee$. Then there exists the least positive
integer $N$ such that $N\mu\in Q^\vee$. One has $M | N$ and $N |
m$.

The collection of points of $T_m$ belonging to $\overline{F}$ (we denote
the set of these points by $F_M$), where $F$ is the fundamental
domain of the Weyl group $W$, coincides with the set of
elements
\begin{equation}\label{fin-orb-1}
s=\tfrac{s_1}M \omega^\vee_1+\cdots +\tfrac{s_l}M \omega^\vee_l,  \qquad
 \omega_i^\vee =\frac{2\omega_i}{\langle \alpha_i,\alpha_i \rangle},
 \end{equation}
where $s_1,s_2,\dots,s_l$ runs over values from $\{ 0,1,2,\dots \}$
and satisfy the following condition: there
exists a non-negative integer $s_0$ such that
\begin{equation}\label{fin-orb-2}
s_0+\sum_{i=1}^l s_im_i=M,
 \end{equation}
where $m_1,m_2,\dots, m_l$ are non-negative integers from formula
\eqref{highestroot} (see \cite{MP87}). (Values of $m_i$ for all simple Lie algebras
can be found in Subsection~\ref{sec2.4}.)

To every positive integer $M$ there corresponds the grid $F_M$ of
points \eqref{fin-orb-1} in $\overline{F}$ which corresponds to some set
$T_m$ such that $M | m$. The precise relation between $M$ and $m$
can be def\/ined by the grid $F_M$ (see~\cite{MP06}) . Acting upon
the grid $F_M$ by elements of the Weyl group $W$ we obtain the
whole set $T_m$. Below, we are interested in grids $F_M$ and do
not need the corresponding numbers $m$.

For studying f\/inite $E$-orbit function transforms we need grids
$F^e_M$ such that $T_m$ is obtained by action by elements of $W^{\rm af\/f}_e$.
In order to obtain $F_M^e$ we f\/ix a positive root $\alpha$ and construct the
the set $F_M\cup r_\alpha F_M$. The set $F_M\cap r_\alpha F_M$
can be non-empty. Taking each point from $F_M\cup r_\alpha F_M$
only  once we obtain the grid $F^e_M$. The set $\cup_{w\in W^{\rm af\/f}_e} wF^e_M$
coincides with $T_m$, where some points are taken several
times. The set $F^e_M$ depends on a choice of a root $\alpha$.

\subsection[Grids $F^e_M$ for $A_2$, $C_2$ and $G_2$]{Grids $\boldsymbol{F^e_M}$ for $\boldsymbol{A_2}$, $\boldsymbol{C_2}$ and $\boldsymbol{G_2}$}

In this section we give some examples of grids $F^e_M$ for the rank
two cases (see \cite{Kash-P}). Since
the long root $\xi$ of $A_2$ is representable in the form
$\xi=\alpha_1+\alpha_2$, where $\alpha_1$ and $\alpha_2$ are simple roots,
that is, $m_1=m_2=1$ (see formula \eqref{fin-orb-2}), then
\[
 F_M(A_2)=\left\{ \tfrac{s_1}M \omega_1+\tfrac{s_2}M \omega_2;\
s_0+s_1+s_2=M,\ s_0,s_1,s_2\in {\mathbb Z}^{\ge 0} \right\}.
 \]
A direct computation shows that in the $\omega$-coordinates we have
\[
F_2(A_2)=\left\{ (0,0),(1,0),(0,1),(\tfrac12,0),(0,\tfrac12),(\tfrac12,\tfrac12)
\right\}.
 \]
We take a root $\alpha$ coinciding with the f\/irst simple root.
Then
\[
r_\alpha F_2(A_2)=\left\{ (0,0),(-1,1),(0,1),(-\tfrac12,\tfrac12),
(0,\tfrac12),(-\tfrac12,1).
\right\}.
 \]
Thus, the grid $F^e_2(A_2)$ consists of dif\/ferent points from
$F_2(A_2) \cup r_\alpha F_2(A_2)$ (9 points).

For $F_3(A_2)$ we have
 \[
F_3(A_2)=\left\{ (0,0),(1,0),(0,1),(\tfrac13,0),(0,\tfrac13),(\tfrac23,0),
(0,\tfrac23),(\tfrac23,\tfrac13),(\tfrac13,\tfrac23),(\tfrac13,\tfrac13)\right\}.
\]
Therefore,{\samepage
 \[
r_\alpha F_3(A_2)=\left\{ (0,0),(-1,1),(0,1),(-\tfrac13,\tfrac13),
(0,\tfrac13),(-\tfrac23,\tfrac23),
(0,\tfrac23),(-\tfrac23,1),(-\tfrac13,1),(-\tfrac13,\tfrac23)\right\}
\]
and $F^e_3(A_2)$ consists of 16 points.}

Since the long root $\xi$ of $C_2$ is representable in the form
$\xi=2\alpha_1+\alpha_2$, where $\alpha_1$ and $\alpha_2$ are simple roots,
that is, $m_1=2,m_2=1$, then
\[
 F_M(C_2)=\left\{ \tfrac{s_1}M \omega^\vee_1+\tfrac{s_2}M \omega^\vee_2; \
s_0+2s_1+s_2=M,\  s_0,s_1,s_2\in {\mathbb Z}^{\ge 0} \right\}.
  \]
A direct computation shows that in the $\omega^\vee$-coordinates we have
\begin{gather*}
F^e_2(C_2)=\left\{ (0,0),(0,1),(\tfrac12,0),(0,\tfrac12), (-\tfrac12,\tfrac12)
\right\},
\\
F^e_3(C_2)=\left\{ (0,0),(0,1),(\tfrac13,0),(0,\tfrac13),
(0,\tfrac23),(\tfrac13,\tfrac13), (-\tfrac13,\tfrac13),(-\tfrac13,\tfrac23)
\right\}.
\end{gather*}

Since the long root $\xi$ of $G_2$ is representable in the form
$\xi=2\alpha_1+3\alpha_2$, where $\alpha_1$ and $\alpha_2$ are simple roots,
that is, $m_1=2$, $m_2=3$, then
\[
 F_M(G_2)=\left\{ \tfrac{s_1}M \omega^\vee_1+\tfrac{s_2}M \omega^\vee_2; \
s_0+2s_1+3s_2=M,\  s_0,s_1,s_2\in {\mathbb Z}^{\ge 0} \right\}.
 \]
A computation shows that in the $\omega^\vee$-coordinates we have
\begin{gather*}
F^e_2(G_2)=\left\{ (0,0),(1,0),(-1,3)\right\},
\\
F^e_3(G_2)=\left\{ (0,0),(0,\tfrac13),(\tfrac13,0),(-\tfrac13,1)\right\},
\\
F^e_4(G_2)=F^e_2(G_2)\bigcup\left\{ (\tfrac14,0),(0,\tfrac14),
(-\tfrac14,\tfrac34\right\},
\\
F^e_5(G_2)=\left\{ (0,0),(0,\tfrac15),(\tfrac15,0),(\tfrac15,\tfrac15),
(\tfrac25,0),(-\tfrac15,\tfrac35),(-\tfrac15,\tfrac45),(-\tfrac25,\tfrac65) \right\},
\\
F^e_8(G_2)=F^e_4(G_2)\bigcup\left\{(\tfrac18,0),(0,\tfrac18),
(\tfrac18,\tfrac18),(\tfrac14,\tfrac18),(-\tfrac18,\tfrac38),(-\tfrac18,\tfrac12),
(-\tfrac14,\tfrac78) \right\}.
\end{gather*}

\subsection[Expansion in $E$-orbit functions through expansion on grids]{Expansion in $\boldsymbol{E}$-orbit functions through expansion on grids}

Let us give an analogue of the f\/inite Fourier transform when instead
of exponential functions we use $E$-orbit functions. This analogue
is not so simple as f\/inite Fourier transform. It is called the
f\/inite $E$-orbit function transform. This transform is used in order
to be able to recover (at least approximately) the expansion
$f(x)=\sum_\lambda a_\lambda E_\lambda(x)$ for continuous values of
$x$ by values of~$f(x)$ on a f\/inite set of point.

Under considering the f\/inite Fourier transform in Section~\ref{sec9.1}, we
have restricted the exponential function to a discrete set.
Similarly, in order to determine f\/inite transform, based on
$E$-orbit functions, we have to restrict $E$-orbit functions
$E_\lambda(x)$ to appropriate f\/inite sets of values of~$x$.
Candidates for such f\/inite sets are sets~$T_m$. However, $E$-orbit
functions $E_\lambda(x)$ with integral~$\lambda$ are invariant with
respect to the af\/f\/ine even Weyl group $W_e^{\rm af\/f}$. For this
reason, we consider $E$-orbit functions $E_\lambda(x)$ on grids
$F^e_M$, which are parts of the sets $T_m$.

On the other side, we have also to choose a f\/inite number of
$E$-orbit functions, that is, a f\/inite number of
integral $\lambda\in P_+^e$. The best choose is when a number of
$E$-orbit functions coincides with the number $|F^e_M|$ of elements
in $F^e_M$. These
$E$-orbit functions must be selected in such a~way that the matrix
\begin{equation}\label{determi}
\left( E_{\lambda_i}(x_j)\right)_{\lambda_i\in \Omega,x_j\in
F_M}
\end{equation}
(where $\Omega$ is our f\/inite set of integral elements
$\lambda$) is not singular.
In order to have
non-singularity of this matrix some conditions must be satisf\/ied.
In general, they are not known. For this reason, we consider some,
more weak, form of the transform (when $|\Omega|\ge |F^e_M|$) and then
explain how the set $|\Omega|$ of $\lambda\in P^e_+$ can be chosen in
such a way that $|\Omega|= |F^e_M|$.

Let $O_e(\lambda)$ and $O_e(\mu)$ be two dif\/ferent $W_e$-orbits for
elements $\lambda$ and $\mu$ of $P^e_+$. We say that the
group $T_m$ {\it separates} $O_e(\lambda)$ and $O_e(\mu)$ if for any two
elements $\lambda_1\in O_e(\lambda)$ and $\mu_1\in O_e(\mu)$ there
exists an element $x\in T_m$ such that $\exp (2\pi{\rm i}\langle
\lambda_1,x \rangle) \ne \exp (2\pi{\rm i}\langle \mu_1,x \rangle)$.
Note that $\lambda$ may coincide with $\mu$.

Let $f_1$ and $f_2$ be two functions on the Euclidean space $E_n$
which are f\/inite linear combinations of $E$-orbit functions. We
introduce a $T_m$-scalar product for $f_1$ and $f_2$ by the formula
\[
\langle f_1,f_2 \rangle_{T_m}=\sum_{x\in T_m}
f_1(x)\overline{f_2(x)} .
\]
Then the following proposition is true (see \cite{MP87} and
\cite{MP06}):

\begin{proposition} \label{proposition7}  If $T_m$ separates the orbits
$O_e(\lambda)$ and $O_e(\mu)$, $\lambda,\mu\in P^e_+$, then
\begin{equation}\label{Fuir}
\langle E_\lambda,E_\mu \rangle_{T_m}=m^n
|O_e(\lambda)|\delta_{\lambda\mu}.
 \end{equation}
 \end{proposition}

\begin{proof} We have
\begin{alignat*}{2}
\langle E_\lambda,E_\mu \rangle_{T_m}=& \sum_{x\in T_m}
\sum_{\sigma\in O_e(\lambda)} \sum_{\tau \in O_e(\mu)}
\exp (2\pi{\rm i}\langle \sigma-\tau,x\rangle)
\notag\\
\qquad\qquad\qquad  =& \sum_{\sigma\in O_e(\lambda)}
\sum_{\tau \in O_e(\mu)}  \left( \sum_{x\in T_m} \exp
(2\pi{\rm i}\langle \sigma-\tau,x\rangle)\right).
\notag
\end{alignat*}
Since $T_m$ separates $O_e(\lambda)$ and $O_e(\mu)$, then none of the
dif\/ferences $\sigma-\tau$ in the last sum vanishes on $T_m$.
Since $T_m$ is a group and $|T_m|=m^n$, one has
\[
\sum_{x\in T_m} \exp (2\pi{\rm i}\langle
\sigma-\tau,x\rangle)=m^n\delta_{\sigma,\tau}.
\]
Therefore, $\langle E_\lambda,E_\mu \rangle_{T_m}=m^n
|O_e(\lambda)|\delta_{\lambda\mu}$. The proposition is proved.
\end{proof}

Let $f$ be an invariant (with respect to $W_e^{\rm af\/f}$) function
on the Euclidean space $E_n$ which is a~f\/inite linear combination of
$E$-orbit functions:
\begin{equation}\label{Fuir-2}
 f(x)=\sum_{\lambda_j\in
P^e_+} a_{\lambda_j}E_{\lambda_j}(x).
 \end{equation}
Our aim is to determine $f(x)$ by its values on a f\/inite subset of
$E_n$, namely, on $T_m$.

We suppose that $T_m$ separate orbits $O_e(\lambda_j)$ with
$\lambda_j$ from the right hand side of \eqref{Fuir-2}. Then taking
the $T_m$-scalar product of both sides of \eqref{Fuir-2} with
$E_{\lambda_j}$ and using the relation \eqref{Fuir} we obtain
\[
a_{\lambda_j}=\left( m^n |O_e(\lambda_j)|\right)^{-1}\langle f,
E_{\lambda_j} \rangle_{T_m}.
\]
Let now $s^{(1)},s^{(2)},\dots,s^{(h)}$ be all elements of
$\overline{F}_e\cap
\frac 1m Q^\vee$. By $W_{s^{(i)}}$ we denote the subgroup of $W_e$
whose elements leave $s^{(i)}$ invariant. Then
 \begin{equation}\label{Fuir-3}
a_{\lambda_j}=m^{-n}|O_e(\lambda_j)|^{-1}\sum_{x\in T_m} f(x)
\overline{E_{\lambda_j}(x)}= m^{-n} |W_{\lambda_j}| \sum_{i=1}^h
|W_{s^{(i)}}|^{-1} f(s^{(i)})
\overline{\varphi_{\lambda_j}(s^{(i)})} .
 \end{equation}

Thus, {\it the finite number of values $f(s^{(i)})$,
$i=1,2,\dots,h$, of the function $f(x)$ determines the
coefficients $a_{\lambda_j}$ and, therefore, determines the
function $f(x)$ on the whole space $E_n$.}

This means that we can reconstruct a $W_e^{\rm af\/f}$-invariant
function $f(x)$ on the whole Euclidean space $E_n$ by its values on
the f\/inite set $F_M$ under an appropriate value of $M$. Namely, we
have to expand this function, taken on $F_M$, into the series
\eqref{Fuir-2} by means of the coef\/f\/icients $a_{\lambda_j}$,
determined by formula \eqref{Fuir-3}, and then to continue
analytically the expansion \eqref{Fuir-2} to the whole fundamental
domain $F_e$ (and, therefore, to the whole space $E_n$), that is, to
consider the decomposition \eqref{Fuir-2} for all $x\in E_n$.

We have assumed that the function $f(x)$ is a f\/inite linear
combination of $E$-orbit functions. If $f(x)$ expands into inf\/inite sum
of orbit functions, then for applying the above procedure we have to
approximate the function $f(x)$ by taking a f\/inite number of terms
in this inf\/inite sum and then to apply the procedure. That is, in this
case we obtain an approximate expression of the function $f(x)$ by
using a f\/inite number of its values.

At last, we explain how to choose a set $\Omega$ in formula
\eqref{determi}. The set $F_M$ consists of the points~\eqref{fin-orb-1}. These points determines the set $\Xi$ of points
 \[
\lambda=s_1\omega_1+s_2\omega_2+\cdots +s_l\omega_l,
 \]
where $s_1,s_2,\dots,s_l$ run over the same values as for the set
$F_M$. The set $\Xi\cup r_\alpha \Xi$, where each point is taken
only once, can be taken as the set $\Omega$ (see \cite{MP06}).

\section[$W_e$-symmetric functions]{$\boldsymbol{W_e}$-symmetric functions}\label{sec10}

$E$-orbit functions are symmetrized
versions of the exponential function, when symmetrization
is fulf\/illed by an even Weyl group $W_e$. Instead of the exponential function we
can take any other set of functions, for example, a set of orthogonal
polynomials or a countable set of functions. Then we obtain a
corresponding set of orthogonal $W_e$-symmetric  polynomials or a
set of $W_e$-symmetric functions. Such sets of polynomials
and functions are considered in this section.


\subsection[Symmetrization by $E$-orbit functions]{Symmetrization by $\boldsymbol{E}$-orbit functions}

$E$-orbit functions can be used for obtaining $W_e$-symmetric sets
of functions. Let $u_m(x)$, $m=0,1,2,\dots$, be a set of continuous
functions of one variables. We create functions of $n$ variables
\[
u_{i_1,i_2,\dots,i_n}(x_1,x_2,\dots,x_n)\equiv u_{i_1}(x_1)
u_{i_2}(x_2)\cdots u_{i_n}(x_n),\qquad i_k=0,1,2,\dots.
\]
Then the functions
\begin{equation}\label{func-s}
{\tilde u}_{i_1,i_2,\dots,i_n}(\lambda_1,\lambda_2,\dots,\lambda_n)=
\int_{F_e} u_{i_1,i_2,\dots,i_n}(x_1,x_2,\dots,x_n)
E_{\lambda}(x_1,x_2,\dots,x_n) dx,
\end{equation}
where $\lambda\equiv (\lambda_1,\lambda_2,\dots,\lambda_n)$,
$E_{\lambda}(x)$ is a $E$-orbit function, and $dx$ is the Euclidean
measure on $E_n$ (that is, $dx=dx_1\cdots dx_n$), is symmetric with
respect to the action of the even Weyl group $W_e$. Indeed, for
$w\in W_e$ we have
\begin{alignat}{2}
{\tilde u}_{i_1,i_2,\dots,i_n}(w\lambda)&= \int_{F_e}
u_{i_1,i_2,\dots,i_n}(x_1,x_2,\dots,x_n)
E_{w\lambda}(x_1,x_2,\dots,x_n)\, d x
\notag\\
&=\int_{F_e} u_{i_1,i_2,\dots,i_n}(x_1,x_2,\dots,x_n)
E_{\lambda}(x_1,x_2,\dots,x_n)\, d x={\tilde
u}_{i_1,i_2,\dots,i_n}(\lambda). \notag \end{alignat}

Formula \eqref{func-s} is used for obtaining $W_e$-symmetric
functions or polynomials.

If $u_m(x)$, $m=0,1,2,\dots$, are orthogonal functions, then the
functions \eqref{func-s}, taken for $i_1\ge i_2\ge \cdots\ge i_n$,
constitute a set of $W_e$-symmetric orthogonal functions on
the domain $D_+^e$.

\subsection[Eigenfunctions of $E$-orbit function transform
for $W_e(A_n)$]{Eigenfunctions of $\boldsymbol{E}$-orbit function transform
for $\boldsymbol{W_e(A_n)}$}

Let $H_n(x)$, $n=0,1,2,\dots$, be the well-known Hermite polynomials.
They are def\/ined by the formula
\[
H_n(x)=n!\sum_{m=0}^{[n/2]} \frac{(-1)^m (2x)^{n-2m}}{m!(n-2m)!},
\]
where $[n/2]$ is an integral part of the number $n/2$.
They satisfy the relation
\[
\frac1{\sqrt{2\pi}} \int_{-\infty}^\infty e^{{\rm
i}px}e^{-p^2/2}H_m(p)dp={\rm i}^{-m}e^{-x^2/2}H_m(x)
\]
(see, for example, Subsection 12.2.4 in \cite{KVII}), which can be
written in the form
 \begin{equation}\label{Herm-1}
 \int_{-\infty}^\infty e^{2\pi {\rm i}px}e^{-\pi p^2}
 H_m(\sqrt{2\pi}p)dp={\rm i}^{m}e^{-\pi x^2}H_m(\sqrt{2\pi}x).
 \end{equation}
This relation shows that the function $e^{-\pi p^2}
 H_m(\sqrt{2\pi}p)$ is an eigenfunction of the Fourier transform of
 one variable with eigenvalue ${\rm i}^{m}$.

Using the Hermite polynomials we create polynomials of many
variables
\begin{equation}\label{Her-m}
H_{\bf m}({\bf  x})\equiv H_{m_1,m_2,\dots,m_n}(x_1,x_2,\dots
,x_n):=H_{m_1}(x_1)H_{m_2}(x_2)\cdots H_{m_n}(x_n).
 \end{equation}
The functions
\begin{equation}\label{Her-2}
 e^{-|{\bf  x}|^2/2} H_{\bf m}({\bf  x}),\qquad
m_i=0,1,2,\dots,\qquad i=1,2,\dots,n,
 \end{equation}
form an orthogonal basis of the Hilbert space $L^2(\mathbb{R}^n)$
with the scalar product{\samepage
\[
\langle f_1,f_2\rangle:=\int_{\mathbb{R}^n}
f_1(\mathbf{x})\overline{f_2(\mathbf{x})}d\mathbf{x} ,
\]
where $d\mathbf{x}=dx_1\,dx_2\cdots dx_n$.}

We make $W_e$-symmetrization of the functions
\[
e^{-\pi |x|} H_{\bf m}(\sqrt{2\pi} {\bf x}),\qquad m_i=0,1,2,\dots ,
\]
(obtained from \eqref{Her-2} by replacing ${\bf x}$ by
$\sqrt{2\pi}{\bf x}$) by means of $E$-orbit functions of $A_{n-1}$:
\begin{equation}\label{Her-3}
\int_{\mathbb{R}^n} E_\lambda(\mathbf{x})
 e^{-\pi|{\bf  x}|^2} H_{\bf m}(\sqrt{2\pi}{\bf  x})=
 {\rm i}^{|{\bf m}|} e^{-\pi|\lambda|^2}
 \mathcal{H}_{\bf m}(\sqrt{2\pi}\lambda),
\end{equation}
where $E_\lambda(\mathbf{x})$ is an $E$-orbit function of $A_{n-1}$
and  $\lambda=(\lambda_1,\lambda_2,\dots,\lambda_n)$.

The polynomials $\mathcal{H}_{\bf m}$  are
symmetric with respect to the even Weyl
group $W_e \equiv S_n/S_2:=S^e_n$ of $A_{n-1}$:
\[
\mathcal{H}_{\bf m}(w\lambda)=\mathcal{H}_{\bf m}(\lambda),\qquad
\mathcal{H}_{w\bf m}(\lambda)=\mathcal{H}_{\bf m}(\lambda), \qquad
 w\in S^e_n.
\]
For this reason, $\mathcal{H}_{\bf m}(\lambda)$ can be considered
for values of $\lambda=(\lambda_1,\lambda_2,\dots,\lambda_n)$ such
that $\lambda_1\ge \lambda_2\ge \cdots \ge \lambda_n$.

The polynomials $\mathcal{H}_{\bf m}$ are of the form{\samepage
 \begin{equation}\label{sym-H}
\mathcal{H}_{\bf m}(\lambda)=\sum_{w\in S^e_n} H_{w\bf m}(\lambda),
 \end{equation}
where the polynomials $H_{w\bf m}(\lambda)$ are of the form
\eqref{Her-m}.}

Now we apply $E$-orbit function transform \eqref{F-3} (we denote this
transform by $\mathfrak{F}$)
to the $W_e$-symmetric function \eqref{sym-H}. Taking into
account formula \eqref{Her-3} we obtain
\begin{alignat*}{2}
\mathfrak{F}\left(  e^{-\pi|{\bf  x}|^2} \mathcal{H}_{\bf m}
(\sqrt{2\pi}{\bf x})\right):=& \; \frac2{|S_n|} \int_{\mathbb{R}^n}
E_\lambda({\bf x}) e^{-\pi|{\bf  x}|^2} \mathcal{H}_{\bf m}
(\sqrt{2\pi}{\bf x}) d{\bf x}
 \notag\\
=& \; {\rm i}^{|{\bf m}|} e^{-\pi|\lambda|^2}
 \mathcal{H}_{\bf m}(\sqrt{2\pi}\lambda),
 \notag
\end{alignat*}
where $|S_n|$ is an order of the permutation group $S_n$, that is,
functions \eqref{sym-H} are eigenfunctions of the $E$-orbit
function transform $\mathfrak{F}$. Since the functions \eqref{sym-H}
for $m_i=0,1,2,\dots$, $i=1,2,\dots,n$, $m_1\ge m_2\ge \cdots \ge m_n$,
form an orthogonal basis
of the Hilbert space $L_{\rm sym}^2(\mathbb{R}^n)$
of functions from~$L^2(\mathbb{R}^n)$ symmetric with respect to $W_e$,
then they constitute a complete set of
eigenfunctions of this transform. Thus, this transform has only four
eigenvalues ${\rm i}$, $-{\rm i}$, $1$, $-1$ in $L_{\rm sym}^2(\mathbb{R}^n)$.
This means that, as in the
case of the usual Fourier transform, we have
\[
\mathfrak{F}^4=1.
\]

\subsection[$W_e(A_n)$-symmetric sets of polynomials]{$\boldsymbol{W_e(A_n)}$-symmetric sets of polynomials}

In the previous subsection we constructed $W_e$-symmetric sets of
functions connected with Hermite polynomials. Other sets of
orthogonal polynomials can be similarly constructed.

Let $p_m(x)$, $m=0,1,2,\dots$, be the set of orthogonal polynomials
in one variable such that
\[
\int_{\mathbb{R}} p_m(x)p_{m'}(x)d\sigma(x) =\delta_{mm'},
\]
where $d\sigma(x)$ is some orthogonality measure, which may be
continuous or discrete.

We create a set of symmetric polynomials of $n$ variables as follows:
 \begin{gather}\label{sym-p}
p^{\rm sym}_{\bf m}({\bf x})= \sum_{w\in S^e_n/S_{\bf m}} p_{m_{w(1)}}(x_1)
p_{m_{w(2)}}(x_2)\cdots p_{m_{w(n)}}(x_n),
\\
 m_i=0,1,2,\dots,\qquad i=1,2,\dots,n,\nonumber
\end{gather}
where ${\bf m}=(m_1,m_2,\dots,m_n)$, $m_1\ge m_2\ge
\cdots \ge m_n\ge 0$, ${\bf x}=(x_1,x_2,\dots,x_n)$, and
$w(1),w(2)$, $\dots,w(n)$ is a set of numbers $1,2,\dots,n$
transformed by the permutation $w\in S^e_n/S_{\bf m}$, where
$S_{\bf m}$ is the subgroup of $S_n$ consisting of elements
leaving ${\bf m}$ invariant.

It is easy to check that the polynomials $p^{\rm sym}_{\bf m}({\bf
x})$ are symmetric with respect to transformations of $S^e_n$:
\[
p^{\rm sym}_{\bf m}(w{\bf x}) = p^{\rm sym}_{\bf m}({\bf x}),\qquad
w\in S_n.
\]
Thus, we may consider the polynomials \eqref{sym-p}
on the closure of the fundamental domain of the transformation group
$W_e(A_{n-1})\equiv S^e_n$. This closure (which is denoted as
$D^e_+$) coincides with the set of points ${\bf
x}=(x_1,x_2,\dots,x_n)$ for which
\[
x_1,x_2\ge \cdots \ge x_n.
\]

The set of polynomials \eqref{sym-p}
is orthogonal with respect to the
product measure $d\sigma({\bf x})\equiv d\sigma(x_1)\,
d\sigma(x_2)\cdots d\sigma(x_n)$. Indeed, we have
\[
\int_{D^e_+} p^{\rm sym}_{\bf m}({\bf x}) \overline{p^{\rm sym}_{{\bf
m}'}({\bf x})}d\sigma({\bf x})=\frac{|O_e({\bf m})|}{|S^e_n|}
\delta_{{\bf mm}'}=\frac1{|S_{\bf m}|} \delta_{{\bf mm}'},
\]
where $O_e({\bf m})$ is the $S^e_n$-orbit of the point ${\bf m}$.

\subsection*{Acknowledgements}

The f\/irst author acknowledges CRM of University of Montreal for
hospitality when this paper was under preparation. We are grateful
for partial support for this work from the National Science and
Engineering Research Council of Canada, MITACS, the MIND Institute
of Costa Mesa, California, and Lockheed Martin, Canada.

\pdfbookmark[1]{References}{ref}
\LastPageEnding

\end{document}